\documentstyle[epsf,12pt]{report}
\input{epsf.tex}
\newcommand{\al}{\alpha'}
\newcommand{\de}{\partial}
\newcommand{\be}{\begin{equation}}
\newcommand{\ba}{\begin{eqnarray}}
\newcommand{\ea}{\end{eqnarray}}
\newcommand{\ee}{\end{equation}}

\newcommand{\ca}{\mathcal}

\newcommand{\f}{\frac}
\newcommand{\s}{\sqrt}
\newcommand{\ti}{\tilde}
\newcommand{\ap}{\alpha}

\newcommand{\mb}{\mathbf}
\newcommand{\ddd}{\cdot\cdot\cdot}
\newcommand{\dd}{\cdot\cdot}
\newcommand{\no}{\nonumber \\}
\newcommand{\la}{\langle}
\newcommand{\lb}{\rangle}
\newcommand{\ep}{\epsilon}
\newcommand{\zz}{{\mathbf Z}_2}
\newcommand{\z}{{\mathbf Z }}

\newcommand{\beq}{\begin{equation}}
\newcommand{\eeq}{\end{equation}}
\newcommand{\beqa}{\begin{eqnarray}}
\newcommand{\eeqa}{\end{eqnarray}}

\newcommand{\pa}{\partial}
\newcommand{\A}{\alpha}
\newcommand{\B}{\beta}

\renewcommand{\ss}{\scriptscriptstyle}

\begin{document}
\begin{titlepage}
\thispagestyle{empty}
\begin{flushright}
{\large December, 2002} 
\end{flushright}

\vspace{3cm}
\begin{center}
\noindent{\Huge \textbf{Worldsheet Description of Tachyon\\
\vspace{8mm}
Condensation in Open String Theory}}\\
\vspace{3cm}
\noindent{\LARGE
          Tadaoki Uesugi}\\

\bigskip
\bigskip
{\Large \it Department of Physics, Faculty of Science \\ University of Tokyo \\
\medskip
Tokyo 113-0033, Japan}\\
\vspace{4cm}
{\large A Dissertation in candidacy for\\
the degree of Doctor of Philosophy}
\end{center}
\begin{abstract}
In this thesis we review the fundamental framework of boundary string field 
theory (BSFT) and 
apply it to the tachyon condensation on non-BPS systems in the 
superstring theory. 
The boundary string field theory can be regarded as a natural 
extension of the nonlinear sigma model. By 
using this theory we can describe the tachyon condensation exactly and also 
obtain the effective actions on non-BPS systems consisting of the 
Dirac-Born-Infeld type action and the Wess-Zumino type action. Especially the 
Wess-Zumino action is written by superconnection and coincides 
with the mathematical argument by K-theory. Moreover we also discuss the 
tachyon condensation keeping the conformal invariance (on-shell). The exact 
argument by using the boundary state formalism gives a good support to the 
conjecture of the tachyon condensation and it is also consistent with boundary 
string field theory.
\end{abstract}
\end{titlepage}

\newpage

\tableofcontents
\clearpage
\chapter{Introduction}
\hspace*{4.5mm}
In recent several years 
there has been tremendous progress in the nonperturbative aspect 
of string theory. One of key ingredients of this is the discovery of 
D-branes \cite{Po2}. The D-branes are nonperturbative objects (solitons) in 
string theory which are considered to play the same role as instantons in 
gauge theory. Another important discovery is the string duality. The 
string duality unifies five supersymmetric string theories in a single 
framework called M-theory \cite{Wi3}. These two important discoveries 
have caused the recent development of string theory such as the matrix 
theory \cite{BFSS} and AdS/CFT \cite{Mal} correspondence. 

The important thing here is that the above progress has often relied on the 
existence of vast degrees of supersymmetry in the target space. 
So, the natural question is what we can 
find in the less or no supersymmetric structure of string theory. In the 
less supersymmetric phases of string theory we can expect that 
some dynamical phenomena happen in the same way as the confinement of quarks 
in less supersymmetric gauge theory. Since we do not understand the 
dynamical aspect of string theory very well, it is meaningful and 
theoretically interesting to study it.

On the other hand we already know that we can realize {\sl spontaneously} 
broken phases of the supersymmetry in superstring theories by 
appropriate compactification or putting D-branes. 
In the latter point of view 
non-BPS D-brane systems can be regarded as one of simple models 
we can consider in order to investigate the structure of string theory without 
supersymmetry. Especially there are two kinds of D-brane systems without any 
supersymmetry in the flat space background of Type II theory. They are the 
brane-antibrane system and the non-BPS D-brane. 
Although of course several other kinds of 
non-BPS systems exist in string theory, D-branes are convenient objects 
we can deal with by using the conventional open string theory \cite{Po2}. If 
we examine the open string spectrum on these non-BPS systems, we can see that 
tachyonic modes appear even though there are no tachyonic modes in the 
closed string spectrum. In the early stage of string theory as a dual
model \cite{dual15}, 
the tachyonic modes were interpreted as the inconsistency of the theory. 
The recent idea about the interpretation of the tachyonic mode is 
that the theory is not inconsistent but stays in an unstable phase and will 
decay to a stable phase where full or some of the supersymmetries are 
restored. 
 
Several years ago Ashoke Sen \cite{sen13} conjectured that these unstable 
D-brane systems 
decay to the vacuum without any D-branes or other stable D-branes. 
This dynamical decay process is called tachyon condensation \cite{sen13}. 
Since then 
there has been a lot of work about the verification of this conjecture in the 
quantitative way (for a review see \cite{sen18}). However, it seems that in 
general we can not examine this by 
the conventional open string theory because this process relates two 
different backgrounds in string theory. One is an unstable background due to 
the existence of unstable D-branes and the other is a stable background where 
all or some supersymmetries are unbroken. What the conventional perturbative 
string theory can do is to examine the fluctuations and to calculate 
their scattering amplitudes around the {\sl fixed} background, and it can not 
predict how two different backgrounds are related with each other. Therefore, 
we can say that this is an important problem of nonperturbative aspects of 
string theory in the same sense as the fact that the string duality has 
related various different string theories (vacuums) with each other. Indeed 
for several years a lot of methods have been developed and by using them Sen's 
conjecture of the tachyon condensation has been checked. 
They include the noncommutaive 
field theory \cite{SeWi,HaKrLa1,witten2}, K-theory 
\cite{Wi1,Ho1,MiMo,HaMi} and string field theory.

In this thesis we will investigate the tachyon condensation mainly by using  
open string field theory. This is a kind of 
``field theory" for open strings. If it has a feature similar to 
the field theory with a kind of scalar potential, we can relate two 
different backgrounds 
in string theory with each other as a maximum point and a 
minimum one of the scalar potential, which can not be done by the conventional 
perturbative string theory. Indeed, string field theory has that kind of 
potential called tachyon potential because there appears 
a scalar field which is a tachyon on unstable D-branes. In this sense this 
theory is well suited to verify Sen's conjecture.

For now two types of covariant formulations of open string field theory are 
known. One is Witten's cubic string field theory (CSFT) 
\cite{WiCSFT,Wisu} or Berkovits' extended formulation for the 
superstring \cite{Ber1}. By using these formulations the conjecture 
of tachyon condensation has been checked \cite{bosonic,super,lump}. 
Especially the approximate method called the level truncation has 
been used to calculate the tachyon potential. These analyses have shown that 
Sen's conjecture is correct in the good accuracy (For a good review see 
\cite{Ohmori} and references there in).

On the other hand there is the other type of string field theory. It is 
called boundary string field theory (BSFT), which is also constructed by 
Witten \cite{Wi2,Sh1,LiWi}. 
This formalism is quite different from the cubic string 
field theory. The most important fact is that this formalism enables 
us to calculate quantities like tachyon potential in the {\sl exact} way 
\cite{GeSh1,KuMaMo1,KuMaMo2,TaTeUe,KrLa}, which might not be done by 
the cubic string field theory. In this sense this formalism has almost 
completed the verification of Sen's conjecture, although there remain other 
related problems which we will mention in the final chapter.   

The most characteristic feature in boundary string field theory is the fact 
that this theory can be regarded as a natural extension 
\cite{AnTs1,Ts1,Ts2,FrTs,An} of the two-dimensional 
nonlinear sigma model. Therefore we can 
apply the worldsheet picture in the conventional perturbative string theory 
even to the boundary string field theory, and in this sense the 
worldsheet description is a key word in the title of this thesis. 

By the way, historically speaking, before the string field 
theory was used the other method 
\cite{sen11,sen12,sen19,sen20,sen14,sen16,senI,MaSe2,BeGa1,bergman3,
bergman1,frau,NaTaUe} had been applied 
to prove some part of Sen's conjecture (for a review see 
the papers \cite{sen18,Gab1}). 
In the other paragraphs we have said that in general we 
can not use the conventional perturbative string theory for the tachyon 
condensation. However, in some 
special situations we can apply it and describe the tachyon condensation 
exactly. We will also explain this method in this thesis.

The organization of this thesis is as follows. In chapter 2 we will introduce 
non-BPS systems in string theory by using the boundary state formalism 
\cite{pol,Gr,Vecchia,Gab1}, and 
explain the notion of tachyon condensation. The boundary state formalism 
has something to do with boundary string field theory and it is also fully 
used in chapter 4. This chapter is rather elementary, thus the reader could 
skip it if one already knows about them. 
 
In chapter 3 we will explain the boundary string field theory and its 
description of tachyon condensation, which is named an off-shell 
description. First we will explain the relation 
\cite{HaKuMa} between the worldsheet renormalization group flow and dynamics 
of D-branes in the target space in section \ref{RGTach}. After that we will 
introduce Witten's original construction \cite{Wi2} of boundary string field 
theory using Batalin-Vilkovisky formulation in section \ref{BVformalism}. 
In section \ref{superaction} we will give the explicit forms of worldsheet 
actions which describe the non-BPS systems in superstring theory.

These are the formal part of chapter 3, while the practical calculations are 
performed from section \ref{exacttach} based on its formal construction. 
In section \ref{exacttach} we will calculate the tachyon potential 
and describe the tachyon condensation by showing that the correct values of 
D-brane tensions are produced. 
Moreover we can also give the effective actions on non-BPS 
systems by using boundary string field theory\cite{KuMaMo2,TaTeUe,KrLa}. 
They consist of the Dirac-Born-Infeld actions (section \ref{WV}) and 
Wess-Zumino terms (section \ref{RRcouple}). Especially we will see the 
mathematically intriguing structure of the Wess-Zumino terms written by 
superconnections \cite{Qu}. In this way a merit of boundary string field 
theory, which is different from other string field theories, is that some of 
the results are exact. 
However, there remain some issues about the meaning of 
``exact" in boundary string field theory. In section \ref{calcu} and 
\ref{meaning} we will explain the meaning of calculations of boundary string 
field theory and how we can justify these results as exact ones.

In chapter 4 we will consider the special case of tachyon condensation 
which we can deal with by the conventional perturbative string theory. 
This kind of analysis is named on-shell description and can be 
performed by using the open string technique \cite{MaSe2}. 
In this thesis we will fully use the boundary state formalism \cite{NaTaUe} 
which is introduced in chapter 2. 
Historically this method \cite{sen12,sen14} had been found before the 
string field theories were applied to the tachyon condensation. The content of 
this chapter will easily be understood after we introduce the idea of 
boundary string field theory.  

In the last chapter we will have the conclusion and mention future problems 
including closed string dynamics. 
In appendixes we also present the conventions of conformal field theory 
which 
we use in this thesis and several detail calculations which are needed in 
various chapters.

\chapter{Non-BPS D-brane Systems in String Theory}
\hspace*{4.5mm}
In this chapter we will define the non-BPS systems in string theory 
\cite{sen18}. In the flat space background two kinds of 
non-BPS systems are known: the brane-antibrane system and the non-BPS 
D-branes. We can deal with these D-branes by the open string theory \cite{Po2}.
The characteristic feature of 
these D-branes is that they include tachyon modes in their open string 
spectra. Therefore, in general, they are unstable and decay to something.  

By the way, instead of the open string theory there is another convenient way 
to express D-branes. It is called the boundary state formalism \cite{pol,Gr}. 
In the next few sections 
we will review a complete description of the boundary 
state formalism for a BPS D-brane and unstable D-brane systems. 

This chapter is rather elementary, thus one who is familiar with materials in 
this area can move on to the next chapter.

\section{Boundary State of BPS D-branes \label{bstate}}
\hspace*{4.5mm}
In general the D-branes are defined by classical open strings. We 
usually regard D-branes as open string boundaries and consider 
various boundary conditions. On the other hand D-branes interact with 
closed strings. If we would like to 
pay attention to the latter aspect of D-branes, 
it is inconvenient to express D-branes by the open string Hilbert space. 
Fortunately, we have a good prescription for expressing them 
by {\sl the closed string Hilbert space}. 
That is the boundary state formalism \cite{pol,Gr}. In this section we will 
review the construction of the boundary state for a BPS D-brane. 

To construct the boundary state we have to specify the boundary condition of 
open strings. The boundary condition is 
given by a variation of the standard action,
\beqa
I_0=\frac{1}{4\pi}\int d^2z (\pa X^{\mu}\bar{\pa}X_{\mu}
-i\psi^{\mu}_{L}\bar{\pa}\psi_{L\mu}+i\psi^{\mu}_{R}
\pa\psi_{{\ss R}\mu}).
\eeqa
Then, we can obtain 
the following boundary condition for a D$p$-brane 
stretching in the direction $X^1,\cdots, X^p$:
\beqa
\label{bc1}
\pa_1 X^{\alpha}(\sigma_1,\sigma_2)|_{\sigma_1=0,\pi}=0,~~&~~ 
\pa_2 X^{i}(\sigma_1,\sigma_2)|_{\sigma_1=0,\pi}=0,\no
\psi^{\alpha}_L(0,\sigma_2)=\gamma_1 \psi^{\alpha}_R(0,\sigma_2),~~&~~
\psi^{i}_L(0,\sigma_2)=-\gamma_1 \psi^{i}_R(0,\sigma_2),\no
\psi^{\alpha}_L(\pi,\sigma_2)=\gamma_2 \psi^{\alpha}_R(\pi,\sigma_2),~~&~~
\psi^i_L(\pi,\sigma_2)=-\gamma_2 \psi^i_R(\pi,\sigma_2),\\
&\no
(\alpha,\beta,\cdots=0,1,\cdots, p)~~&~~(i,j,\cdots=p+1,\cdots,9)\nonumber
\eeqa
where the condition $\gamma_1=-\gamma_2~~(\gamma_1=\gamma_2)$ corresponds to 
NS(R)-sector of open strings and $\gamma_{1,2}$ takes $\pm 1$. 
The boundary condition (\ref{bc1}) can simply be obtained by 
performing T-duality transformation to that of a D9-brane.

Now we are ready to define the boundary state for a D$p$-brane. 
Throughout this thesis we take the lightcone gauge formalism 
\cite{bergman1,pol} of the boundary states because it is convenient that we do 
not have to consider the ghost contributions. The drawback is that we 
can not consider all D$p$-branes because we throw away not only the 
contribution of ghosts but also that of nonzero modes for two directions in 
the total ten-dimensional space. 
Moreover in the lightcone gauge we consider the space where the freedom of 
$X^{+}$ direction is fixed, namely, the time direction in the lightcone gauge 
obeys the Dirichlet boundary condition. Therefore the D$p$-branes we consider 
are $(p+1)$-dimensional instanton-like objects and the range of $p$ is $-1$ to 
$7$. However, we can obtain usual D$p$-branes by performing the Wick 
rotation.

\begin{figure} [tbhp]
\begin{center}
\epsfbox{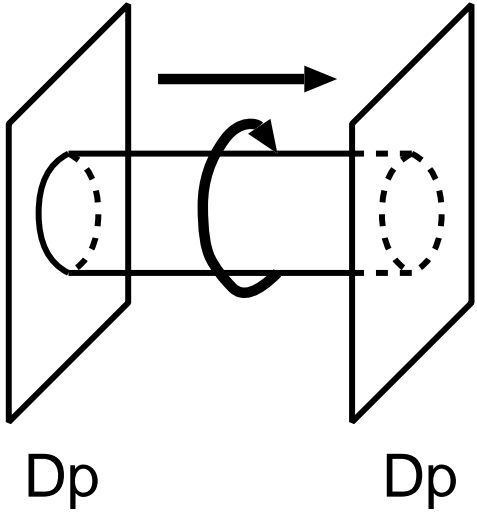}
\caption{Open-closed duality  \label{duality}}
\epsfxsize=70mm
\end{center}
\end{figure}

The boundary state $|Dp,\gamma\lb$ belongs to the closed string Hilbert 
space and satisfies the following constraint equations which correspond to 
equations in (\ref{bc1})
\beqa
\label{bsdef1}
(\pa_w X^{\mu}(\sigma_1,0)-S^{\mu}_{~\nu} \bar{\pa}_w 
X^{\nu}(\sigma_1,0))|Dp,\gamma_1\lb_{\ss NSNS(RR)}=0,\no
(\psi^{\mu}_L(\sigma_1,0)-i\gamma_1 
S^{\mu}_{~\nu}\psi^{\nu}_R(\sigma_1,0))|Dp,\gamma_1\lb_{\ss NSNS(RR)}=0,
\eeqa
where we have defined a matrix $S^{\mu}_{~\nu}$ as 
$S^{\mu}_{~\nu}=(\delta^{\alpha}_{~\beta},-\delta^i_{~j})~(\alpha,\beta
=0,\cdots, p~|~i,j=p+1,\cdots, 9)$, which is just the data of boundary 
condition of the D$p$-brane. Here note that the roles of the worldsheet space 
($\sigma_1$) and time ($\sigma_2$) are interchanged. This can be understood 
from figure \ref{duality} which represents the open-closed duality. The factor 
$i$ in the second equation comes from the conformal transformation 
corresponding to the interchange of $\sigma_1$ and $\sigma_2$. 

Before solving this constraint equation we have to notice that the 
closed string Hilbert space which is spanned by boundary states are only 
limited to NSNS and RR-sector. The proof of this is as follows. First, if we 
consider the open string picture in figure \ref{duality}, we can see that open 
strings should have some kind of periodicity in the direction of the 
worldsheet time $\sigma_2$ because the cylinder represents an open string one 
loop diagram. If we set this periodicity to $T$, this statement can be 
represented by
\beqa
\label{cylperiod}
\left\{\begin{array}{c}
\psi^{\mu}_L(\sigma_1,0)=\gamma_3 \psi^{\mu}_L(\sigma_1,T),\\
\psi^{\mu}_R(\sigma_1,0)=\gamma_4 \psi^{\mu}_R(\sigma_1,T).
\end{array}\right.
\eeqa 
Here the periodicity of bosonic fields $X^{\mu}(\sigma_1,\sigma_2)$ is 
irrelevant in this proof, thus we have omitted it. By using the above 
relations we can prove the following identities:
\beqa
\psi^{\mu}_L(0,0)\mathop{=}_{(\ref{cylperiod})}\gamma_3\psi^{\mu}_L(0,T)
\mathop{=}_{(\ref{bc1})}\gamma_3\gamma_1 S^{\mu}_{~\nu}\psi^{\nu}_R(0,T),
\eeqa
and
\beqa
\psi^{\mu}_L(0,0)\mathop{=}_{(\ref{bc1})}\gamma_1S^{\mu}_{~\nu}
\psi^{\nu}_R(0,T)\mathop{=}_{(\ref{cylperiod})}\gamma_1\gamma_4 
S^{\mu}_{~\nu}\psi^{\nu}_R(0,T).
\eeqa
From these equations the following relation holds:
\beqa
\gamma_3=\gamma_4. 
\eeqa
If we see figure \ref{duality} in the closed string picture, equation 
(\ref{cylperiod}) can be regarded as the boundary condition of closed strings 
emitted by the D$p$-brane. It indicates that there is no NS-R or R-NS sector 
in the boundary state. 

Now we have come to the stage of solving constraint equation 
(\ref{bsdef1}). By inserting mode expansions of $X^{\mu},\psi^{\mu}_L$ and 
$\psi^{\mu}_R$ (see eq.(\ref{modeexpand})) into eq.(\ref{bsdef1}) we can 
obtain the following equations:
\beqa
\label{Xmodebc}
(\alpha^{\mu}_m+S^{\mu}_{~\nu}\ti{\alpha}_{-m}^{\nu})
|Dp,\gamma\lb_{\ss NSNS(RR)}=0,\\
\label{psimodebc}
(\psi^{\mu}_r-i\gamma S^{\mu}_{~\nu}\ti{\psi}^{\nu}_{-r})
|Dp,\gamma\lb_{\ss NSNS(RR)}=0,
\eeqa
where we have rewritten $\gamma_1$ with $\gamma$ and the value of the 
subscript label $r$ takes half-integers(integers) for NSNS(RR)-sector. 
\\\\
{\bf For nonzero modes}
\\\\
\hspace*{4.5mm}
For nonzero modes 
($\alpha^{\mu}_m,~\psi^{\mu}_r,~\ti{\psi}^{\mu}_r~(m,r\neq 0)$) the above 
equations can be easily solved and the result is equal to the following 
coherent state:
\beqa
\label{nonzerosol}
|Dp,\gamma\lb_{\ss NSNS(RR)}&=&
{\cal N}_p\exp\left[-\sum^{\infty}_{m=1}m^{-1}\alpha^{\mu}_{-m}S_{\mu\nu}
\ti{\alpha}^{\nu}_{-m}+i\gamma\sum_{r>0}\psi^{\mu}_{-r}S_{\mu\nu}
\ti{\psi}^{\nu}_{-r}\right]\no
&&~~~~~~~~~~~~~~~~~~~~~~~~~~~~~~~~~~~~~~~\times~|Dp,\gamma\lb^0_{\ss NSNS(RR)},
\eeqa
where $|Dp,\gamma\lb^0_{\ss NSNS(RR)}$ is the contribution from zero modes 
including the conformally invariant vacuum $|0\lb$, and ${\cal N}_p$ is the 
normalization constant of $|Dp,\gamma\lb_{\ss NSNS(RR)}$, which can be 
determined by Cardy's condition \cite{cardy} explained later. 
\\\\
{\bf For zero modes}
\\\\
\hspace*{4.5mm}
Equations (\ref{Xmodebc}) and (\ref{psimodebc}) for zero modes are rewritten as
\beqa
\label{bstzero1}
(p^{\alpha}_L+p^{\alpha}_R)|Dp,\gamma\lb^0_{\ss NSNS(RR)}&=&0,
~~~~~(p^i_L-p^i_R)|Dp,\gamma\lb^0_{\ss NSNS(RR)}=0,\\
\label{bstpsizero}
(\psi^{\A}_0-i\gamma\ti{\psi}^{\A}_0)|Dp,\gamma\lb^0_{\ss RR}&=&0,
~~~~~(\psi^i_0+i\gamma\ti{\psi}^i_0)|Dp,\gamma\lb^0_{\ss RR}=0,
\eeqa 
where $\alpha,\beta,\cdots=0,\cdots, p$ and $i,j,\cdots=p+1,\cdots, 9$. 
Note that equations (\ref{bstpsizero}) are defined only in RR-sector of 
the boundary state.
 
For equations (\ref{bstzero1}) the solution depends on whether we consider 
a non-compactified direction or a compactified direction. For a 
non-compactified direction the second equation in eq.(\ref{bstzero1}) becomes 
a trivial equation and the first equation in 
eq.(\ref{bstzero1}) can be rewritten as 
$p^{\alpha}|Dp,\gamma\lb^0_{\ss NSNS(RR)}=0$. 
Moreover if the position of the D$p$-brane is $x^i=a^i$, 
the Hilbert space for zero modes can be written as 
\beqa
\label{delta}
\delta^{9-p}(x^i-a^i)|0\lb
=\int\frac{d^{9-p}k}{(2\pi)^{9-p}}e^{ik_i(x-a)^i}|0\lb
\equiv\int\frac{d^{9-p}k}{(2\pi)^{9-p}}e^{-ik_ia^i}|{\bf 0},{\bf k}\lb,
\eeqa
where 
\beqa
\label{knormal}
\la {\bf 0},{\bf k}|{\bf 0},{\bf k}^{\prime}\lb=(2\pi)^{9-p}\delta^{9-p}
({\bf k-k}^{\prime}).
\eeqa
Note that this state satisfies $p^{\alpha}|Dp,\gamma\lb^0_{\ss NSNS(RR)}=0$.

For a compactified direction the equations in eq.(\ref{bstzero1})  
can be rewritten as 
\beqa
n^{\alpha}|Dp,\gamma\lb^0_{\ss NSNS(RR)}&=&0,\no
w^i|Dp,\gamma\lb^0_{\ss NSNS(RR)}&=&0,
\eeqa
where $n^{\alpha}$ and $w^i$ denote the momentum number and the winding number 
(see eq.(\ref{discretep})). If we consider the ten-dimensional space 
where all of the directions are compactified, the Hilbert space of this state 
can be spanned by the linear combination of 
$|{\bf w},{\bf n}\lb$ such as 
${\displaystyle \sum_{w^{\alpha}\in{\bf Z}^{p+1},n^i\in{\bf Z}^{9-p}}
f_{{\bf w},{\bf n}}|{\bf w},{\bf n}\lb}$, 
where $f_{{\bf w},{\bf n}}$ is an coefficient for a given sector with 
$w^{\alpha},n^i$. To determine it we make eq.(\ref{delta}) discrete like
\beqa
\label{momentum}
\int\frac{d^{9-p}k}{(2\pi)^{9-p}}e^{ik_i(x-a)^i}|0\lb
&\rightarrow&(V_{9-p})^{-1}
\sum_{n^i\in{\bf Z}^{9-p}}e^{i\frac{n^i}{R_i}(x-a)^i}|0\lb\no
&\equiv&
~(V_{9-p})^{-1}\sum_{n^i\in{\bf Z}^{9-p}}e^{-i\frac{n^i}{R_i}a^i}
|{\bf 0},{\bf n}\lb.
\eeqa
Corresponding to the last equation we can consider a state with winding 
numbers given by
\beqa
\label{Wilsonline}
\sum_{w^{\alpha}\in{\bf Z}^{p+1}}e^{i\frac{w^{\alpha}R_{\alpha}}{2}
(x^{\alpha}_L-x^{\alpha}_R-a^{\alpha})}|0\lb
~\equiv~\sum_{w^{\alpha}\in{\bf Z}^{p+1}}e^{-i
\frac{w^{\alpha}R_{\alpha}}{2}a^{\alpha}}|{\bf w},{\bf 0}\lb,
\eeqa
where $a^{\alpha}$ is called the Wilson line. By T-duality 
($R\leftrightarrow\frac{2}{R},~n\leftrightarrow w$) this corresponds to 
the parameter $a^i$ in eq.(\ref{momentum}), which represents the position of 
the D$p$-brane.  In terms of open strings these parameters $a^{\A}$ and $a^i$ 
can be interpreted as expectation values of open string backgrounds
$A^{\A}(X(z,\bar{z}))~(\mbox{gauge field})$ and 
$\Phi^i(X(z,\bar{z}))~(\mbox{transverse scalar})$. 

After all in compactified directions the zero mode part of boundary 
states can be written as
\beqa
(V_{9-p})^{-1}\sum_{w^{\alpha}\in{\bf Z}^{p+1},n^i\in{\bf Z}^{9-p}}
e^{-i\left(\frac{w^{\alpha}R_{\alpha}}{2}a^{\alpha}+
\frac{n^i}{R_i}a^i\right)}|{\bf w},{\bf n}\lb.
\eeqa
Corresponding to (\ref{knormal}) the normalization of $|{\bf n},{\bf w}\lb$ 
becomes
\beqa
\la {\bf n}^{\prime},{\bf w}^{\prime}|{\bf n},{\bf w}\lb
=V\delta_{{\bf n}^{\prime},{\bf n}}\delta_{{\bf w}^{\prime},{\bf w}},
\eeqa
where $V$ is the volume of the compactified space.

The equation left to solve is eq.(\ref{bstpsizero}), 
which determines the contribution from zero modes of worldsheet fermions 
$\psi^{\mu}_L$ and $\psi^{\mu}_R$ in the RR-sector. To solve this we first 
define the following new valuables:
\beqa
\psi^{\mu}_{\pm}=\frac{1}{\sqrt{2}}(\psi^{\mu}_0\pm i\ti{\psi}^{\mu}_0),
\eeqa
and also define $|\Omega p,\gamma\lb^0_{\ss RR}$ as the limited Hilbert space 
of $|Dp,\gamma\lb^0_{\ss RR}$ which is spanned only by $\psi^{\mu}_0$ and 
$\ti{\psi}^{\mu}_0$. By using these we can rewritten eq.(\ref{bstpsizero}) 
simply as 
\beqa
\label{bstpsizero2}
\psi^{\A}_{-}|\Omega p,+\lb^0_{\ss RR}=0~&,&
~\psi^i_{+}|\Omega p,+\lb^0_{\ss RR}=0~,\no
\psi^{\A}_{+}|\Omega p,-\lb^0_{\ss RR}=0~&,&
~\psi^i_{-}|\Omega p,-\lb^0_{\ss RR}=0.
\eeqa
Since we can obtain the anticommuting relations between these 
valuables as
\beqa
\label{zeropsisol1}
\{\psi^{\mu}_{\pm},\psi^{\nu}_{\pm}\}=0~,~\{\psi^{\mu}_{+},\psi^{\nu}_{-}\}
=\eta^{\mu\nu},
\eeqa
we can construct the zero mode part of boundary states by using the 
representation theory of Clifford Algebra. 

First we can define ``the ground state" $|\Omega\lb^0_{\ss RR}$ as 
$|\Omega\lb^0_{\ss RR}\equiv|\Omega 7,+\lb^0_{\ss RR}$ because of 
\\
$\psi^{\mu}_{-}|\Omega 7,+\lb^0_{\ss RR}=0$ for all $\mu~(\mu=-1,0,\cdots,7)$.
Next we can define ``the excited states" by 
multiplying creation operators $\psi^{\mu}_{+}$ by this ground state. 
Especially the state which satisfies eq.(\ref{bstpsizero2}) is given by
\beqa
\label{zeropsisol2}
|\Omega p,+\lb^0_{\ss RR}=\prod_i \psi^i_{+}|\Omega\lb^0_{\ss RR},\no
|\Omega p,-\lb^0_{\ss RR}=\prod_{\A} \psi^{\A}_{+}|\Omega\lb^0_{\ss RR}.
\eeqa
The ambiguity left in this solution is phase factors $\pm 1$ and the 
normalization of ``the ground state" $|\Omega\lb^0_{\ss RR}$. The first
ambiguity can be gotten rid of by fixing the convention. Here we have chosen 
the following convention: 
\beqa
|\Omega p,+\lb^0_{\ss RR}&=&\psi^7_{+}\psi^6_{+}\cdots \psi^{p+1}_{+}
|\Omega\lb^0_{\ss RR},\no
|\Omega p,-\lb^0_{\ss RR}&=&\psi^0_{+}\psi^1_{+}\cdots \psi^{p}_{+}
|\Omega\lb^0_{\ss RR},
\eeqa
for a D$p$-brane which extends in directions $X^0,X^1,\cdots X^p$. 
The second ambiguity is not related to the convention. The method to fix this 
is the same as that of fixing the total normalization ${\cal N}_p$ in 
eq.(\ref{nonzerosol}). Here, to avoid the complicated situation we first 
give the answer to this:
\beqa
\la\Omega|\Omega\lb^0_{\ss RR}=16.
\eeqa
You can check that this result is consistent with the later calculation of 
the interaction between two D$p$-branes. By using this the 
normalization of the zero mode part $|Dp,\gamma\lb^0_{\ss RR}$ can be 
calculated and the result is
\beqa
\la \Omega p,\gamma|\Omega p,\gamma^{\prime}\lb^0_{\ss RR}
=16\delta_{\gamma,\gamma^{\prime}}.
\eeqa
\\\\
{\bf GSO projection in the NSNS-sector} 
\\\\
\hspace*{4.5mm}
To obtain the modular invariant spectrum 
in closed string theory we have to perform GSO-projection. This can be applied 
to boundary states too because these states also belong to the closed string 
Hilbert space. In the NSNS sector the worldsheet fermion number $F$ is
defined by 
\beqa
F=\sum^{\infty}_{r=1/2}\psi^{\mu}_{-r}\psi_{\mu r}-1,
\eeqa
and $\ti{F}$ is defined by replacing $\psi_r$ with $\ti{\psi}_r$. 
By a simple calculation we can see the relations
\beqa
\label{GSONSNS}
(-1)^F|Dp,\gamma\lb_{\ss NSNS}=-|Dp,-\gamma\lb_{\ss NSNS},\no
(-1)^{\ti{F}}|Dp,\gamma\lb_{\ss NSNS}=-|Dp,-\gamma\lb_{\ss NSNS},
\eeqa
where we have included the ghost contribution in the fermion 
number\footnote{Note the relation $(-1)^F|0\lb=-|0\lb$.}. 

Therefore we can obtain the following linearly combined boundary state which 
is GSO-invariant $\left((-1)^F=1,~(-1)^{\ti{F}}=1\right)$ in the NSNS-sector
\beqa
\label{NSNS}
|Dp\lb_{\ss NSNS}=\frac{1}{2}(|Dp,+\lb_{\ss NSNS}-|Dp,-\lb_{\ss NSNS}),
\eeqa 
where we have included $\frac{1}{2}$ for later convenience.
\\\\
{\bf GSO projection in the RR-sector}
\\\\
\hspace*{4.5mm}
In the RR-sector the worldsheet fermion number $F$ is defined by
\beqa
(-1)^F~\equiv~2^4\psi^0_0\cdots\psi^7_0(-1)^{\sum^{\infty}_{m=1}
\psi^{\mu}_{-m}\psi_{\mu m}}.
\eeqa
By using eq.(\ref{nonzerosol}) and (\ref{zeropsisol2}) we can obtain the 
following relations
\beqa
\label{GSORR}
(-1)^F|Dp,\gamma\lb_{\ss RR}=|Dp,-\gamma\lb_{\ss RR},\no
~(-1)^{\ti{F}}|Dp,\gamma\lb_{\ss RR}=(-1)^{7-p}|Dp,-\gamma\lb_{\ss RR},
\eeqa
where the phase factor $(-1)^{7-p}$ comes from the zero mode contribution. 
From the 
first equation we can see that the GSO-invariant state in the left-sector 
$((-1)^F=1)$ is 
\beqa
\label{RR}
|Dp\lb_{\ss RR}=\frac{1}{2}(|Dp,+\lb_{\ss RR}+|Dp,-\lb_{\ss RR}). 
\eeqa

This state also has to satisfy the GSO-projection in the right-sector
\\ 
$((-1)^{\ti{F}}=\mp 1~~(-1~\mbox{for IIA},~+1~\mbox{for IIB}))$. From this 
constraint we can see the correct spectrum of D$p$-branes with even(odd) 
integer $p$ for IIA(IIB). 
\\\\\\
{\bf Normalization of the boundary state}
\\\\
\hspace*{4.5mm}
At this stage we have the complete form of boundary state $|Dp\lb$ 
written by
\beqa
\label{total}
|Dp\lb~\equiv~|Dp\lb_{\ss NSNS}+|Dp\lb_{\ss RR},
\eeqa
where $|Dp\lb_{\ss NSNS}$ and $|Dp\lb_{\ss RR}$ 
are defined in eq.(\ref{NSNS}) and 
eq.(\ref{RR}), while we have not yet determined the normalization constant 
${\cal N}_p$, which appears in eq.(\ref{nonzerosol}). This constant can be 
determined by the consistency (``Cardy Condition'' \cite{cardy}) which comes 
from open-closed duality in figure \ref{duality}. 

First we can see that figure \ref{duality} can be regarded as a diagram of 
interaction between two parallel D$p$-branes. It is represented by 
${\cal A}=\la Dp|\Delta|Dp\lb$, where $\Delta$ is a closed string propagator
\beqa
\label{closedprop}
\Delta&=&\frac{1}{2\pi}\int_{|z|\leq1}\frac{d^2z}{|z|^2}z^{L_0}
\bar{z}^{\ti{L}_0}\no
&\rightarrow& \int^{\infty}_0 ds \exp[-s(L_0+\ti{L}_0)].
\eeqa 
Here we have used the relation $(L_0-\ti{L}_0)|Dp\lb=0$, which we can easily 
check from the explicit expression of $L_0$ and $\ti{L}_0$ 
(see eq.(\ref{L0})). The explicit form of this amplitude can easily be 
calculated and its result becomes
\beqa
\label{tree}
{\cal A}&=&\frac{1}{2}|{\cal N}_p|^2 V_{p+1}\int^{\infty}_0 ds 
(4\pi s)^{-\frac{9-p}{2}}\left
[\frac{f_3(q)^8-f_4(q)^8-f_2(q)^8}{f_1(q)^8}\right]\no
&=&0,
\eeqa
where $V_{p+1}$ denotes the (p+1)-dimensional 
volume, and we have used modular functions $f_i(q)~(i=1,2,3,4)$ defined by
\beqa
\label{func}
f_1(q)=q^{\f{1}{12}}\prod_{n=1}^\infty (1-q^{2n})&,&f_2(q)=\s{2}q^{\f{1}{12}}\prod_{n=1}^\infty (1+q^{2n}), \no
f_3(q)=q^{-\f{1}{24}}\prod_{n=1}^\infty (1+q^{2n-1})&,&f_4(q)=q^{-\f{1}{24}}\prod_{n=1}^\infty (1-q^{2n-1}),
\eeqa
with $q=e^{-s}$. 
In the second line of eq.(\ref{tree}) 
we have used the famous Jacobi's abstruse identity:
\beqa
\label{Jacobi}
f_3(q)^8-f_2(q)^8-f_4(q)^8=0,
\eeqa
and this result represents the fact that the force between 
two parallel D-branes is zero and D-branes are stable. In other words the 
gravitational attractive force from the NSNS-sector exactly cancels the 
repulsive Coulomb force from the RR-sector.  

On the other hand we can see that figure \ref{duality} can be regarded as an 
one-loop diagram of open strings. The partition function 
(cylinder amplitude) of this diagram is represented by
\beqa
\label{oneloop0}
{\cal Z}_{{\ss C}_2}=V_{p+1}\int^{\infty}_0 \frac{dt}{2t}{\rm Tr}_{\ss NS-R}
\left[\frac{1+(-1)^F}{2}\ti{q}^{2L_0}\right],
\eeqa
where ${\rm Tr}_{\ss NS-R}\equiv{\rm Tr}_{\ss NS}-{\rm Tr}_{\ss R}$ and 
$\ti{q}=e^{-\pi t}$. its explicit result is given by
\beqa
\label{oneloop}
{\cal Z}_{{\ss C}_2}&=&2V_{p+1}\int^{\infty}_0\frac{dt}{4t}
(16\pi^2t)^{-\frac{p+1}{2}}
\left[\frac{f_3(\ti{q})^8-f_4(\ti{q})^8-f_2(\ti{q})^8}{f_1(\ti{q})^8}\right]\no
&=&\frac{1}{2}V_{p+1}(16\pi^2)^{-\frac{p+1}{2}}\pi^{\frac{7-p}{2}}
\int^{\infty}_0~ds~s^{-\frac{9-p}{2}}
\left[\frac{f_3(q)^8-f_2(q)^8-f_4(q)^8}{f_1(q)^8}\right]\no
&=&0,
\eeqa
where the extra factor $2$ in the first line comes from the Chan-Paton factor.
In the second line we have used coordinate 
transformation with $s=\frac{\pi}{t}$ and modular properties of modular 
functions $f_i(q)$ given by 
\ba
\label{modular}
f_1(e^{-\f{\pi}{t}})=\s{t}f_1(e^{-\pi t}) &,& f_2(e^{-\f{\pi}{t}})=f_4(e^{-\pi t}), \no
f_3(e^{-\f{\pi}{t}})=f_3(e^{-\pi t}) &,& f_4(e^{-\f{\pi}{t}})=f_2(e^{-\pi t}).
\ea
The third line in eq.(\ref{oneloop}) comes from Jacobi's 
abstruse identity (\ref{Jacobi}), and this result represents the fact that 
there are some target space supersymmetries left. 

If we compare this 
result (\ref{oneloop}) with eq.(\ref{tree}) we can determine the coefficient 
${\cal N}_p$ of the boundary state, which is given 
by\footnote{Correctly speaking we can 
not determine the phase factor of this coefficient, while we can support this 
result by calculating other amplitudes between different kinds of D-branes.}
\beqa
\label{normalization}
{\cal N}_p=\frac{\sqrt{\pi}(2\sqrt{2}\pi)^{3-p}}{2}\equiv\frac{T_p}{2}.
\eeqa
Here we can see that 
$\kappa^{-1}T_p$ is equal to the tension of a D$p$-brane~($\kappa$ is 
the ten dimensional gravitational constant).

\section{Generalized Boundary State}
\hspace*{4.5mm}
In the previous section 
we have defined the boundary states which describe various 
D$p$-branes. In this section we will extend these boundary states 
into more general ones \cite{Nappi2,Alwis5}. To explain this let us consider 
an example of a quantum mechanical amplitude which is given by
\beqa
\la \beta|e^{-HT}|\alpha\lb,
\eeqa 
where $H$ is the Hamiltonian of the system of harmonic oscillators, and 
$|\alpha\lb$ and $|\beta\lb$ correspond to the states in the Schroedinger 
picture which represent some boundary conditions at $t=0$ 
and $t=T$. This is an transition amplitude from $|\alpha\lb$ to $|\beta\lb$.

On the other hand if we re-express this amplitude by using the path-integral 
formalism, it becomes
\beqa
\int {\cal D}q~e^{-\int^T_0 dt {\cal L}(q(t))}e^{-I_{\beta}(q(T))}
e^{-I_{\alpha}(q(0))},
\eeqa
where $q(t)$ and ${\cal L}(q)$ are the variable and its Lagrangian which 
describe the system of harmonic oscillators. What we would like to note here 
is that there appear boundary terms $e^{-I_{\beta}(q(T))}$ and 
$e^{-I_{\alpha}(q(0))}$ in the path-integral. 
These factors come from the internal products
\beqa
\la q|\alpha\lb=e^{-I_{\alpha}(q)},
\eeqa 
where $|q\lb$ is the coherent state defined by
\beqa
\label{1coh}
\hat{q}|q\lb=q|q\lb.
\eeqa
The most important feature of the coherent state is that this 
satisfies the completeness condition which is represented by
\beqa
\int {\cal D}q~|q\lb\la q|=1.
\eeqa
Because of this fact we can expand any quantum mechanical state by using the 
coherent state. Especially the state $|\alpha\lb$ can be expressed in the 
following way:
\beqa
|\alpha\lb=\int {\cal D}q~ e^{-I_{\alpha}(q)}|q\lb.
\eeqa
This is the one-dimensional example of the generalized state when we assign 
the weight $e^{-I_{\alpha}(q)}$ to the boundary ($t=0)$. 

Now we would like to extend this formalism to that for string theory. 
First of all we will define the coherent state $|x,\varphi\lb_{\ss NSNS(RR)}$ 
as the two-dimensional extension of the state (\ref{1coh}). 
Its definition in the NSNS-sector is the direct product of $|x\lb$ and 
$|\varphi\lb_{\ss NSNS}
=\frac{1}{2}\left(|\varphi,+\lb_{\ss NSNS}-
|\varphi,-\lb_{\ss NSNS}\right)$ which are defined by
\beqa
\label{defff}
X^{\mu}(\sigma_1,0)|x\lb&=&x^{\mu}(\sigma_1)|x\lb,\no
\left(\psi^{\mu}_L(\sigma_1,0)+i\gamma\psi^{\mu}_R(\sigma_1,0)\right)
|\varphi,\gamma\lb_{\ss NSNS}
&=&\varphi^{\mu}(\sigma_1)|\varphi,\gamma\lb_{\ss NSNS}~~~~~(\gamma=\pm 1), 
\eeqa  
where 
\beqa
x^{\mu}(\sigma_1)&=&~\sum_{m\in{\bf Z}}~x^{\mu}_m e^{im\sigma_1},\no
\varphi^{\mu}(\sigma_1)&=&
\sum_{r\in{\bf Z}+1/2}\varphi^{\mu}_r e^{ir\sigma_1}.
\eeqa
The way to solve these equation is almost the same as in the previous section. 
The answer is given by
\beqa
|x\lb&=&\exp\left[\sum^{\infty}_{m=1}\left\{-\frac{m}{2}x^{\mu}_{-m}x_{m\mu}
+\frac{1}{m}\alpha^{\mu}_{-m}\tilde{\alpha}_{-m\mu}
-\alpha^{\mu}_{-m}x_{m\mu}+x^{\mu}_{-m}\tilde{\alpha}_{-m\mu}\right\}\right]
|0\lb,\no
|\varphi,\gamma\lb_{\ss NSNS}
&=&\exp\left[\sum^{\infty}_{r=1/2}\left\{-\frac{1}{2}
\varphi^{\mu}_{-r}\varphi_{r\mu}-i\gamma\psi^{\mu}_{-r}\tilde{\psi}_{-r\mu}
+\psi^{\mu}_{-r}\varphi_{r\mu}+i\gamma\varphi^{\mu}_{-r}\tilde{\psi}_{-r\mu}
\right\}\right]|0\lb.\no
\eeqa
Here we have defined the total normalization to satisfy the completeness 
condition
\beq
\label{complete}
\begin{array}{c}
{\displaystyle \int {\cal D}x|x\lb\la x|=1},\\
{\displaystyle \int {\cal D}\varphi|\varphi,\gamma\lb_{\ss NSNS}
\la \varphi,\gamma|_{\ss NSNS}=1.}
\end{array}
\eeq
This is the result for the NSNS-sector, 
while we can define this kind of state in the RR-sector in the similar way.

Because of this completeness condition we can expand any boundary states 
in the following way
\beqa
\label{geneb}
|B~\lb_{\ss NSNS(RR)}
=\int {\cal D}x{\cal D}\varphi~ e^{-I_B(x,\varphi)}
|x,\varphi\lb_{\ss NSNS(RR)},
\eeqa
where $e^{-I_B(x,\varphi)}$ is the wave function corresponding to 
some boundary condition. If we use the definition of 
$|x,\varphi\lb_{\ss NSNS(RR)}$ in eq.(\ref{defff}) we can also rewritten the 
above equation as
\beqa
\label{geneb2}
|B~\lb_{\ss NSNS(RR)}&=&\frac{1}{2}\exp\left[-I_B\left(X,
\psi_L+i\psi_R\right)\right]\int 
{\cal D}x{\cal D}\varphi~|x,\varphi,+\lb_{\ss NSNS(RR)}\no
&&~~~~~~\mp\frac{1}{2}\exp\left[-I_B\left(X,
\psi_L-i\psi_R\right)\right]\int
{\cal D}x{\cal D}\varphi~|x,\varphi,-\lb_{\ss NSNS(RR)}\no
&=&\frac{1}{2}\exp\left[-I_B\left(X,
\psi_L+i\psi_R\right)\right]|D9,+\lb_{\ss NSNS(RR)}\no
&&~~~~~~\mp\frac{1}{2}\exp\left[-I_B\left(X,
\psi_L-i\psi_R\right)\right]|D9,-\lb_{\ss NSNS(RR)},
\eeqa
where $|D9~,\pm\lb_{\ss NSNS}$ is the boundary state for a D9-brane defined in 
eq.(\ref{nonzerosol}), and $X^{\mu}$ and 
$\psi^{\mu}_{L(R)}$ act on the boundary state as operators. 
If we set $I_B(x,\varphi)$ to zero, this boundary 
state is proportional to $|D9~\lb_{\ss NSNS(RR)}$. If we would like to 
consider the boundary state for a D$p$-brane, we have only to perform the 
T-duality transformation. 

In general we can identify $I_B(x,\varphi)$ as some boundary action on 
the disk like
\beqa
I_B(x,\varphi)&=&\int^{\pi}_{-\pi}d\sigma_1 d\theta 
~D_{\theta}{\bf X}^{\mu}A_{\mu}({\bf X})\no
&=&\int^{\pi}_{-\pi}d\sigma_1\left[-i\frac{\pa x^{\mu}}{\pa\sigma_1}A_{\mu}(x)
+\frac{1}{2}\varphi^{\mu}\varphi^{\nu}F_{\mu\nu}(x)\right], 
\eeqa 
where 
\begin{eqnarray}
\left\{\begin{array}{lcl}
{\bf X}^{\mu} &=& x^{\mu}+i\theta\varphi^{\mu},\\
D_{\theta} &=& \frac{\partial}{\partial\theta}
+i\theta\frac{\partial}{\partial\sigma_1}.\end{array}\right.
\end{eqnarray}  
If the gauge field $A_{\mu}(x)$ satisfies its equation of motion, 
the boundary state (\ref{geneb}) represents a BPS D9-brane with 
the external gauge field on it. In this case the conformal invariance in the 
worldsheet is preserved. 

In chapter 3 and 4 we will assign more general actions to 
$I_B(x,\varphi)$ in eq.(\ref{geneb}) or eq.(\ref{geneb2}). 
Especially in chapter 3 we will start with the action which does {\sl not} 
satisfy the conformal invariance. In this way we can generalize the usual 
boundary conditions for D-branes into more general ones.

\section{The Definition of Non-BPS Systems}
\hspace*{4.5mm}
In the previous section we have defined the familiar BPS D-branes by using 
the boundary state formalism. The 
characteristic property of these D-branes is the fact that these are 
supersymmetric and stable. On the other hand we can also define unstable 
D-brane systems in Type II string theories. The perturbative 
instability\footnote{The term ``perturbative" means that the instability comes 
from the existence of a perturbatively negative mode (tachyon). We have to 
note that there is also the other cause of instability. It is the 
nonperturbative instability we will mention in the final chapter.} of these 
D-branes indicates that 
these systems are nonsupersymmetric and decay to something stable. It is 
important to study the unstable non-BPS systems because the decay of these 
unstable D-brane systems can be considered as one of dynamical processes in 
string theory, which have not been studied and have not figured out for years. 
Therefore in this section we will define the non-BPS D-brane systems by using 
the boundary state method. 

\subsection{The brane-antibrane system}
\hspace*{4.5mm}
The important non-BPS systems consist of two kinds of D-brane 
systems. These are called the brane-antibrane system and the non-BPS 
D-branes. First we will explain the brane-antibrane system. The antibrane is 
defined by a D-brane which has the opposite RR-charge to an usual D-brane. 
The boundary state of this kind of D$p$-branes is written by
\beqa
\label{DDbar}
|\overline{Dp}\lb~\equiv~|Dp\lb_{\ss NSNS}-|Dp\lb_{\ss RR}.
\eeqa
If we compare this boundary state with eq.(\ref{total}), we can see 
that the sign in front of the RR-sector is opposite to 
that in eq.(\ref{total}). We can easily see that this D$p$-brane has the 
opposite RR-charge if we compute the coupling with a (p+1)-form RR-gauge 
field $C_{p+1}$, which is given by $\la C_{p+1}|\overline{Dp}\lb$. 

The important thing here is to 
consider a D-brane and an anti D-brane at the same time. If we put these two 
kinds of D$p$-branes in the parallel position, the only attractive force 
remains between these D$p$-branes, because not only the gravitational force 
but also the Coulomb force is attractive \cite{BaSu}. We can check this fact 
quantitatively by using the boundary state method. As we saw in 
eq.(\ref{tree}) the interaction is represented by 
$\la \overline{Dp}|\Delta|Dp\lb$, and its explicit result is 
\beqa
\label{DDbaramp1}
\la \overline{Dp}|\Delta|Dp\lb&=&\frac{T_p^2}{8} V_{p+1}
\int^{\infty}_0 ds (4\pi s)^{-\frac{9-p}{2}}\left
[\frac{f_3(q)^8-f_4(q)^8+f_2(q)^8}{f_1(q)^8}\right],
\eeqa
with $q=e^{-s}$. In this result the sign in front of $f_2(q)$ is opposite to 
that in eq.(\ref{tree}), thus this amplitude is not zero and indicates that 
this system is unstable. Moreover, if we perform the coordinate transformation 
$s=\frac{\pi}{t}$ and use the modular properties (\ref{modular}) of functions 
$f_i(q)$, we can find that this equation can be rewritten in the form of 
partition function of open strings as
\beqa
\label{DDbaramp2}
V_{p+1}\int^{\infty}_0\frac{dt}{2t}{\rm Tr}_{\ss NS-R}
\left[\frac{1-(-1)^F}{2}\ti{q}^{2L_0}\right]~~~(\ti{q}=e^{-\pi t}).
\eeqa
Here note that the sign in front of $(-1)^F$ is minus. This fact 
indicates that open strings between a D$p$-brane and an anti D$p$-brane obey 
the opposite GSO projection, which leaves an open string tachyon in the 
spectrum. 
If we admit the fact that the tachyon signals instability of the system, 
we can consider that this system decays to something. 

Finally note that the total tension and total mass of a brane-antibrane system 
is modified by quantum corrections. The classical tension of a brane-antibrane 
is given by $\frac{2T_p}{\kappa}=\frac{2\tau_p}{g}$, where $g$ is the string 
coupling constant and $\tau_p$ is given by
\beqa
\label{tens}
\tau_p=\frac{1}{\sqrt{2}(2\sqrt{2}\pi)^p}.
\eeqa
If we consider the quantum theory, the tension is modified as
\beqa
\label{masscorrection}
\frac{2\tau_p}{g}+a_0+a_1g+a_2g^2+\cdots\cdots,
\eeqa 
where $a_0,a_1,\cdots$ represent unknown constants. 
On the other hand the tension and the mass of a BPS D-brane is not modified by 
quantum corrections due to the target space supersymmetry.

\subsection{The non-BPS D-brane}
\hspace*{4.5mm}
We can define the other unstable system which is called the non-BPS D-brane. 
It is easy to define this D-brane by using the boundary state too. The 
boundary state for a non-BPS D$p$-brane is defined by
\beqa
\label{NDp}
|NDp\lb~\equiv~\sqrt{2}|Dp\lb_{\ss NSNS}.
\eeqa
Note that the RR-sector of the boundary state does not exist and that 
the coefficient $\sqrt{2}$ appears. The first fact indicates that non-BPS 
D-branes do not have any 
RR-charges because the coupling with a (p+1)-form RR-gauge 
field $\la C_{p+1}|NDp\lb$ is equal to zero. The second fact means that the 
tension of a non-BPS D$p$-brane is $\sqrt{2}$ 
times as large as that of a BPS D$p$-brane. We can easily verify this by 
calculating the coupling $\la 0|NDp\lb$ with the vacuum. 

The latter fact stems from Cardy's condition \cite{cardy}. If we calculate the 
interaction $\la NDp|\Delta|NDp\lb$ between two 
non-BPS D$p$-branes and transform it to the open string partition function, 
it becomes
\beqa
\label{NDppa}
V_{p+1}\int^{\infty}_0\frac{dt}{2t}{\rm Tr}_{\ss NS-R}\ti{q}^{2L_0}~~
~(\ti{q}=e^{-\pi t}).
\eeqa
This is just the partition function {\sl without} GSO projection. If we remove 
the $\sqrt{2}$ 
factor from the boundary state (\ref{NDp}), its partition function becomes 
half of eq.(\ref{NDppa}), and this is inconsistent with the open 
string theory. Therefore, we have to include $\sqrt{2}$ factor in the 
boundary state.

The important point about non-BPS D-branes is that there exists 
a tachyon in the open string spectrum, which can be seen from 
eq.(\ref{NDppa}). In the same way as the brane-antibrane system, this is the 
signal that non-BPS D-branes are unstable and decay to something stable. 

Now we have to notice that not all kinds of non-BPS D$p$-branes can exist in 
Type IIA(B) theory. In fact, in Type IIA theory non-BPS D$p$-branes with only 
odd integer $p$ can exist, while in Type IIB theory those with only even 
integer $p$ can. This is complementary to the fact that in Type IIA(B) there 
exist BPS D$p$-branes with even(odd) integer $p$. Its proof is simple. First 
we suppose that a BPS D$p$-brane and a non-BPS D$p$-brane could exist at the 
same time. Next we calculate the interaction between them and transform it 
into the form of partition function of open strings. The result is equal to 
$\frac{1}{\sqrt{2}}$ times as large as eq.(\ref{NDppa}) and is not a 
consistent one as we said.

Finally note that the tension of a non-BPS D-brane is also modified by quantum 
corrections in the same way as eq.(\ref{masscorrection}).

\subsection{The Chan-Paton factors and $(-1)^{F^{\ss S}_{\ss L}}$ twist 
\label{CPt}}
\hspace*{4.5mm}
Here we add Chan-Paton degrees of freedom to open strings on two kinds 
of non-BPS systems. For example, to describe the complete Hilbert space of 
open strings in a pair of brane and antibrane, we should prepare all 
two-by-two matrices 
which are spanned by an unit element and three Pauli matrices 
$\{1,\sigma_1,\sigma_2,\sigma_3\}$. To obtain the correct spectrum of all kinds
of open strings on a pair of brane and antibrane, we attach the fermion 
number $\{+,-,-,+\}$ to $\{1,\sigma_1,\sigma_2,\sigma_3\}$ respectively, and 
we perform GSO-projection $(-1)^{F_{tot}}
=(-1)^{F}\times(-1)^{F_{CP}}=1$, where 
$(-1)^{F}$ acts on the oscillator Hilbert space and $(-1)^{F_{CP}}$ on 
the Chan-Paton factors. As a result we can see that open string modes with 
Chan-Paton factor $1$ or $\sigma_3$ belong to the sector with $(-1)^{F}=1$, 
which includes SUSY multiplets at all stringy levels. Especially the most 
important one is the massless SUSY multiplet consisting of a gauge field and 
a gaugino.  On the other hand, open string modes with $\sigma_1$ or $\sigma_2$
belong to the sector with $(-1)^{F}=-1$, which is the same description as 
eq.(\ref{DDbaramp2}) and includes the tachyon etc. From these facts we 
can see that the Chan-Paton factors $1$ and $\sigma_3$ count the degrees of 
freedom of open strings both of whose ends are on the same D-brane, 
while $\sigma_1$ 
and $\sigma_2$ count those of open strings which stretches between a brane 
and an antibrane. 
Equation (\ref{DDbaramp2}) represents only the open string spectrum 
between a brane and an antibrane.
 
In this way we have obtained the following fields on a D9-brane and an anti 
D9-brane:
\beq
\label{DDbarfield}
\begin{array}{cl}
T(x),~\bar{T}(x)~~ &(\mbox{a complex tachyon}),\no
A_{\mu}^{(1)}(x),~A_{\mu}^{(2)}(x)~~ &(\mbox{two massless gauge fields}),\no
&\mbox{all massive fields},
\end{array}
\eeq
where we have omitted target space fermions, and we have defined 
$A^{(1)}_{\mu}(x)$ and $A^{(2)}_{\mu}(x)$ as the gauge fields on the brane and 
the anti D-brane respectively. If we consider a D$p$-brane and 
an anti D$p$-brane we have only to perform T-duality transformation 
to these fields like $A_9^{(1)}(x)~\rightarrow~\Phi_9^{(1)}(x)$, where 
$\Phi_9^{(1)}(x)$ 
is a transverse scalar representing the degree of freedom 
for a D$p$-brane to move in $x^9$ direction. Here note that the tachyon $T(x)$ 
is a {\sl complex} scalar field because in type II theory open strings have 
two orientations between a brane and an antibrane, which is 
related to the existence of two kinds of Chan-Paton factors $\sigma_1$ and 
$\sigma_2$. Two gauge fields $A_{\mu}^{(1)}(x)$ and 
$A_{\mu}^{(2)}(x)$ also have 
correspondences to Chan-Paton factors. $A_{\mu}^{(1)}+A_{\mu}^{(2)}$ 
belongs to unit element $1$, and $A_{\mu}^{(1)}-A_{\mu}^{(2)}$ to 
$\sigma_3$. We can interpret that the sector with $1$ represents the freedom 
to move two D-branes together, while the sector with $\sigma_3$ represents 
the freedom to separate a D-brane from an anti D-brane. 

In the case of a non-BPS D-brane we do not always have to consider Chan-Paton 
degrees of freedom because there exists only one D-brane and
only one kind of gauge field on it. On the other hand, there is a convenient 
description of the Hilbert space of open strings on {\sl one} non-BPS D-brane 
by considering Chan-Paton degrees of freedom. That is to consider 
$\{1,\sigma_1\}$ as degrees of freedom of open strings on a non-BPS D-brane. 
To obtain the correct spectrum on a non-BPS D-brane we attach the worldsheet 
fermion number $\{+,-\}$ to $\{1,\sigma_1\}$ respectively and 
perform usual GSO-projection 
$(-1)^{F_{tot}}\equiv (-1)^{F}\times(-1)^{F_{CP}}=1$ 
on the open strings. We can see that this prescription is equivalent to the 
other description in eq.(\ref{NDppa}) : the spectrum includes both the 
GSO-even sector and the GSO-odd sector, which have the Chan-Paton factor $1$ 
and $\sigma_1$, respectively. 
The total bosonic spectrum of open strings is obtained as
\beq
\begin{array}{cl}
\label{NDpfield}
T(x)&(\mbox{a real tachyon}),\no
A^{\mu}(x)&(\mbox{one massless gauge field}),\no
~~~~~~~~~~&\mbox{all massive fields}.
\end{array}
\eeq 
Here note that the tachyon $T(x)$ is a {\sl real} scalar field because 
there is only one degree of freedom $\sigma_1$ unlike a brane-antibrane system.
 
Now we have completed to describe the full Hilbert space of open strings on 
two kinds of non-BPS systems. {}From here we will state the important fact 
that these two kinds of D-brane systems are related to each other by the 
projection $(-1)^{F^{\ss S}_{\ss L}}=1$, where $F^{\ss S}_{\ss L}$ is the 
left-moving {\sl spacetime} fermion number \cite{sen16}. 

Firstly, we will prove that 
this projection changes IIA(B) theory into IIB(A) theory in the closed string 
Hilbert space. Let us start with the Green-Schwarz formalism of type IIB 
string theory, in which the valuables are 
$X^{i},~S^a_L$ and $S^a_R$ in the light-cone gauge ($i,a=1,\cdots,8)$. 
The operator $(-1)^{F^{\ss S}_{\ss L}}$ 
changes the valuable $S^a_L$ into $-S^a_L$ and the boundary condition of the 
twisted sector is 
$S^a_L(\sigma_1+2\pi)=-S^a_L(\sigma_1),~S^a_R(\sigma_1+2\pi)=S^a_R(\sigma_1)$.
Therefore, the total massless spectrum with the untwisted sector and 
the twisted sector is given by
\beqa
|i\lb~\otimes~(|j\lb~\oplus~|\dot{a}\lb)~~&~~(\mbox{untwisted sector}),\no
S^a_{-\frac{1}{2}}|0\lb
~\otimes~(|j\lb~\oplus~|\dot{a}\lb)~~&~~(\mbox{twisted sector}).
\eeqa
This is exactly the massless spectrum in the Type IIA theory. Moreover we can 
check that all the massive spectra are also the same as those of the Type IIA 
theory. This is the complete proof.

Now let us go back to the argument about D-branes. We can see that a BPS 
D$p$-brane becomes an anti BPS D$p$-brane under this projection because of
\beqa 
\label{projec}
(-1)^{F^S_L}~:~|Dp\lb_{\ss RR}~\rightarrow~-|Dp\lb_{\ss RR}.
\eeqa
Moreover, we have to consider Chan-Paton degrees of freedom to describe a 
brane-antibrane system completely. 
Chan-Paton matrices $\Lambda$ have to transform under 
$(-1)^{F^{\ss S}_{\ss L}}$ in the following way
\beqa
\label{CPtrans}
\Lambda=\left(\begin{array}{cc}
a & b\\
c & d
\end{array}\right)~\rightarrow~\sigma_1\Lambda(\sigma_1)^{-1}
=\left(\begin{array}{cc}
d & c\\
b & a
\end{array}\right),
\eeqa
because $(-1)^{F^{\ss S}_{\ss L}}$ interchanges a brane and an antibrane. 
From this equation we can see that the open string degrees of 
freedom with $\sigma_2$ and $\sigma_3$ are projected out, while those with 
$1$ and $\sigma_1$ remains. Therefore, we can conclude that a pair of 
D$p$-brane and anti D$p$-brane changes into a non-BPS D$p$-brane under this 
projection. If we use the fact that the twist $(-1)^{F^{\ss S}_{\ss L}}$  
interchanges IIA and IIB, we can see that this is consistent 
with the fact that in Type IIA(B) theory BPS D$p$-branes with even (odd) 
integer p exist, while non-BPS D$p$-branes with odd (even) integer p exist.  
We have shown this relation in figure \ref{del}.

Moreover, according to this projection the fields on a brane-antibrane system 
reduce to those on a non-BPS D-brane in the following way
\beqa
\label{descentfield}
T(x),~\bar{T}(x)&\rightarrow&\frac{1}{2}T(x),\no
\left(A^{\mu}_{(1)}(x)+A^{\mu}_{(2)}(x)\right)
&\rightarrow&2A^{\mu}(x),\no
\left(A_{\mu}^{(1)}(x)-A_{\mu}^{(2)}(x)\right)&\rightarrow&0.
\eeqa
Here the last equation is consistent with the fact that a non-BPS D-brane 
does not consist of two kinds of D-branes like a brane-antibrane system, 
because it indicates that there exist no degrees of 
freedom to separate one D-brane from the other D-brane. 
 
Next let us consider the further projection on a non-BPS D$p$-brane. As we saw 
in eq.(\ref{NDp}), the boundary state of a non-BPS D$p$-brane does not have 
its RR-sector, thus it seems invariant under the projection (\ref{projec}) 
about the spacetime fermion number $F_{\ss L}^{\ss S}$. 
However, the Chan-Paton factors transform in the nontrivial way
\beqa
\label{CPtr}
1~\rightarrow~1~~,~~\sigma_1~\rightarrow~-\sigma_1.
\eeqa
This can be understood if we know the following couplings on a non-BPS 
D$p$-brane
\beqa
\int C_{p}\wedge dT~,~\int B_{\ss NSNS}\wedge (*dA),
\eeqa  
where $C_{p},~B_{\ss NSNS},~T$ and $A$ are a RR p-form, a NSNS B-field, a 
tachyon and a gauge field, respectively. In fact these results 
will be obtained 
in chapter 3 by calculating the worldvolume action 
(see eq.(\ref{nonBPSaction11}) and eq.(\ref{wz-n})) on non-BPS systems. 
Here $B_{\ss NSNS}$ is invariant under the 
projection, while $C_{p}$ is not invariant. $T$ and $A$ have $\sigma_1$ and 
$1$ as Chan-Paton factors. Thus, in order that these actions are totally 
invariant, we have to assign the transformation rule (\ref{CPtr}) to $1$ 
and $\sigma_1$. As a result only the sector with the 
Chan-Paton factor $1$ remains 
under the projection, and a BPS D$p$-brane appears. This result is also 
consistent with the fact that $(-1)^{F^{\ss S}_{\ss L}}$ twist exchanges IIA 
and IIB (see figure \ref{del}).

\section{The Tachyon Condensation and the Descent Relation \label{tachcon}}
\hspace*{4.5mm}
Until now we have defined the brane-antibrane system and the non-BPS D-brane 
as unstable objects in Type II string theory. The instability stems from 
the tachyon particle which exists on these D-brane systems. Thus, we can expect
that these kinds of D-branes decay to something stable where tachyons do not
exist. Indeed this expectation turns out to be correct. 
Moreover, it is also known what type of objects those D-branes decay to. 
Especially two kinds of decay channels are known.

One decay channel is the case that these D-branes decay to the vacuum without 
any D-branes. In other words the vacuum does not include any open 
string degrees of freedom and it is described only by the closed string theory.
In this sense this vacuum is called closed string vacuum. The situation is 
similar to the Hawking radiation of a nonextremal blackhole\footnote{If we 
regard a Schwarzshild blackhole as a brane-antibrane pair or something, the 
quantitative argument of decay of D-branes, 
which will be explained in the later chapters, 
might also be applied to the Hawking radiation.}. 
Unstable D-branes also radiate various closed strings, 
lose their energy(mass) and finally decay to nothing. 

\begin{figure} [tbhp]
\begin{center}
\epsfxsize=100mm
\epsfbox{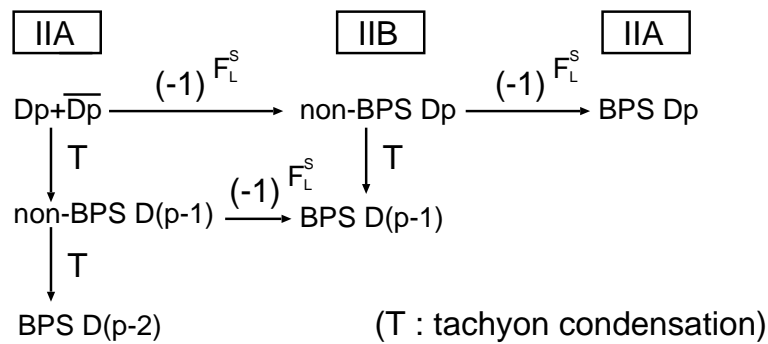}
\caption{The descent relation \label{del}}
\end{center}
\end{figure}

Moreover, there is a more interesting decay channel. A 
$Dp+\overline{Dp}$ system can decay to a non-BPS D$(p-1)$ brane or to a 
BPS $D(p-2)$ brane. Of course, the non-BPS $D(p-1)$ brane is an intermediate 
state, and finally it also decays to the closed string vacuum or 
a BPS D$(p-2)$-brane, which is stable. 
By getting together with the $(-1)^{F^{\ss S}_{\ss L}}$ twist in the last 
subsection we can draw the figure of the relation between 
various D-brane systems. It is called the descent relation \cite{sen16,sen27} 
(see figure \ref{del}). 

Moreover if we start with multiple non-BPS systems 
we can consider the decay of them to D-branes with lower 
codimensions more than two. This can be stated as
\beqa
\label{multiple}
2^{m-1}~\mbox{pairs of}~Dp+\overline{Dp}~&\rightarrow&~\left\{\begin{array}{l} 
\mbox{one non-BPS}~D(p-2m+1)\mbox{-brane},\\
\mbox{one BPS}~D(p-2m)\mbox{-brane},
\end{array}\right.\no
\no
2^{m-1}~\mbox{Non-BPS}~Dp\mbox{-brane}
&\rightarrow&~\mbox{one BPS}~D(p-2m+1)\mbox{-brane}.
\eeqa
The reason why we have
to prepare $2^{m-1}$ pairs is related to K-theory, which is a mathematical 
theory as an extension of cohomology \cite{Wi1,Ho1}. In this thesis we do not 
explain it.

Now the problem is how we can understand these decay channels. The easiest 
and the most intuitive way is to consider the field theory on these D-branes. 
In general it is known that the Yang-Mills theory exists on BPS D-branes 
because of the massless gauge field $A_{\mu}(x)$ on them. As we saw in the 
last subsection the massless gauge fields also exist in both 
brane-antibranes and non-BPS D-branes, therefore we expect that some kind 
of gauge theory exists on these D-brane systems, too. Indeed this expectation 
turns out to be correct and 
it is known that those field theories can be roughly 
approximated by scalar-QED or Yang-Mills-Higgs theories \cite{pe}. 

\begin{figure} [tbhp]
\begin{center}
\epsfxsize=100mm
\epsfbox{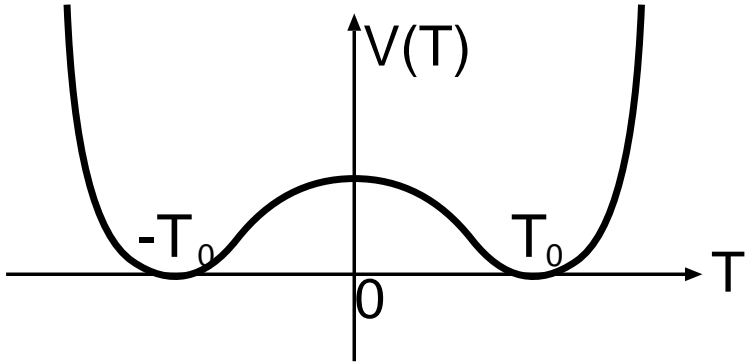}
\caption{A typical shape of the tachyon potential \label{potential}}
\end{center}
\end{figure}

As an example let us consider the decay of a non-BPS D9-brane and the field 
theory on it. For simplicity we ignore massive fields and all fermionic 
fields on these D-brane systems and we pay attention to the massless gauge 
field $A_{\mu}(x)$ and the tachyon field $T(x)$. 
The most important thing here is that the 
tachyon field is a scalar field, thus it has a scalar potential, which is 
called tachyon potential. In figure \ref{potential} we have drawn a typical 
form of tachyon potential. There are two kinds of extremum points $T=0$ 
and\footnote{There is the other minimum point $T=-T_0$, while the physical 
situation at this point is indistinguishable from $T=T_0$.} $T=T_0$. The 
former is the maximum point where the non-BPS D9-brane exists, while at the 
latter point there are no D-branes and it represents the vacuum with only 
closed strings. The first decay channel of a non-BPS D9-brane 
corresponds to rolling down the tachyon potential from the top to the bottom, 
and the tachyon field $T(x)$ obtains the expectation value $T_0$.

\begin{figure} [tbhp]
\begin{center}
\epsfxsize=100mm
\epsfbox{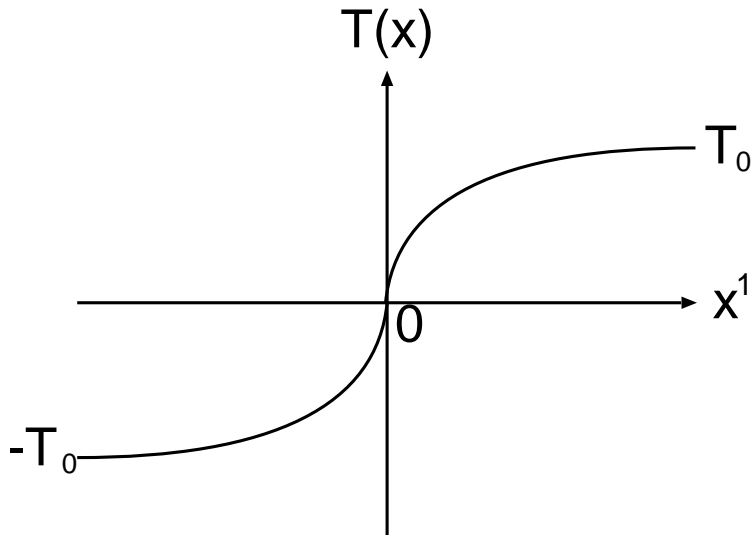}
\caption{The kink profile of the tachyon $T(x)$ \label{kink}}
\end{center}
\end{figure}

Now let us assume that the action for a non-BPS D9-brane is given by
\beqa
\label{action5}
S=\int d^{10}x \left(F^{\mu\nu}F_{\mu\nu}+\pa^{\mu}T\pa_{\mu}T+V(T)\right),
\eeqa
where $V(T)$ is the tachyon potential like figure \ref{potential}. In this 
action we can consider a topologically nontrivial 
solution to the equation of motion of the tachyon field $T(x)$. 
The profile of the tachyon can be given by
\beq
T(x)\rightarrow\left\{\begin{array}{ll}
T_0~~&(\mbox{at}~x^1\rightarrow\infty),\\  
-T_0~~&(\mbox{at}~x^1\rightarrow-\infty).
\end{array}\right.
\eeq
This type of configuration is called a kink (see figure \ref{kink}). From 
fig. \ref{kink} we can see that the tachyon field deletes the energy density 
of a non-BPS D9-brane almost everywhere except around $x^1=0$. 
Therefore we can 
guess that a codimension one D-brane remains at $x^1=0$ 
after the decay of a non-BPS 
D9-brane. This is just a BPS D8-brane, and this process corresponds to the 
second decay channel of non-BPS D9-branes.

In both decay channels 
the tachyon field obtains the expectation value and this kind 
of process is called tachyon condensation. However, the conjecture of decay of 
D-branes can not exactly 
be verified by using the conventional field theory action 
like eq.(\ref{action5}). Indeed, 
this action can not explain the fact that at $T=T_0$ there does not remain any 
matter\footnote{After the tachyon condensation all the fields on the D-brane 
should get the {\sl infinite} mass and totally decouple from the theory.} from 
open strings because of the absence of any D-branes. According to the 
conjecture, the action $S$ should be equal to zero at $T=T_0$, while in the 
conventional field theory the kinetic term $F^{\mu\nu}F_{\mu\nu}
+\pa^{\mu}T\pa_{\mu}T$ remains. 

To verify these correctly we have to proceed to a more extended field 
theory for strings, which includes $\alpha^{\prime}$ corrections. It is 
called string field theory. Indeed the action of string field theory 
includes the 
infinite number of derivatives, and it is {\sl nonlocal}. The nonlocality of 
that action explains the complete vanishing of the matter at the minimum 
point of tachyon potential. In the next chapter we will verify the conjecture 
of tachyon condensation by using string field theory. 

\section{Are Non-BPS Systems Physical?}
\hspace*{4.5mm}
Now we have explained all kinds of unstable D$p$-brane systems 
in Type II theories, while 
the situation in Type I theory, which we do not deal with in this thesis, is 
a little more complicated. 

In Type I theory there appear several {\sl stable} 
non-BPS D$p$-branes. The reason why they become stable comes from the fact 
that the orientifold projection $\frac{1+\Omega}{2}$ in the open string 
partition function projects out the tachyon field. It is known that the 
stable non-BPS D0-brane plays an important role in the discussion about the 
Type I and Heterotic 
duality \cite{Po3}. According to the prediction of string duality between 
Type I and Heterotic theory, there should exist an object in the Heterotic 
theory which corresponds to the non-BPS D0-brane in Type I theory. 
The duality between these 
two theories is known to be a strong-weak duality (S-duality), thus we can 
predict that the corresponding object in the 
Heterotic theory is not a solitonic 
object but a fundamental object because the non-BPS D-brane is a solitonic 
object in Type I theory. In fact, the author in \cite{sen14} showed that the 
non-BPS D0-brane (non-BPS D-particle) in Type I theory corresponds to the 
firstly-excited massive state of a fundamental closed string in the Heterotic 
theory, by the discussion about the comparison of quantum numbers between 
them.
 
Since then, there have appeared 
other similar arguments about non-BPS systems by 
using string dualities like Type II/$K3$ and Hetero$/T^4$ duality 
\cite{bergman3}. In this way we started to recognize that the non-BPS D-branes 
(and also other non-BPS systems) are important {\sl physical} objects in 
string theory. 

\chapter{Off-Shell Description of Tachyon Condensation \label{OFFSHELL}}
\hspace*{4.5mm}
In the previous chapter we have introduced the non-BPS D-brane systems in 
string theory and have explained the decay of those D-branes called 
tachyon condensation. In this chapter we would like to 
prove that conjecture in the quantitative way. The best and 
natural way to describe the tachyon condensation is to use string field 
theory. Especially since we deal with the decay of D-branes, 
we need open string field theory. In this theory we usually choose and fix 
the closed string background and we can obtain the ``stringy" type of field 
theory picture of open strings like the tachyon potential which was explained 
in section \ref{tachcon}.  

Now several kinds of formulations of open string field 
theory are known. The most famous one is Witten's cubic string field 
theory \cite{WiCSFT,Wisu} or its nonpolynomial extension by Berkovits 
\cite{Ber1}. Of course, these string field theories have 
given good support \cite{bosonic,super,lump} to the conjecture by 
calculating the tachyon potential, although its analysis called 
the level truncation method is approximate one. On the 
other hand, one of string field theories called boundary string field 
theory (BSFT) \cite{Wi2,Sh1,LiWi,Ma,NiPr} 
is appropriate to describe the tachyon 
condensation {\sl exactly} \cite{Wi2,GeSh1,KuMaMo1,KuMaMo2,TaTeUe,KrLa}. 
In this chapter we will introduce the formulation of boundary string field 
theory, and by using it we will calculate the action of string field theory 
and the tachyon potential. In addition we will prove the conjecture about the 
tachyon condensation.    

\section{The Relation between the Worldsheet RG Flow and the Tachyon 
Condensation \label{RGTach}}
\hspace*{4.5mm}
In this and next sections we will introduce the formalism of boundary string 
field theory (BSFT). Here we will discuss it in both of the 
bosonic string \cite{Wi2,Sh1,LiWi} and the superstring \cite{Ma,NiPr} in the 
parallel way because there is not much formal 
difference between two. 

BSFT is one type of string field theories which is defined on 
{\sl the total boundary space} of the two dimensional sigma model. We 
start with the two dimensional sigma model action $I_0$ with the one 
dimensional boundary action $I_B$ on the disk. 
$I_0$ is defined inside the disk as the bulk action whose general form is 
written in the following way:
\beqa
\label{action0}
I_0&=&\frac{1}{4\pi}\int d^2z d^2\theta[G_{\mu\nu}({\bf X})
+B_{\mu\nu}({\bf X})]D_{\bar{\theta}}{\bf X}^{\nu}D_{\theta}{\bf X}^{\mu}\no
&=&\frac{1}{4\pi}\int d^2z\Bigl\{[G_{\mu\nu}(X)+B_{\mu\nu}(X)]\pa X^{\mu}
\bar{\pa}X^{\nu}\no
&&~~~~~~~~~~~+G_{\mu\nu}(X)(-i\psi^{\mu}_L\bar{{\cal D}}\psi^{\nu}_L
+i\psi^{\mu}_R{\cal D}\psi^{\nu}_R)+\frac{1}{2}R_{\mu\nu\rho\sigma}(X)
\psi^{\mu}_L\psi^{\nu}_L\psi^{\rho}_R\psi^{\sigma}_R\Bigr\},
\eeqa
where
\beqa
D_{\theta}&=&\pa_{\theta}+\theta\pa~~,~~D_{\bar{\theta}}=\pa_{\bar{\theta}}
+\bar{\theta}\bar{\pa},\no
{\bf X}^{\mu}&=&X^{\mu}+i^{\frac{1}{2}}\theta\psi^{\mu}_L
-i^{-\frac{1}{2}}\bar{\theta}\psi^{\mu}_R+\theta\bar{\theta}F^{\mu},\no
\bar{{\cal D}}\psi^{\nu}_L&=&\bar{\de}\psi^{\nu}_L
+\left[\Gamma^{\nu}_{\rho\sigma}(X)+\frac{1}{2}H^{\nu}_{~\rho\sigma}(X)\right]
\bar{\de}X^{\rho}\psi_L^{\sigma},\no
{\cal D}\psi^{\nu}_R&=&\de\psi^{\nu}_R
+\left[\Gamma^{\nu}_{\rho\sigma}(X)-\frac{1}{2}H^{\nu}_{~\rho\sigma}(X)\right]
\de X^{\rho}\psi_R^{\sigma},\no
\eeqa
$F^{\mu}$ is an auxiliary field in the worldsheet 
and $H^{\nu}_{~\rho\sigma}(X)$ is the 
field strength of NSNS B-field $B_{\mu\nu}(X)$. Here we have omitted all terms 
with ghosts $(b,c~;~\beta,\gamma)$. 
If we want to consider the bosonic string theory we have only to remove 
the terms including worldsheet fermions $(\psi^{\mu}_L,~\psi^{\mu}_R)$ and 
superconformal ghosts $(\beta,~\gamma)$ from the above action.  

On the other hand, the action $I_B$ is defined on the boundary of disk. 
This boundary part is the most important part in BSFT and it consists 
of {\sl all kinds} of fields on the boundary of disk. For example, the 
most general form of boundary action $I_B$ for the bosonic string theory 
can be written by
\beqa
\label{action1.9}
I_B=
\int^{\pi}_{-\pi} \frac{d\tau}{2\pi}~{\cal V}(X^{\mu}(\tau),b(\tau),c(\tau)),
\eeqa
where
\beqa
\label{action2}
{\cal V}(X^{\mu},b,c)&=&T(X)+\dot{X}^{\mu}A_{\mu}(X)
+\dot{X}^{\mu}\dot{X}^{\nu}C_{\mu\nu}(X)+\ddot{X}^{\mu}D_{\mu}(X)+\cdots\cdots
\no
&\equiv&\sum_i\lambda^i{\cal V}_i(X^{\mu},b,c).
\eeqa
Here in the first line of eq.(\ref{action2}) we have omitted terms with 
$b,~c$ ghosts, and $\lambda^i$ are coupling constants in the two 
dimensional sigma model, which represent various target space fields. For 
example if we expand the first term of eq.(\ref{action2}) like 
\beqa
T(X)=a_0+a_{1\mu}X^{\mu}+a_{2\mu\nu}X^{\mu}X^{\nu}+\cdots\cdots,
\eeqa
various coefficients $a_0,a_{1\mu},a_{2\mu\nu},\cdots$ are a part of 
coupling constants $\lambda^i$. 
The boundary action (\ref{action2}) is for the bosonic string,
while there are some subtle points to write down the explicit form of boundary 
action for the superstring as we will see later in section \ref{superaction}. 
For this reason we will explain only the summary of the idea of BSFT by using 
the action (\ref{action2}). The idea is completely the same as that of the 
superstring. 

In the action (\ref{action2}) $T(X)$ represents an open string tachyon, 
$A_{\mu}(X)$ a gauge field and the other fields like $C_{\mu\nu}(X)$ and 
$D_{\mu}(X)$ massive modes of open strings on a bosonic D-brane. In the 
language of statistical mechanics 
the tachyon field has a dimension less than 1, 
which corresponds to a relevant operator, the gauge field has dimension 1 
(marginal), and the other fields have dimensions greater than 1 (irrelevant). 
This fact means that we have included all operators which break the conformal 
invariance. Note that we have broken the conformal invariance only on 
the boundary of disk and not inside the disk. This means that we 
fix a closed string background to satisfy the equation of motion of 
(super)gravity (``on-shell"), while we do not require the equation of motion 
of (super)Maxwell theory (``off-shell"). In general the off-shell string theory
is described by (open) string field theory. Therefore, to consider the sigma 
model action like eq.(\ref{action2}) is one of descriptions of extending an 
on-shell theory to an off-shell one, and boundary string field theory (BSFT) 
is based on this idea. Here, the name ``boundary" comes from the boundary 
perturbation in action (\ref{action2}). Moreover, another feature of this 
theory is that we can see the explicit background independence of closed 
strings in action (\ref{action0}). For this reason, the original name of this 
formalism in the paper \cite{Wi2} was ``background independent open string 
field theory"\footnote{However, because of the limitation of 
calculability, we can not avoid choosing one of simple 
closed string backgrounds like the flat space or orbifolds.}.

\begin{figure} [tbhp]
\begin{center}
\epsfxsize=100mm
\epsfbox{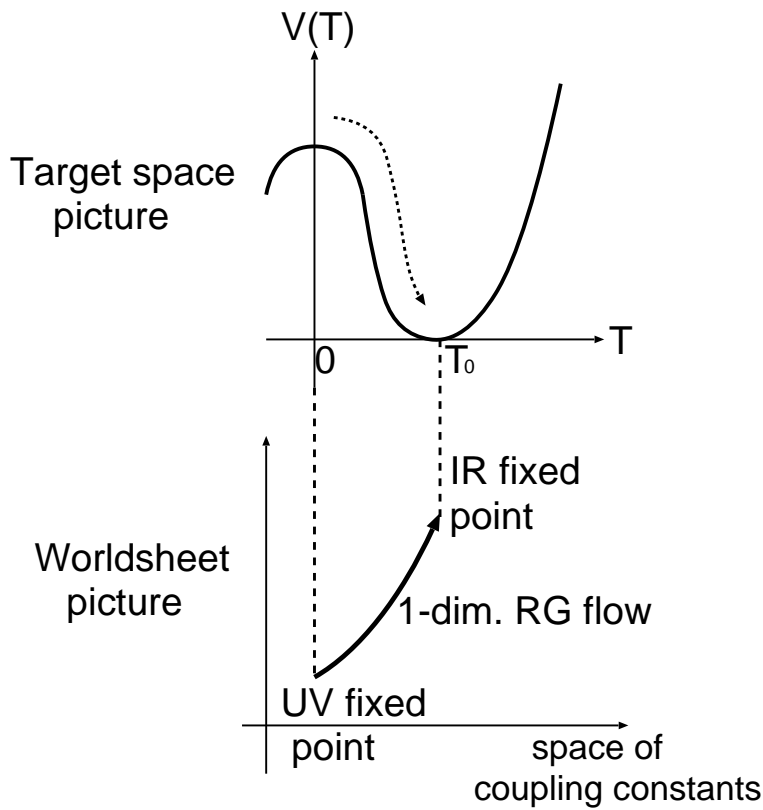}
\caption{The relation between one-dimensional renormalization group flow and 
the tachyon potential. 
On top of the potential unstable D-branes live. After its tachyon condensation 
those D-branes disappear at the bottom of the potential. \label{Rg}}
\end{center}
\end{figure}

Before the construction of BSFT, we will explain the relation between the 
renormalization group of the sigma model and the physical picture of 
decay of D-branes. When we consider the general nonlinear sigma model which 
breaks the conformal invariance like eq.(\ref{action2}), the 
renormalization group flow starts and the theory flows out of a conformal 
fixed point and leaves for another fixed point. Especially relevant 
operators play the most important role to cause the renormalization group flow,
 and these relevant operators are tachyons which can be seen in 
eq.(\ref{action2}). Therefore, in the language of the target space of string 
theory, we can identify the renormalization group flow as the process of 
tachyon condensation, which means that the tachyon fields have their 
expectation values. 
In string theory (open string) tachyon fields appear when a D-brane is 
perturbatively unstable, thus we can also say that the renormalization group 
flow represents the decay process of the D-brane. The UV fixed point where the 
renormalization group starts corresponds to the situation that there is an 
unstable D-brane in the target space. 
When that flow reaches the final IR fixed point, 
the D-brane completely decays and a stable vacuum or a stable D-brane 
appears. However, to discuss the decay of D-branes in the quantitative way we 
need some physical quantity. That is just the target space 
action $S$ (action of string field theory) for fields 
$T(x),~A_{\mu}(x),~\cdots$ in eq.(\ref{action2}) or tachyon potential $V(T)$, 
which is derived from $S$. Indeed there is a relation between the tachyon 
potential and the renormalization group flow \cite{HaKuMa}, and it is drawn in 
figure \ref{Rg}. 

In figure \ref{Rg} the conformal 
invariance is kept only at the extremum points $T=0$ and $T=T_0$. In the 
conventional first-quantized string theory we fix the open string background 
at either $T=0$ or $T=T_0$ and usually investigate the scatterings 
(S-matrices) of the fluctuating fields around $T=0$ or $T=T_0$. Therefore 
from the first-quantized string theory we can not see how two vacuums $T=0$ and
$T=T_0$ are related or connected with each other. Its relation can not be 
understood without the tachyon potential. In this chapter we will explain that 
BSFT determines the action of string field theory and the tachyon potential.

\section{The Batalin-Vilkovisky Formulation of Boundary String Field Theory 
\label{BVformalism}}   
\hspace*{4.5mm}
In this section we will introduce the Batalin-Vilkovisky(BV) formalism 
\cite{BV} because the construction of BSFT is based on it \cite{Wi2}. 
The BV formalism has been used in 
order to construct various string field theories like cubic open string field 
theory and closed string field theories. 
   
First we start with a supermanifold $M$ equipped with U(1) 
symmetry that we will call the ghost number.  The essential element in the BV 
formalism is the symplectic structure and we can define a fermionic 
non-degenerate two-form $\omega$ with U(1) charge $-1$ which is closed, 
$d\omega=0$ . If we 
define the local coordinate $\lambda^i~(i=1,2,\cdots,\mbox{dim}M)$ 
to express the manifold, then we can 
consider the Poisson bracket given by
\beqa
\{F,G\}=\frac{\de_r F}{\de \lambda^i}\omega^{ij}\frac{\de_l G}{\de \lambda^j},
\eeqa
where $\omega^{ij}$ is the inverse matrix of $\omega_{ij}$, and 
$\frac{\de_r}{\de \lambda^i},~\frac{\de_l}{\de \lambda^j}$ 
denote the right-derivative and the left-derivative respectively. By using 
these elements we can define the 
{\sl classical} master equation for the action $S$ of string field theory 
as follows:
\beqa
\label{master}
\{S,S\}=0.
\eeqa

This is the standard description of the BV formalism, while we can 
find a more convenient form of equation which is equivalent to 
eq.(\ref{master}). To do it let us introduce a fermionic vector field 
$V=V^i\frac{\de}{\de \lambda^i}$ with U(1) charge $1$. If we consider the 
infinitesimal diffeomorphism $\lambda^i\rightarrow \lambda^i+\epsilon V^i$, 
the symplectic form $\omega$ transforms as $\omega\rightarrow\omega
+\epsilon({\cal L}_V\omega)$,
where ${\cal L}_V$ is the Lie derivative along the direction of the vector 
field $V$. It can also be defined by 
\beqa
\label{defcalL}
{\cal L}_V=di_V+i_Vd,
\eeqa
where $i_V$ is the operation of contraction with $V$. If we assume that $V$ 
is a killing vector field, it generates a symmetry of $\omega$ and satisfies
\beqa
\label{sym10}
{\cal L}_V \omega=0.
\eeqa
{}From the definition of ${\cal L}_V$ in eq.(\ref{defcalL}) and the fact that 
$\omega$ is closed ($d\omega=0$), it can also be written as
\beqa
\label{sym11}
d(i_V\omega)=0.
\eeqa

Here we claim that if the vector field $V$ is nilpotent ($V^2=0$) and 
satisfies the killing condition (${\cal L}_V\omega=0$), the 
following equation is equivalent to the master equation (\ref{master}):
\beqa
\label{master2}
i_V\omega=dS.
\eeqa 
The component form of this equation is given by
\beqa
\label{master2.1}
V^j\omega_{ji}=\frac{\de_l S}{\de \lambda^i}.
\eeqa
Note that the killing condition (\ref{sym11}) guarantees that $S$ exists as a 
local solution to equation (\ref{master2}).

Let us prove the above statement. First we can easily check that the following 
identity holds 
\beqa
{\cal L}_Vi_W+i_W{\cal L}_V=i_{VW+WV},
\eeqa
where $V$ and $W$ are any fermionic vector fields. By setting $W=V$ and using 
the nilpotency of $V$ and the killing condition (\ref{sym10}) we can obtain
\beqa
{\cal L}_Vi_V\omega=0.
\eeqa
From the definition of ${\cal L}_V$ in eq.(\ref{defcalL}) and equation 
(\ref{sym11}), the above equation becomes
\beqa
di_Vi_V\omega=d\left(V^i\frac{\de_l S}{\de \lambda^i}\right)=d\{S,S\}=0.
\eeqa
Here we have used equation (\ref{master2.1}). Thus, we can find that the 
nilpotency $V^2=0$ and the killing condition (\ref{sym10}) lead equation 
(\ref{master2}) to $\{S,S\}=\mbox{constant}$. If we consider that this 
equation should also hold at on-shell regions ($dS=0$), 
the value of $\{S,S\}$ becomes equal to zero. This is the proof of the 
equivalence between the classical master equation (\ref{master}) and 
eq.(\ref{master2}).

Now we are ready to apply this BV-formalism to the sigma model action for 
constructing BSFT. In BSFT we regard $\lambda^i$ and $S(\lambda)$ in 
eq.(\ref{master2.1}) as the target space fields and the action of string field 
theory, respectively. Thus, the supermanifold $M$ is the infinite dimensional 
space ($\mbox{dim}M=\infty$) spanned by all kinds of target space fields. The 
difficult thing here is to construct $V,\omega$ in eq.(\ref{master2}) by 
using the language of valuables in the two-dimensional sigma model. 
First we can identify the fermionic killing vector field $V$ as the bulk BRST 
operator $Q_B$ in the worldsheet since this operator satisfies the nilpotency 
condition and has U(1) charge $1$ if we define U(1) charge by the worldsheet 
ghost number. Next we express the closed two-form $\omega$ by using boundary 
fields in action (\ref{action2}) in the following 
way
\beqa
\omega(\lambda)&=&
\frac{1}{2}\omega_{ij}(\lambda)d\lambda^i\wedge d\lambda^j,\no
&&\no
\omega_{ij}(\lambda)&=&\left\{\begin{array}{ll}
{\displaystyle \frac{1}{2}\int^{\pi}_{-\pi} \frac{d\tau_1}{2\pi}
\int^{\pi}_{-\pi} \frac{d\tau_2}{2\pi} 
~\la c{\cal V}_i(\tau_1)c{\cal V}_j(\tau_2)
\lb_{\lambda}}~~&(\mbox{bosonic string}),\\
&\\
{\displaystyle \frac{1}{2}\int^{\pi}_{-\pi}  \frac{d\tau_1}{2\pi}
\int^{\pi}_{-\pi}  \frac{d\tau_2}{2\pi} 
~\la ce^{-\phi}{\cal W}_i(\tau_1)ce^{-\phi}{\cal W}_j
(\tau_2)\lb_{\lambda}}~~&(\mbox{superstring}),
\end{array}\right.
\eeqa
where $c(\tau)$ and $e^{-\phi}(\tau)$ are the ghost and the bosonized 
superconformal ghost, respectively. ${\cal V}_i(\tau_1)$ and 
${\cal V}_j(\tau_2)$ are vertex operators of fields in the boundary action 
(\ref{action2}), while ${\cal W}_i(\tau_1)$ and ${\cal W}_j(\tau_2)$ are 
those with $-1$ picture\footnote{If 
we want to use only vertex operators with $0$ picture, 
the two-form $\omega$ can be written as follows \cite{NiPr}
\beqa
\omega_{ij}(\lambda)=\frac{1}{2}
\int^{\pi}_{-\pi} \frac{d\tau_1}{2\pi}
\int^{\pi}_{-\pi} \frac{d\tau_2}{2\pi} 
~\la Y(\tau_1)c{\cal V}_i(\tau_1)Y(\tau_2)
c{\cal V}_j(\tau_2)\lb_{\lambda},
\eeqa
where $Y(\tau)$ is the inverse picture changing operator.}. The correlator 
$\la\cdots\lb_{\lambda}$ is evaluated with a non-conformal 
boundary action like eq.(\ref{action2}), and in the path-integral formalism 
it is represented by
\beqa
\la\cdots\lb_{\lambda}=\left\{\begin{array}{ll}
{\displaystyle \int DX~\cdots~\exp[-I_0(X)-I_B(X)]}~~
&(\mbox{bosonic string}),\no
&\no
{\displaystyle \int DX D\psi~\cdots~\exp[-I_0(X,\psi)-I_B(X,\psi)]}
~~&(\mbox{superstring}).
\end{array}\right.
\eeqa
where $\lambda^i$ is included in the boundary action $I_B$ like 
eq.(\ref{action2}). This definition of $\omega$ is quite 
desirable. Indeed it has the correct U(1) charge because the worldsheet ghost 
number\footnote{Note that the vacuum has ghost number $-3$.} of $\omega$ is 
$-1$. Moreover, it satisfies the necessary identities 
$d\omega={\cal L}_V\omega=0$ and the non-degeneracy (for proof, see 
\cite{Wi2,NiPr}). 

Therefore we can rewrite equation (\ref{master2.1}) in terms 
of the boundary fields as follows
\beqa
\label{defBSFT}
\frac{\de S}{\de \lambda^i}=\left\{\begin{array}{ll}
{\displaystyle \frac{1}{2}\int^{\pi}_{-\pi}\frac{d\tau_1}{2\pi}
\int^{\pi}_{-\pi}\frac{d\tau_2}{2\pi}
~\la c{\cal V}_i(\tau_1)\{Q_B,c{\cal V}(\tau_2)\}
\lb_{\lambda}}~~&(\mbox{bosonic string}),\\
&\\
{\displaystyle \frac{1}{2}\int^{\pi}_{-\pi}\frac{d\tau_1}{2\pi}
\int^{\pi}_{-\pi}\frac{d\tau_2}{2\pi}
~\la ce^{-\phi}{\cal W}_i(\tau_1)\{Q_B,ce^{-\phi}
{\cal W}(\tau_2)\}\lb_{\lambda}}~~&(\mbox{superstring}),
\end{array}\right.
\eeqa
where ${\cal V}=\lambda^i{\cal V}_i$ and 
${\cal W}=\lambda^i{\cal W}_i$. Note that we have used the following 
relation 
\beqa
\left\{Q_B,c{\cal V}(\tau_2)\right\}=\left\{Q_B^j\frac{\pa}{\pa\lambda^j},
\lambda^kc{\cal V}_k(\tau_2)\right\}=Q_B^j\left(c{\cal V}_j(\tau_2)\right).
\eeqa   

The next step is to solve the above equation and to obtain the explicit form 
of action $S$. It is known that the solution to 
this equation can be obtained if we restrict possible boundary 
terms in action (\ref{action1.9}) to those which do not include the worldsheet 
ghosts $(b,~c)$ and superghosts $(\beta,~\gamma)$. 
In this case\footnote{In reality we have already used this 
assumption in the expressions before. If we do not assume this the situation 
becomes a little complicated \cite{Wi2,NiPr}. What this assumption means is 
explained in section \ref{meaning}.} the action $S$ is given by
\beqa
\label{generalsol}
S(\lambda)=\left\{\begin{array}{cl}
{\displaystyle 
\left(\beta^i(\lambda)\frac{\de}{\de\lambda^i}+1\right)Z(\lambda)}~
~&(\mbox{bosonic string}),\\
&\\
{\displaystyle Z(\lambda)}~~&(\mbox{superstring}),
\end{array}\right. 
\eeqa
where $\beta^i(\lambda)$ is the beta function of the coupling constant 
$\lambda^i$ and $Z(\lambda)$ is the partition function of the matter part on 
the disk, which is given by
\beqa
Z(\lambda)=\left\{\begin{array}{ll}
{\displaystyle \int DX~\exp[-I_0(X)-I_B(X)]}~~&(\mbox{bosonic string}),\no
&\no
{\displaystyle \int DX D\psi~\exp[-I_0(X,\psi)-I_B(X,\psi)]}
~~&(\mbox{superstring}).
\end{array}\right.
\eeqa
The important point is that $I_B$ is a non-conformal boundary action like 
eq.(\ref{action1.9}) which does not include the worldsheet ghosts $(b,~c)$ and 
superghosts $(\beta,~\gamma)$. 

Now let us prove this relation. First we can easily calculate the 
anticommutators in eq.(\ref{defBSFT}) and its result is
\beqa
\label{anticom}
\{Q_B,c{\cal V}_i(\tau_2)\}&=&(1-\Delta_i)c\dot{c}{\cal V}_i(\tau_2),\no
\{Q_B,c{\cal W}_i(\tau_2)\}&=&\left(\frac{1}{2}-\Delta_i\right)
c\dot{c}{\cal W}_i(\tau_2),
\eeqa
where $\Delta_i$ is the conformal dimension of the primary operator 
${\cal V}_i(\tau)$ or ${\cal W}_i(\tau)$. 
From this we can obtain the following expression 
\beqa
\label{noncovS}
\frac{\de S}{\de \lambda^i}=\left\{\begin{array}{ll}
{\displaystyle 
-(1-\Delta_j)\lambda^j G_{ij}(\lambda)}~~&(\mbox{bosonic string}),\\
&\\
{\displaystyle -\left(\frac{1}{2}-\Delta_j\right)\lambda^j G_{ij}(\lambda)}~~&
(\mbox{superstring}),
\end{array}\right.
\eeqa
where
\beqa
\label{G}
G_{ij}(\lambda)=\left\{\begin{array}{ll}
{\displaystyle 
2\int^{\pi}_{-\pi}\frac{d\tau_1}{2\pi}\int^{\pi}_{-\pi}\frac{d\tau_2}{2\pi} 
\sin^2
\left(\frac{\tau_1-\tau_2}{2}
\right)\la {\cal V}_i(\tau_1){\cal V}_j(\tau_2)\lb_{\lambda}}~~&
(\mbox{bosonic string}),\\
&\\
{\displaystyle 
\int^{\pi}_{-\pi}\frac{d\tau_1}{2\pi}\int^{\pi}_{-\pi}\frac{d\tau_2}{2\pi} 
\sin\left(\frac{\tau_1-\tau_2}{2}\right)\la 
{\cal W}_i(\tau_1){\cal W}_j(\tau_2)\lb_{\lambda}}~~&(\mbox{superstring}).
\end{array}\right.
\eeqa 
Here we have used a three-point correlation function of ghosts and a two-point 
one of superconformal ghosts 
\beqa
\la c(\tau_1)c\dot{c}(\tau_2)\lb&=&-4\sin^2
\left(\frac{\tau_1-\tau_2}{2}\right),\no
\la e^{-\phi}(\tau_1)e^{-\phi}(\tau_2)\lb&=&\frac{1}{2}
\sin^{-1}\left(\frac{\tau_1-\tau_2}{2}\right).
\eeqa
The ghost part and superconformal ghost part appear only in the above 
correlation functions because we have imposed the assumption that the boundary 
action $I_B$ and the vertex operators ${\cal V}_i(\tau),~{\cal W}_i(\tau)$ do 
not include any terms with $b,c,\beta$ and $\gamma$. 
Therefore the later calculation depends only on the matter part.  

Here we will use a trick to prove eq.(\ref{generalsol}) in the formal 
way. First we start with the case of the bosonic string theory. We assume that 
the matter vertex operators in eq.(\ref{action2}) consist of two kinds of 
operators which are completely decoupled from each other. 
We denote one by ${\cal V}_{1i}$ and the other by ${\cal V}_{2i}$, and the 
vertex operator ${\cal V}$ is written as
\beqa
{\cal V}=\sum_i\lambda_1^i{\cal V}_{1i}
+\sum_j\lambda_2^j{\cal V}_{2j}.
\eeqa
Inserting this into eq.(\ref{G}) we can see that there appear three 
kinds of two point functions : $\la{\cal V}_{1i}(\tau_1){\cal V}_{1j}(\tau_2)
\lb_{\lambda},~
\la{\cal V}_{2i}(\tau_1){\cal V}_{2j}(\tau_2)\lb_{\lambda}$ and 
$\la{\cal V}_{1i}(\tau_1){\cal V}_{2j}(\tau_2)\lb_{\lambda}$. The third type 
of two-point functions splits into products of one-point functions 
$\la{\cal V}_{1i}(0)\lb_{\lambda_1}\times\la{\cal V}_{2j}(0)
\lb_{\lambda_2}$, where the dependence of $\tau$ is dropped because of the 
rotational invariance of $\tau$ on the boundary of disk. If we use an 
identity of a simple integral
\beqa
2\int^{\pi}_{-\pi}\frac{d\tau_1}{2\pi}\int^{\pi}_{-\pi}\frac{d\tau_2}{2\pi}
\sin^2\left(\frac{\tau_1-\tau_2}{2}\right)=1,
\eeqa
we can obtain the following expression of $dS\left(=\sum_i d\lambda^i
\frac{\de S}{\de \lambda^i}\right)$:
\beqa
\label{proof1}
dS&=&
\left(a_2^j\frac{\de Z_2}{\de\lambda^{j}_2}(\lambda_2)\right)dZ_1(\lambda_1)
+\left(a_1^j\frac{\de Z_1}{\de \lambda^{j}_1}(\lambda_1)\right)dZ_2(\lambda_2)
\no
&&~~~~~~~~~~+Z_2(\lambda_2)\left(d\lambda^{i}_1V_{1i}(\lambda_{1})\right)
+Z_1(\lambda_1)\left(d\lambda^{i}_2V_{2i}(\lambda_{2})\right),
\eeqa
where
\beqa
a_1^j(\lambda_1)&=&-(1-\Delta_{1j})\lambda^{j}_1,\no
a_2^j(\lambda_2)&=&-(1-\Delta_{2j})\lambda^{j}_2,\no
V_{1i}(\lambda_1)&=&-(1-\Delta_{1j})\lambda^{j}_1G_{ij}(\lambda_1),\no
V_{2i}(\lambda_2)&=&-(1-\Delta_{2j})\lambda^{j}_2G_{ij}(\lambda_2).
\eeqa
Here $Z_1(\lambda_1)~\left(\mbox{or}~Z_2(\lambda_2)\right)$ 
is the disk partition function whose boundary action $I_B$ 
is given by $\sum_i\lambda_1^i{\cal V}_{1i}
~\left(\mbox{or}~\sum_j\lambda_2^j{\cal V}_{2j}\right)$.
$dZ_1(\lambda_1)~\left(\mbox{or}~dZ_2(\lambda_2)\right)$ is defined by 
$\sum_i d\lambda_1^i\frac{\pa Z_1}{\pa \lambda_1^i}
~\left(\mbox{or}~\sum_j 
d\lambda_2^j\frac{\pa Z_2}{\pa \lambda_2^j}\right)$, and 
$\frac{\pa Z_1}{\pa\lambda_1^i}
~\left(\mbox{or}~\frac{\pa Z_2}{\pa\lambda_2^j}\right)$ 
is equal to the one-point function $\la{\cal V}_{1i}\lb_{\lambda_1}
~\left(\mbox{or}~\la{\cal V}_{2j}\lb_{\lambda_2}\right)$. 
Since we have assumed 
that the matter sector is split into two parts which are completely decoupled 
from each other, we can easily see that the total disk partition function 
$Z(\lambda)$ is equal to $Z_1(\lambda_1)Z_2(\lambda_2)$.

Next, due to $ddS=0$ the right-hand side of eq.(\ref{proof1}) obeys some 
constraint equation. 
Especially the term including $d\lambda_1^i\wedge d\lambda_2^j$ should be
\beqa
0&=&dZ_1(\lambda_1)\wedge\left\{d\lambda^{i}_2
V_{2i}(\lambda_{2})
-d\left(a_2^j\frac{\de Z_2}{\de\lambda^{j}_2}(\lambda_2)\right)\right\}\no
&&+\left\{d\left(a_1^j\frac{\de Z_1}{\de \lambda^{j}_1}(\lambda_1)\right)
-d\lambda^{i}_1V_{1i}(\lambda_{1})\right\}\wedge dZ_2(\lambda_2).
\eeqa
This equation implies 
that we can divide this equation into two parts which depend 
only on either $\lambda_1$ or $\lambda_2$ and we can obtain the following 
equations:
\beqa
d\left(a_1^j\frac{\de Z_1}{\de \lambda^{j}_1}(\lambda_1)\right)
-d\lambda^{i}_1V_{1i}(\lambda_{1})&=&-kdZ_1(\lambda_1),\no
d\lambda^{i}_2V_{2i}(\lambda_{2})
-d\left(a_2^j\frac{\de Z_2}{\de\lambda^{j}_2}(\lambda_2)\right)&=&
kdZ_2(\lambda_2),
\eeqa
where $k$ is some constant. By using these equations and erasing 
$V_{1i}$ and $V_{2i}$, equation (\ref{proof1}) becomes 
\beqa
dS=d\left\{\left(a^i_1(\lambda_1)\frac{\de}{\de\lambda_1^i}
+a^i_2(\lambda_2)\frac{\de}{\de\lambda_2^i}+k\right)Z_1(\lambda_1)
Z_2(\lambda_2)\right\},
\eeqa
and up to a constant the string field action $S$ is determined as
\beqa
S=\left(a^i_1(\lambda_1)\frac{\de}{\de\lambda_1^i}
+a^i_2(\lambda_2)\frac{\de}{\de\lambda_2^i}+k\right)Z_1(\lambda_1)
Z_2(\lambda_2).
\eeqa

Until now we have assumed that the matter system consists of two 
parts which are decoupled from each other, while the above result can 
naturally be extended to the general one as 
\beqa
\label{noncovsol}
S(\lambda)=\left(a^i(\lambda)\frac{\de}{\de\lambda^i}+k\right)Z(\lambda),
\eeqa
where
\beqa
a^i(\lambda)=-(1-\Delta_{i})\lambda^{i}.
\eeqa
Here the parameter $k$ is related to overall normalization of the partition 
function $Z$ which depends on the convention, and it can not be determined 
from eq.(\ref{G}). Thus, we fix it by hand to $k=1$. 

By the way, the result (\ref{noncovsol}) does not look covariant under the 
reparametrization of $\lambda$ like 
$\lambda^i\rightarrow f^i(\lambda)$. To solve this issue 
let us recall that the 
general form of beta function of $\lambda^i$ is given by 
\beqa
\beta^i(\lambda)=-(1-\Delta_i)\lambda^i+O(\lambda^2).
\eeqa   
Next notice that in the renormalization group theory the beta function 
$\beta^i(\lambda)$ can be set to the linear function of $\lambda^i$ if we take 
the appropriate redefinition of $\lambda^i$. Therefore we can also write 
equation (\ref{noncovsol}) as
\beqa
\label{covsol}
S(\lambda)=\left(\beta^i(\lambda)\frac{\de}{\de\lambda^i}+1\right)Z(\lambda).
\eeqa
In the same way we can replace eq.(\ref{noncovS}) with:
\beqa
\label{covS}
\frac{\de S}{\de \lambda^i}=\beta^j(\lambda)G_{ij}(\lambda).
\eeqa
This is the covariant form, and we have found that
BSFT chooses the special coordinate of $\lambda^i$ so that their beta 
functions become linear. 

Now we have completed the proof of eq.(\ref{generalsol}) in the case of 
bosonic string. This method of proof is also available in the case of 
superstring if we replace ${\cal V}_i$ by ${\cal W}_i$. The only difference 
is that one point function $\la {\cal W}_i(0)\lb_{\lambda}$ becomes zero 
because ${\cal W}_i$ is a vertex operator with $(-1)$ picture and its 
worldsheet statistics is fermionic. 
Thus, the first two terms in eq.(\ref{proof1}) become 
zero, and as a result the term with the derivative of $\lambda^i$ in 
eq.(\ref{covsol}) vanishes. Therefore, the action of superstring field theory 
becomes equal to just the partition function like 
eq.(\ref{generalsol}). Moreover, the same relation as eq.(\ref{covS}) holds if 
we notice that in the superstring the beta function of $\lambda^i$ is given by 
\beqa 
\beta^i(\lambda)=-\left(\frac{1}{2}-\Delta_i\right)\lambda^i+O^2(\lambda). 
\eeqa

By the way, from eq.(\ref{generalsol}) we can claim that the action 
$S(\lambda)$ is an off-shell (non-conformal) generalization of 
the boundary entropy $g$ \cite{AffLud}. The boundary entropy is defined in the 
statistical mechanics as the internal product of the boundary 
state and the conformal vacuum, and it is just equal to the disk partition 
function:
\beqa
\label{555}
g=\la 0|B\lb=Z_{disk}(\Sigma_B),
\eeqa
where $|B\lb$ can be any boundary state like eq.(\ref{geneb}), and the proof 
of equivalence of $\la 0|B\lb$ and $Z_{disk}(\Sigma_B)$ is shown in appendix 
\ref{Expression}. Moreover, if we use the 
explicit form of boundary state, we can find out that the 
boundary entropy is equal to the mass of the D-brane. This is the physical meaning of 
the boundary entropy in on-shell string theory, and we can regard the 
action $S$ as an off-shell (non-conformal) generalization of the 
boundary entropy. This is obvious in the superstring because $S$ is completely 
equal to the partition function even in the off-shell region like 
eq.(\ref{generalsol}), while we can also find out that this statement is 
meaningful even in the bosonic string too. 

One property supporting the above statement 
is that it is equal to the disk partition function at on-shell 
(conformal) points 
\beqa
S(\lambda^*)=Z(\lambda^*),
\eeqa
where $\lambda^*$ is a solution to $\beta^i(\lambda)=0$, which is the 
equation of motion of target space fields. The second is that the action $S$ 
of string field theory satisfies the famous g-theorem \cite{AffLud}. 
The g-theorem is the boundary
analog of the c-theorem, which appears in the bulk conformal field 
theory\footnote{Here 
$c$ is equal to the central charge at fixed points and the theorem says that 
$c$ monotonically decreases along the two-dimensional renormalization group 
flow.}, and there are two statements. The first statement 
is that the boundary entropy at the UV 
fixed point is always greater than at the IR fixed point:
\beqa
g_{UV}>g_{IR}.
\eeqa 
This means that the mass of an unstable D-brane is greater than that of a 
D-brane which appears after the decay of the unstable D-brane. This 
is a physically desirable result. 
 
The second statement of the g-theorem is that the boundary entropy $g$ at the 
UV conformal fixed point monotonically 
decreases along the renormalization group flow 
and it takes the minimal value at the IR fixed point. In fact, we can 
show that the action $S$ of string field theory 
satisfies this statement. This is 
because the dependence of the action $S$ on the worldsheet 
cut-off scale $x$ is determined in the following way:
\beqa
\label{dependx}
\frac{dS}{d\log|x|}=-\beta^i(\lambda)\frac{\de}{\de \lambda^i}S
=-\beta^i\beta^jG_{ij}(\lambda),
\eeqa
where we have used equation (\ref{covS}) and the renormalization group 
equation:
\beqa
\frac{d\lambda^i}{d\log|x|}=-\beta^i(\lambda).
\eeqa
If we notice that $G_{ij}(\lambda)$ is positive 
definite\footnote{$G_{ij}(\lambda)$ is the same as the Zamolchikov metric.} 
from the explicit form of it in eq.(\ref{G}), we can see that 
the action $S$ monotonically decreases along the renormalization group flow. 
This is the proof of the second statement of the g-theorem. 

From these facts we have found out that $S$ is equal to the mass 
of the unstable D-brane which decays and it also seems to be the same as 
the potential energy of the system. 
Therefore by investigating the action $S$ of string field theory
we can verify the decay of unstable D-brane systems in the quantitative 
way. Moreover, these features of the action $S$ completely 
coincide with the content of section \ref{RGTach} and figure \ref{Rg}.  

\section{The Worldsheet Actions for Non-BPS Systems in the Superstring Theory 
\label{superaction}}
\hspace*{4.5mm}
In the last section we have constructed the general framework of BSFT, 
and in section \ref{RGTach} 
we have explained its idea by using the boundary action
$I_B$ (\ref{action2}) for the bosonic string. The reason why we have not 
written down the boundary action $I_B$ for the superstring 
theory is that its form is a little complicated and it also depends on which 
non-BPS system we would like to consider. In this section we will give the 
explicit form of boundary actions for the non-BPS D-brane and the 
brane-antibrane system in the superstring theory, and in section 
\ref{exacttach} 
we will concretely calculate the actions $S$ of string field theory 
for these systems by using the relation (\ref{generalsol}). 

The two-dimensional bulk action is common in both of these systems and is 
given by eq.(\ref{action0}). This action is written by the two-dimensional 
superfield ${\bf X}^{\mu}$, thus it is natural to 
require the one-dimensional worldsheet 
supersymmetry\footnote{The original idea of BSFT is to include 
{\sl all kinds} of fields in the boundary action $I_B$ as an off-shell 
generalization of the first quantized string theory. For this reason there 
seems to be no need to require the one-dimensional worldsheet supersymmetry in 
the boundary action $I_B$ and to restrict possible terms. However, by 
requiring this we can determine the boundary actions for the non-BPS systems 
which we are interested in and we can see in the next few sections that those 
boundary actions describe the tachyon condensation correctly.} in 
the boundary action $I_B$. The superspace representation in the 
one-dimensional theory is defined by:
\begin{eqnarray}
\left\{\begin{array}{lcl}
{\bf X}^{\mu} &=& X^{\mu}+2i\theta\psi^{\mu},\\
D_{\theta} &=& \frac{\partial}{\partial\theta}
+\theta\frac{\partial}{\partial\tau}.\end{array}\right.
\end{eqnarray} 
If we assign the worldsheet dimensions $0,1$ to $X^{\mu},~D_{\theta}$ and
write all possible terms in the boundary action $I_B$ by 
using them, then it becomes as follows:
\beqa
\label{BPSboundary}
\int_{\pa\Sigma} d\tau 
d\theta\left[-iD_{\theta}{\bf X}^{\mu}A_{\mu}({\bf X})+\cdots\cdots\right],
\eeqa 
where the field $A_{\mu}({\bf X})$ represents the gauge field and the terms 
$\cdots$ do all massive fields. 
Unfortunately in RNS formalism we can 
not include target space fermions such a gaugino in the sigma model 
action. This might be resolved by considering BSFT formalism in the 
Green-Schwarz formalism \cite{GrSc} 
or Berkovits' pure spinor formalism \cite{Ber4}. We 
do not consider this problem here and will discuss it in the final chapter. 

The more serious problem here is that in the above sigma model action we can 
not see any tachyonic fields. Indeed boundary action 
(\ref{BPSboundary}) is for a BPS D-brane. This means that if we write a 
one-dimensional supersymmetric action by using only the superfield 
${\bf X}^{\mu}$, we can not include any relevant terms which represent 
tachyonic modes. To describe tachyon condensation on non-BPS systems we have 
to include relevant terms in the action. 

\subsection{For a brane-antibrane system}
\hspace*{4.5mm}
First we will consider the boundary action for a brane-antibrane system 
according to the description in subsection \ref{CPt}. In that subsection we 
have required the GSO-projection 
$(-1)^{F_{tot}}\equiv (-1)^F\times (-1)^{F_{CP}}=1$, where $F_{CP}$ gives the 
fermion number to the Chan-Paton factors, and assigns $\{+,-,-,+\}$ to 
$\{1,\sigma_1,\sigma_2,\sigma_3\}$. However, these Pauli matrices are not 
fermionic quantities but bosonic ones, and this statement is not precise. To 
describe this situation exactly we have to prepare some fermionic quantities 
which play the role similar to the Chan-Paton factors. To do this 
we should include some fermionic variable in the boundary worldsheet action 
$I_B$ instead of including the Chan-Paton degrees of freedom. Indeed in the 
next subsection we can see that this fermionic variable plays the same role as 
the Chan-Paton factors and obeys the GSO-projection which is described in 
subsection \ref{CPt}. 

Now let us consider a complex {\sl fermionic} superfield 
${\bf \Gamma},~{\bf \bar{\Gamma}}$ defined by \cite{TaTeUe,KrLa,Wi1} 
\beqa
\label{compo}
{\bf \Gamma}=\eta+\theta F.
\eeqa
If we write {\sl the most general} one-dimensional action by using 
${\bf X}^{\mu},~{\bf \Gamma},~{\bf \bar{\Gamma}}$ and $D_{\theta}$ with 
their dimension $0,0,0,1$ respectively, then we can obtain the following 
action\footnote{At first glance a general functional $g({\bf X})$ appears to 
be missing as the coefficient of the second term 
$-{\bf \bar{\Gamma}}D_{\theta}{\bf \Gamma}$, but if we use degrees of field 
redefinition of ${\bf \Gamma}$ and ${\bf \bar{\Gamma}}$ we can fix its 
coefficient to $1$.}   
\begin{eqnarray}
\label{eq3.5}
I_B &=& \int_{\partial\Sigma}d\tau 
d\theta\Biggl[\frac{1}{\sqrt{2\pi}}
{\bf \bar{\Gamma}}T({\bf X})+\frac{1}{\sqrt{2\pi}}\bar{T}({\bf X}){\bf \Gamma}
\no
&&~~~~~~~~~~~~-{\bf \bar{\Gamma}}D_{\theta}{\bf \Gamma}
-\frac{i}{2}D_{\theta}{\bf X}^{\mu}A^{(+)}_{\mu}({\bf X})
+iA^{(-)}_{\mu}({\bf X})D_{\theta}{\bf X}^{\mu}{\bf \bar{\Gamma}}{\bf \Gamma}
\no
&&~~~~~~~~~~~~~~~~~~+\cdots\cdots\Biggr],
\end{eqnarray}
where in each line of the above action relevant, marginal and irrelevant 
terms are shown. Here notice that several functionals 
$A^{\pm}_{\mu}({\bf X}),~T({\bf X})$ and $\bar{T}({\bf X})$ appear in the 
action. We can identify these functionals with two gauge fields and a 
complex tachyon, and from (\ref{DDbarfield}) we can find out that this 
spectrum is the same as that of a $D9+\overline{D9}$ system. 

Therefore, we claim that this action describes a $D9+\overline{D9}$ system 
in the off-shell region. This conjecture is also supported if we rewrite the 
above action in the following form
\begin{eqnarray}
\label{eq3}
I_B &=& \int_{\partial\Sigma}d\tau d\theta[-{\bf \bar{\Gamma}}
(D_{\theta}-iA^{(-)}_{\mu}({\bf X})D_{\theta}{\bf X}^{\mu}){\bf \Gamma}
+\frac{1}{\sqrt{2\pi}}{\bf \bar{\Gamma}}T({\bf X})\no
&&+\frac{1}{\sqrt{2\pi}}\bar{T}({\bf X}){\bf \Gamma}
-\frac{i}{2}D_{\theta}{\bf X}^{\mu}A^{(+)}_{\mu}({\bf X})+\cdots\cdots].
\end{eqnarray}
The form of the first term reminds us of the covariant derivative in 
gauge theory. Indeed, this action is invariant under the following gauge 
transformation:\footnote{Correctly speaking this symmetry is not 
two-dimensional gauge symmetry but non-linear global symmetry in the sigma 
model action. It is famous in usual sigma models (Type I, Heterotic) that 
world-sheet global symmetry corresponds to target space gauge symmetry.}
\begin{eqnarray}
\label{eq2.1} 
{\bf \Gamma}&\rightarrow&e^{i\Lambda({\bf X})}{\bf \Gamma},\no
T({\bf X})&\rightarrow& e^{i\Lambda({\bf X})}T({\bf X}),\no
A^{(-)}_{\mu}({\bf X})&\rightarrow& A^{(-)}_{\mu}({\bf X})
+\pa_{\mu}\Lambda({\bf X}),\no
A^{(+)}_{\mu}({\bf X})&\rightarrow& A^{(+)}_{\mu}({\bf X})
+\pa_{\mu}\Lambda^{\prime}({\bf X}).
\end{eqnarray}
This is a desirable property because it is known that there exists a gauge 
theory in the world-volume theory of D-branes. 

If we write $I_B$ in the component form and integrate out auxiliary 
fields $F$ and $\bar{F}$ in eq.(\ref{compo}), it becomes 
\begin{eqnarray}
\label{eq6}
I_B &=& \int_{\de\Sigma} d\tau 
[\bar{\eta}\dot{\eta}+2i\bar{\eta}\eta\psi^{\mu}
\psi{^\nu}F^{(-)}_{\mu\nu}-i\bar{\eta}\eta \dot{X^\mu}A^{(-)}_{\mu}
+\frac{1}{2\pi}\bar{T}T \no
& &-i\sqrt{\frac{2}{\pi}}\bar{\eta}\psi^{\mu}D_{\mu}T
+i\sqrt{\frac{2}{\pi}}\psi^{\mu}\eta \overline{D_{\mu}T}
-\f{i}{2}\dot{X}^{\mu}A^{(+)}_{\mu}+i\psi^{\mu}
\psi^{\nu}F^{(+)}_{\mu\nu}],
\end{eqnarray}
where we have employed the following definition:
\begin{eqnarray}
\label{covdef}
\left\{\begin{array}{lcl}
D_{\mu}T &=& \partial_{\mu}T-iA^{(-)}_{\mu}T,\\
F^{(\pm)}_{\mu\nu} &=& \partial_{\mu}A^{(\pm)}_{\nu}
-\partial_{\nu}A^{(\pm)}_{\mu},\end{array}\right.
\end{eqnarray}
and we have omitted irrelevant terms ($\cdots\cdots$ part in eq.(\ref{eq3.5}) 
and eq.(\ref{eq3})).

As we have said, the action (\ref{eq3}) is just for a $D9+\overline{D9}$ 
system, but it is easy to obtain the action for a $Dp+\overline{Dp}$ system 
since we have only to replace $A^{(\pm)}_i({\bf X})$ in the transverse 
direction with transverse scalars 
$\Phi^{(\pm)}_i({\bf X})$ following the T-duality. 

\subsection{The matrix form of worldsheet action \label{matrixform}}
\hspace*{4.5mm}
The worldsheet action (\ref{eq3}) is a little complicated, while if we 
integrate the fields $\eta$ and $\bar{\eta}$ in eq.(\ref{eq6}) 
we can rewrite it in the simple matrix form \cite{TaTeUe,KrLa}.

In BSFT what we would like to calculate is the disk partition function because 
the action of string field theory 
is related to it by equation (\ref{generalsol}). 
The partition function $Z$ is defined by
\begin{eqnarray}
\label{eq5.1}
Z &=&
\int DX D\psi D\eta D\bar{\eta} \exp[-I_0(X,\psi)-I_B(X,\psi,\eta,\bar{\eta})],
\end{eqnarray}
where $I_0$ and $I_B$ are given by eq.(\ref{action0}) and eq.(\ref{eq6}). 
In this subsection we will omit irrelevant terms in $I_B$.

Firstly, we integrate only $\eta(\tau)$ and $\bar{\eta}(\tau)$. To do this we 
rewrite the above expression in the following way:
\begin{eqnarray}
\label{eq7}
Z=\int DX D\psi \exp[-I_0(X,\psi)] 
\int D\eta D\bar{\eta} \exp[-I_B(X,\psi,\eta,\bar{\eta})].  
\end{eqnarray} 

The simplest way to integrate $\eta(\tau)$ and $\bar{\eta}(\tau)$ is to 
transform this path integral into the Hamiltonian 
formalism.\footnote{Of course, we can obtain the final result 
(\ref{eq9}) by calculating the path integral perturbatively in the following 
way. Firstly, in the open string NS sector $\psi(\tau)$ obeys the 
anti-periodic boundary condition on the boundary of disk 
so that $\psi(\tau)$ has half-integer 
Fourier modes. Thus, $\eta(\tau)$ and $\bar{\eta}(\tau)$ should obey the 
anti-periodic boundary condition in order for (\ref{eq6}) to be locally 
well-defined. The Green function of $\eta$ and $\bar{\eta}$ which obeys 
anti-periodic boundary condition is given by
\begin{eqnarray}
\langle\eta(\tau)\bar{\eta}(\tau^{\prime})\rangle
&=&\frac{1}{2}\epsilon(\tau-\tau^{\prime})
=\sum_{r \in \z + \frac{1}{2}>0} \frac{{\rm sin(r(\tau-\tau'))}}{r},\\
&&\epsilon(\tau)\equiv
\left\{\begin{array}{c}
1~(\tau>0)\\
0~(\tau=0)\\
-1~(\tau<0)\end{array}.\right.
\end{eqnarray}
We can check order by order 
that equation (\ref{eq7}) is equal to (\ref{eq9}).
Note that this identity holds when we take 
$\epsilon(0)=0$ regularization.}
According to the standard canonical formalism 
the fields $\eta$ and 
$\bar{\eta}$ are quantized and from (\ref{eq6}) these obey 
the usual anti-canonical quantization condition:
\begin{eqnarray}
\{\eta,\bar{\eta}\}=1.
\end{eqnarray}
Note that both the fields $\eta$ and $\bar{\eta}$ live on the boundary 
of the disk, which is a circle. Thus, the path integral of $\eta(\tau)$ and 
$\bar{\eta}(\tau)$ in eq.(\ref{eq7}) corresponds to a one-loop calculation, 
and it is expressed by the path ordering trace in the Hamiltonian formalism 
as follows:
\begin{eqnarray}
\label{eq8}
Z &=&\int DX D\psi \exp[-I_0(X,\psi)]\nonumber\\ 
&\times&{\rm Tr~P}\exp\int^{\pi}_{-\pi}d\tau
\biggl[i\frac{[\bar{\eta},\eta]}{2}
\dot{X}^{\mu}A^{(-)}_{\mu}(X)-2i\frac{[\bar{\eta},\eta]}{2}\psi^{\mu}\psi^{\nu}
F^{(-)}_{\mu\nu}(X)+i\sqrt{\frac{2}{\pi}}\bar{\eta}\psi^{\mu}D_{\mu}T(X)
\nonumber\\
&&-i\sqrt{\frac{2}{\pi}}\psi^{\mu}\eta\overline{D_{\mu}T(X)}
-\frac{1}{2\pi}\bar{T}(X)T(X)+\frac{i}{2}\dot{X}^{\mu}A^{(+)}_{\mu}(X)
-i\psi^{\mu}\psi^{\nu}F^{(+)}_{\mu\nu}(X)\biggr],
\end{eqnarray}
where ``${\rm P}$" represents the path ordering and ``${\rm Tr}$"(trace) 
implies that we should sum expectation values in the two-state Hilbert space 
$(|\downarrow\lb,~|\uparrow\lb)$:
\begin{eqnarray}
\eta|\downarrow\rangle=0~&,&~\bar{\eta}|\downarrow\rangle=|\uparrow\rangle,
\nonumber\\
\eta|\uparrow\rangle=|\downarrow\rangle~&,&~\bar{\eta}|\uparrow\rangle=0.
\end{eqnarray}
Note that constructing Hamiltonian from Lagrangian 
we have set the operator ordering by an antisymmetrization of 
$\eta$ and $\bar{\eta}$ in eq.(\ref{eq8}).

Here we can check that the operator commutation and anti-commutation 
relations of $\bar{\eta}, \eta$ and $[\bar{\eta},\eta]$ are the same as those 
of the Pauli matrices, $\sigma_{+}, \sigma_{-}$ and $\sigma_3$ (where 
$\sigma_{\pm}\equiv\frac{1}{2}(\sigma_1\pm i\sigma_2)$). Therefore we can 
replace $\bar{\eta}, \eta$ and $[\bar{\eta},\eta]$ with 
$\sigma_{+}, \sigma_{-}$ 
and $\sigma_3$ respectively\footnote{Strictly speaking we need cocycle factors 
to keep the correct worldsheet statistics.}. In this way we can see that the 
superfield ${\bf \Gamma}$ or its lowest component $\eta $ plays the same 
role\footnote{From the action (\ref{eq3.5}) we can understand the 
meaning of the statement which was mentioned in the paragraphs before 
eq.(\ref{DDbarfield}) and (\ref{NDpfield}).} as Chan-Paton factors. 
    
After this replacement $Z$ becomes:
\begin{eqnarray}
\label{eq9}
& &Z~=~\int DX D\psi \exp[-I_0(X,\psi)]\times{\rm Tr~P}\exp\int^{\pi}_{-\pi} 
d\tau M(\tau),
\end{eqnarray}
where
\begin{eqnarray}
& &M(\tau)~=~\left(\begin{array}{cc}
i\dot{X}^{\mu}A^{(1)}_{\mu}-2i\psi^{\mu}\psi^{\nu}F^{(1)}_{\mu\nu}
-\frac{1}{2\pi}T\bar{T}~&~ i\sqrt{\frac{2}{\pi}}\psi^{\mu}D_{\mu}T
\nonumber\\
-i\sqrt{\frac{2}{\pi}}\psi^{\mu}\overline{D_{\mu}T}~&~i\dot{X}^{\mu}
A^{(2)}_{\mu}-2i\psi^{\mu}\psi^{\nu}F^{(2)}_{\mu\nu}
-\frac{1}{2\pi} \bar{T}T \nonumber
\end{array}\right).\\
\end{eqnarray}
Here the fields $A^{(1)}_{\mu}$ and $A^{(2)}_{\mu}$ are defined by
\beqa
A^{(1)}_{\mu}&=&\frac{1}{2}(A^{(+)}_{\mu}+A^{(-)}_{\mu}),\no
A^{(2)}_{\mu}&=&\frac{1}{2}(A^{(+)}_{\mu}-A^{(-)}_{\mu}),
\eeqa
and $F^{(1)}_{\mu\nu}$ and $F^{(2)}_{\mu\nu}$ are their field strengths.

This is one expression of the partition function $Z$. This action is simpler 
than the original one (\ref{eq6}) because it is expressed by a simple matrix 
$M(\tau)$. Moreover, from this matrix $M(\tau)$ we can read off the physical 
situation in the brane-antibrane system. The diagonal blocks 
represent the degrees of freedom of open strings both of whose ends are on the 
same brane. 
The upper diagonal block is for the brane, and the lower one is for the 
anti-brane. On the other hand, the non-diagonal blocks come from open strings 
which stretches between the brane and the anti-brane. From these facts we can 
conclude that $A^{(1)}_{\mu}$ and $A^{(2)}_{\mu}$ are the gauge fields on the 
brane and the anti-brane respectively, and from eq.(\ref{covdef}) we can see 
that the tachyon belongs to 
the bifundamental representation of the gauge group $U(1)\times U(1)$. 

Moreover, 
this action can be naturally generalized to multiple branes and antibranes. 
If we consider parallel $M$ branes and $N$ antibranes with zero relative 
distance among them, there appears a $U(M)\times U(N)$ gauge 
theory in this system and the tachyon couples the gauge fields with the 
representation $M\times\bar{N}$ or $\bar{M}\times N$. 
To realize this situation we have only to replace abelian gauge fields 
with non-abelian ones in eq.(\ref{eq9}). 
However, action (\ref{eq9}) has one fault 
that the gauge symmetry and the worldsheet supersymmetry can not be seen 
explicitly. Moreover, this action is nonlocal due to the path ordering factor 
and is regarded as an effective sigma model action, not the fundamental 
one because fields $\eta$ and $\bar{\eta}$ are already integrated out. In this
thesis we do not use the fundamental action for multiple branes-antibranes, 
while it is given in the papers \cite{Ho,KrLa} only for $2^{m-1}$ pairs of 
branes and antibranes.

\subsection{For a non-BPS D-brane}
\hspace*{4.5mm}
It is easy to obtain the action for a non-BPS D-brane because the 
brane-antibrane system and the non-BPS D-brane are related to each other by 
the descent relation (\ref{descentfield}). According to this rule the complex 
tachyon field changes into a real tachyon field, thus it is natural to replace 
the complex field ${\bf \Gamma},~{\bf \bar{\Gamma}}$ by a real field 
${\bf \Gamma}$.  By this reduction we can obtain the following action 
\cite{KuMaMo2,HaKuMa,Wi1}
\beqa
\label{eq1}
I_B &=& \int_{\partial\Sigma}d\tau 
d\theta\left[\frac{1}{\sqrt{2\pi}}
T({\bf X}){\bf \Gamma}-{\bf \Gamma}D_{\theta}{\bf \Gamma}
-iD_{\theta}{\bf X}^{\mu}A_{\mu}({\bf X})+\cdots\cdots\right],\\
&=& \int_{\partial\Sigma}d\tau 
\left[\frac{1}{8\pi}T(X)^2+\eta\dot{\eta}
+i\sqrt{\frac{2}{\pi}}\psi^{\mu}\eta\partial_{\mu}T\right.\no
&&~~~~~~~~~~~~~~~~~~~~~~~~~~~~~~~\left.-i\dot{X}^{\mu}A_{\mu}
+2iF_{\mu\nu}\psi^{\mu}\psi^{\nu}+\cdots\cdots\right], 
\label{bin}
\end{eqnarray}
where in the second line we have written down the component form of action 
by integrating out the auxiliary field $F$ in eq.(\ref{compo}), 
and $\cdots$ part represents 
irrelevant (massive) terms. This action can also be obtained 
by writing down all possible terms with ${\bf X}^{\mu},~{\bf \Gamma}$ and 
$D_{\theta}$ in the same way as a brane-antibrane system.

We claim that this action describes a non-BPS D9-brane. Indeed, from 
eq.(\ref{NDpfield}) we can see that the functionals $T(X)$ and 
$A_{\mu}(X)$ are target space fields on a non-BPS D9-brane. 
Here we can see that the covariant derivative in the first 
term of brane-antibrane action (\ref{eq3}) disappears. In the later 
sections we will find out that the concrete calculation of the action $S$ 
for a non-BPS D-brane is easier than that for a brane-antibrane system 
because of the absence of terms with covariant derivatives.

Furthermore this action is invariant under the following gauge transformation 
\beqa
\label{eq2.11}
{\bf \Gamma}&\rightarrow&{\bf \Gamma},\no
T({\bf X})&\rightarrow& T({\bf X}),\no
A_{\mu}({\bf X})&\rightarrow& A_{\mu}({\bf X})
+\pa_{\mu}\Lambda({\bf X}). 
\eeqa 
This means that the tachyon field $T({\bf X})$ is gauge transformed in the 
U(1) adjoint representation (that is equal to the gauge singlet).

Note that action (\ref{eq1}) is for a non-BPS D9-brane. If we would like to 
obtain the action for a non-BPS D$p$-brane, we have only to replace the field 
$A^i({\bf X})$ in the transverse direction with the transverse scalar 
$\Phi^i({\bf X})$ following the T-duality transformation.

Finally we make a brief comment that we can also obtain 
the matrix form of action for a non-BPS D-brane or the action for 
multiple non-BPS D-branes. We forgo doing it here. 

\section{Calculability of Boundary String Field Theory 
\label{calcu}}
\hspace*{4.5mm}
Until now we have given the formal description of boundary string field 
theory (BSFT) for the bosonic string and superstring theories. In the next few 
sections we move on to concrete calculations in order to 
investigate the tachyon condensation or to obtain the target space actions 
for non-BPS systems in the superstring theory. For computations in 
the bosonic string theory see the papers \cite{Wi2,KuMaMo1,GeSh1}.

In BSFT the fundamental quantity is the action $S$ of string field theory 
and it is related to the disk partition function $Z$, which is defined by 
eq.(\ref{eq5.1}).
However, there remain two serious problems about concrete calculations of BSFT.
The first problem is renormalizability of the sigma model on the 
disk. As we said in section \ref{RGTach} one of principles of BSFT 
is to include 
all kinds of boundary fields in the boundary action $I_B$, while we can not 
calculate the disk partition function because it is {\sl non-renormalizable}. 
It seems that we can not use BSFT in the concrete calculation. 

For this reason 
we will try restricting terms in the boundary action only to 
renormalizable terms and calculating the disk partition function. According to 
one of principles 
of quantum field theory we have to include all renormalizable 
terms which keep the symmetry we impose, otherwise the renormalization group 
does not close. Fortunately the worldsheet actions (\ref{eq3}) and 
(\ref{eq1}) for a brane-antibrane and for a non-BPS D-brane are the most 
general actions which are constructed from the superfields 
${\bf X}^{\mu}$ and ${\bf \Gamma}~(\mbox{and}~{\bf \bar{\Gamma}})$, thus they 
are well-defined in the renormalization group theory. As we will see in the 
next few sections we can verify that the results obtained from these 
worldsheet actions give the correct answers which coincide with 
the prediction of 
tachyon condensation. Therefore, we can believe that such a prescription is 
available in various physical situations. Unfortunately there is no proof of 
it, and in section \ref{meaning} we will give some consideration 
about what it means.

The second problem is about the quantum theory of BSFT. The action $S$ which 
we have defined in the previous sections is the classical action of string 
field theory. 
The ``quantum theory" means by considering 
the quantum corrections of the action $S[\lambda]$, where 
$\lambda^i~(i=1,2,\cdots)$ denote target space fields 
like a tachyon, a gauge field. Let us imagine that we started with the sigma 
model action including all massive fields and that we could obtain the 
classical action $S[\lambda]$. By using this action the quantum effective 
action $S_{\ss eff}[\lambda]$ is defined by the path integral in the target 
space sense
\beqa
\label{secondquant}
e^{-S_{\ss eff}[\lambda]}=\int_{\ss 1PI}
\left(\prod_i D\lambda^{\prime i}\right) e^{-S[\lambda+\lambda^{\prime}]},
\eeqa
where $\int_{\ss 1PI}$ denotes the path integral which picks up only 
1PI Feynman diagrams. 
This is just the second quantization of string theory, which is the 
central property 
of string field theory. At first glance this path integral looks 
ill-defined because the classical action $S[\lambda]$ is a ten-dimensional 
action and non-renormalizable, and it includes the infinite number of fields 
and derivatives, which indicates that the action of string field theory is 
{\sl nonlocal}. However, the 
framework of string field theory makes eq.(\ref{secondquant}) 
well-defined because string field theory controls the target space action 
and suppresses its quantum divergences by the worldsheet techniques in 
the same way 
as the conventional first-quantized string theory. Several methods to calculate
the quantum corrections of the target space action 
$S[\lambda]$ (see \cite{Taylor} and references there in) are known 
in cubic string field theory (CSFT) \cite{WiCSFT}.

The formal meaning of the quantum theory of BSFT is also the same as 
eq.(\ref{secondquant}). However, any practical methods to calculate quantum 
corrections are not known in BSFT\footnote{There are some attempts to define 
quantum corrections in BSFT. For example see the paper \cite{CrKrLa}.}. 
Here, it is important to distinguish the meaning of ``quantum" between the 
worldsheet theory and the target space theory. 
The renormalization group flow which we have discussed in the previous sections
comes from quantum calculations in the worldsheet theory. 
Indeed the renormalization in the worldsheet level is needed (for example see 
equation (\ref{fa})) when we even calculate the disk amplitude, because the 
boundary of disk is a circle, which corresponds to a one-loop diagram of 
the one-dimensional worldsheet theory. 
In BSFT we perform quantum calculations in 
the worldsheet sense, while what is obtained as a result is a classical 
action in the target space sense. 

At this moment 
we do not know the quantum theory of BSFT, and thus in the later sections 
we consider only the classical theory of BSFT. 

\section{Tachyon Condensation on Non-BPS Systems \label{exacttach}}
\setcounter{equation}{0}
\hspace*{4.5mm}

In this section we will investigate the tachyon condensation on non-BPS 
systems \cite{TaTeUe,KrLa,KuMaMo2,Ts2,An}. 
All the results obtained in this section are regarded as {\sl exact} ones. 
However, in BSFT there remains a problem about the meaning of ``exact". 
We will explain its meaning in section \ref{meaning}, and we would like to 
ignore it in this section.
 
As we have said in the previous sections the result for a non-BPS D-brane 
can be obtained from that for a brane-antibrane system by the descent 
relation (\ref{descentfield}), thus we will start from the analysis of a 
brane-antibrane system.

\subsection{On brane-antibrane systems}
\hspace*{4.5mm}

Firstly, we will calculate the tachyon potential for a $Dp+\overline{Dp}$ 
system \cite{TaTeUe,KrLa}. It can be obtained from boundary action (\ref{eq9}) 
by setting the gauge fields $A^{\pm}_{\mu}(X)$ to zero and dropping 
derivative terms of the tachyon field $T(X)$. 
This means that we consider the condensation of a 
constant tachyon field $T(X)=T$. After that equation (\ref{eq9}) becomes
\beqa
\label{bkindependent}
S=Z=2\int DXD\psi \exp[-I_0(X,\psi)] e^{-|T|^2}\propto \int d^{p+1}x 
e^{-|T|^2}.
\eeqa
From this equation we can see that the exact tachyon potential for a 
$Dp+\overline{Dp}$ system is given by $V(T)=C_pe^{-|T|^2}$, where the 
coefficient\footnote{The measure $d^{9-p}x$ in eq.(\ref{bkindependent}) is 
included in this coefficient.} 
$C_p$ is a constant which can not be determined by the above path integral.
 
\begin{figure} [tbhp]
\begin{center}
\epsfxsize=70mm
\epsfbox{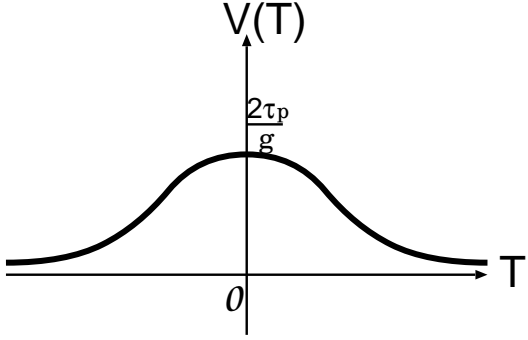}
\caption{The potential $V(T)=\frac{2\tau_p}{g}e^{-|T|^2}$ for a 
$Dp+\overline{Dp}$ system \label{potential3}}
\end{center}
\end{figure}

Here let us assume the first conjecture of the tachyon condensation 
(see section \ref{tachcon}).  It states that
non-BPS systems at the top of the potential $(T=0)$ decay to nothing at 
the minimum of the potential $(T=\infty)$ 
by condensation of a constant tachyon field. If we admit 
this conjecture, the most natural candidate of the value of $C_p$ 
is the tension of a $Dp+\overline{Dp}$ system\footnote{This statement is 
nontrivial in 
cubic string field theory \cite{WiCSFT,Wisu} or Berkovits's formalism 
\cite{Ber1}, thus the calculation of the height of potential can be a 
nontrivial check of the first conjecture of tachyon condensation. Indeed 
various analyses have supported this conjecture in very good accuracy 
\cite{KoSa,bosonic,super}. Unfortunately in BSFT the normalization 
of tachyon potential can not be determined.}. 
Here the tension means the leading part of eq.(\ref{masscorrection}) 
because as we have said in the last section the 
action $S$ of string field theory which we are considering now is the 
classical action in the target space sense. 
This assumption accords with the physical 
picture of decay of a $Dp+\overline{Dp}$ system into nothing.  
Therefore the tachyon potential is given by
\beqa
\label{ponormal}
V(T)=\frac{2\tau_p}{g}e^{-|T|^2},
\eeqa   
where $\tau_p$ is given by eq.(\ref{tens}) and $g$ is the string coupling 
constant. The schematic shape of this potential is 
given in figure \ref{potential3}. This looks different from 
figure \ref{potential} in section \ref{tachcon}. This is because the kinetic 
term of the tachyon field is given by $e^{-|T|^2}\pa^{\mu}\bar{T}\pa_{\mu}T$ 
according to the later calculation of BSFT. Thus if we take the appropriate 
field redefinition $T^{\prime}=f(T)$ which sets the kinetic term to the 
standard form $\pa^{\mu}\bar{T}^{\prime}\pa_{\mu}T^{\prime}$, 
we can map the potential in fig. 
\ref{potential3} to one in fig. \ref{potential} and can shift the minimum 
point of the potential from $T=\infty$ to some finite value 
\cite{MiZw,MiZw2,HaNa}.
    
By the way 
the most interesting fact here is that this form of tachyon potential does 
not depend on what kind of closed string background we choose. This can be 
understood from eq.(\ref{bkindependent}) because we have not used any 
information about the closed string background, which is included in $I_0$ in 
eq.(\ref{bkindependent}). This fact is called universality of the tachyon 
potential \cite{sen23}. The information of the closed string background is 
only included in the coefficient of the tachyon potential in 
eq.(\ref{bkindependent}). This coefficient gives the mass of 
a $Dp+\overline{Dp}$ system, and this fact is physically desirable because the 
mass depends on the closed string background through the measure 
$\sqrt{-\det(G_{\mu\nu}-B_{\mu\nu})}$, which is just the Dirac-Born-Infeld 
factor. 

Now let us consider the second conjecture of tachyon condensation 
\cite{TaTeUe,KrLa}. It is about 
the condensation of topologically nontrivial tachyons, not constant tachyons. 
In this case, as we have said in section \ref{tachcon}, lower dimensional 
D-branes appear after tachyon condensation. Here we start from a 
$Dp+\overline{Dp}$ system, thus we expect that a non-BPS $D(p-1)$ brane or 
a BPS $D(p-2)$ brane is generated after the tachyon condensation.

In the same way as the verification of the first conjecture we set all the 
massive fields and two gauge fields $A^{\pm}_{\mu}(X)$ to zero in boundary 
action (\ref{eq6}), and boundary action (\ref{eq6}) becomes
\beqa
I_B &=& \int_{\de\Sigma} d\tau 
\left[\bar{\eta}\dot{\eta}-i\sqrt{\frac{2}{\pi}}\bar{\eta}\psi^{\mu}
\pa_{\mu}T(X)+i\sqrt{\frac{2}{\pi}}\psi^{\mu}\eta \overline{\pa_{\mu}T}(X)
+\frac{1}{2\pi}\bar{T}T(X)\right]\no
&=& \int_{\de\Sigma} d\tau \left[-\frac{2}{\pi}(\psi^{\mu}\pa_{\mu}
\bar{T}(X))\frac{1}{\pa_{\tau}}(\psi^{\nu}\pa_{\nu}T(X))
+\frac{1}{2\pi}\bar{T}T(X)\right],
\label{action10}
\end{eqnarray}
where in the second line we have eliminated $\eta(\tau)$ 
and $\bar{\eta}(\tau)$ by the Gaussian path integral, and 
$\frac{1}{\pa_{\tau}}$ is defined by
\beqa
\frac{1}{\pa_{\tau}}f(\tau)=\int d\tau^{\prime}\epsilon(\tau-\tau^{\prime})
f(\tau^{\prime}).
\eeqa 
From the above expression it does not seem that we can exactly perform the 
path integral (\ref{eq5.1}) to obtain the disk partition function 
because the boundary action is nonlinear. 
However if we restrict the tachyon field $T(X)$ to the linear order of 
$X^{\mu}$
this becomes the Gaussian path integral and is exactly solvable. 
The meaning of this restriction will be mentioned in section \ref{meaning}.

To perform the path integral 
let us expand $T(X)$ around the zero mode $x^{\mu}$ by using the 
relation $X^{\mu}(\tau)=x^{\mu}+\xi^{\mu}(\tau)$. $T(X)$ becomes
\beqa
T(X)=T(x)+\xi^{\mu}\pa_{\mu}T(x)
+\frac{1}{2}\xi^{\mu}\xi^{\nu}\pa_{\mu}\pa_{\nu}T(x)+\cdots\cdots.
\eeqa
If we restrict the tachyon field to the linear order, boundary action 
(\ref{action10}) becomes 
\beqa
\label{action11}
I_B&=& \int_{\de\Sigma} d\tau \left[\frac{1}{2\pi}\bar{T}T(x)
-\frac{2}{\pi}(\pa_{\mu}\bar{T}
\pa_{\nu}T(x))\psi^{\mu}\frac{1}{\pa_{\tau}}\psi^{\nu}
+\frac{1}{2\pi}\xi^{\mu}\xi^{\nu}\pa_{\mu}\bar{T}\pa_{\nu}T(x)\right],
\eeqa
where we have used the equation $\int d\tau \xi^{\mu}(\tau)=0$ due to  
$\int d\tau X^{\mu}(\tau)=x^{\mu}$. This action includes only the second 
order of the fields $\xi^{\mu}(\tau)$ and $\psi^{\mu}(\tau)$, and path 
integral (\ref{eq5.1}) for the disk partition function 
can exactly be performed as the Gaussian path integral. 

Now let us move on to the concrete calculation of the disk 
partition function $Z$, 
which is the same as the action $S$ of string field theory. 
The difference from eq.(\ref{bkindependent}) is that the calculation 
here depends on what kind of closed string background we choose. In 
generally curved closed string backgrounds it is difficult to perform the 
path integral (\ref{eq5.1}) with bulk sigma model action (\ref{action0}). 
Therefore, 
from now on we will consider the flat background. The meaning of ``flat" here 
is to set the closed string metric $G_{\mu\nu}$ and the NSNS B-field 
$B_{\mu\nu}$ to constant values, which satisfy the equation of motion of 
supergravity. Under this restriction the bulk action $I_0$ in 
eq.(\ref{action0}) is drastically simplified as
\beqa
I_0&=&\frac{1}{4\pi}\int d^2z\Bigl[(G_{\mu\nu}+B_{\mu\nu})\pa X^{\mu}
\bar{\pa}X^{\nu}\no
&&~~~~~~~~~~~~~~-i(G_{\mu\nu}+B_{\mu\nu})\psi^{\mu}_L\bar{\pa}\psi^{\nu}_L
+i(G_{\mu\nu}-B_{\mu\nu})\psi^{\mu}_R\pa\psi^{\nu}_R\Bigr].
\eeqa
In this case the action includes only the second powers of fields 
$(X^{\mu},~\psi^{\mu}_L,~\psi^{\mu}_R)$. Therefore the path integral of 
internal part of disk in eq.(\ref{eq5.1}) can be performed as the Gaussian 
integral and the total expression of eq.(\ref{eq5.1}) reduces to the path 
integral only on the boundary of disk, which is a one-loop diagram in the 
one-dimensional theory:
\beqa
\label{pathint}
Z &=&\int [d^{10}xD\xi D\psi]\exp[-I^{\prime}_B(x,\xi,\psi)],\\
\mbox{where}\no
\label{functional0}
I^{\prime}_B&=&I_B+\frac{1}{8\pi}\int d\tau_1 \int d\tau_2 
\xi^{\mu}(\tau_1)\xi^{\nu}(\tau_2)K^{-1}_{\mu\nu}(\tau_1,\tau_2)\no
&&~~~~~~~~+\frac{1}{2\pi}\int d\tau_1 \int d\tau_2 \psi^{\mu}(\tau_1)
\psi^{\nu}(\tau_2)L^{-1}_{\mu\nu}(\tau_1,\tau_2),
\eeqa
and
\beqa
\label{inverseprop0}
K^{-1}_{\mu\nu}(\tau_1,\tau_2)&=&\frac{1}{2\pi}(G_{\mu\nu}-B_{\mu\nu})
\sum^{\infty}_{m=1}me^{im(\tau_1-\tau_2)}+
\frac{1}{2\pi}(G_{\mu\nu}+B_{\mu\nu})
\sum^{\infty}_{m=1}me^{-im(\tau_1-\tau_2)},\no
L^{-1}_{\mu\nu}(\tau_1,\tau_2)&=&\frac{i}{2\pi}(G_{\mu\nu}-B_{\mu\nu})
\sum^{\infty}_{r=\frac{1}{2}}e^{ir(\tau_1-\tau_2)}-
\frac{i}{2\pi}(G_{\mu\nu}+B_{\mu\nu})
\sum^{\infty}_{r=\frac{1}{2}}e^{-ir(\tau_1-\tau_2)}.\no
\eeqa 
The proof of this expression is given in appendix \ref{Expression}. 
Note that $\psi^{\mu}$ obeys the anti-periodic boundary condition around 
the boundary of disk. 

Inserting the explicit form of boundary action $I_B$ given in 
eq.(\ref{action11}) into eq.(\ref{functional0}), the action $I^{\prime}_B$ 
becomes 
\beqa
\label{functional1}
I^{\prime}_B&=&\bar{T}T(x)+\frac{1}{8\pi}\int d\tau_1 \int d\tau_2 
\xi^{\mu}(\tau_1)\xi^{\nu}(\tau_2)K^{\prime -1}_{\mu\nu}(\tau_1,\tau_2)\no
&&~~~~~~~~+\frac{1}{2\pi}\int d\tau_1 \int d\tau_2 \psi^{\mu}(\tau_1)
\psi^{\nu}(\tau_2)L^{\prime -1}_{\mu\nu}(\tau_1,\tau_2),
\eeqa
where
\beqa
\label{inverseprop1}
K^{\prime -1}_{\mu\nu}(\tau_1,\tau_2)&=&\frac{1}{2\pi}\sum^{\infty}_{m=1}
\{(G_{\mu\nu}-B_{\mu\nu})m+4\pa_{(\mu}\bar{T}\pa_{\nu)}T\}
e^{im(\tau_1-\tau_2)}\no
&&~~~~~~~~+\frac{1}{2\pi}
\sum^{\infty}_{m=1}\{(G_{\mu\nu}+B_{\mu\nu})m+4\pa_{(\mu}\bar{T}\pa_{\nu)}T
\}e^{-im(\tau_1-\tau_2)},\no
L^{\prime -1}_{\mu\nu}(\tau_1,\tau_2)&=&\frac{i}{2\pi}
\sum^{\infty}_{r=\frac{1}{2}}
\{(G_{\mu\nu}-B_{\mu\nu})+\frac{4}{r}\pa_{(\mu}\bar{T}\pa_{\nu)}T\}
e^{ir(\tau_1-\tau_2)}\no
&&~~~~~~~~-\frac{i}{2\pi}\sum^{\infty}_{r=\frac{1}{2}}
\{(G_{\mu\nu}+B_{\mu\nu})+\frac{4}{r}\pa_{(\mu}\bar{T}\pa_{\nu)}T\}
e^{-ir(\tau_1-\tau_2)},\no
\eeqa 
and $\pa_{(\mu}\bar{T}\pa_{\nu)}T$ denotes the symmetric part of 
$\pa_{\mu}\bar{T}\pa_{\nu}T$. From eq.(\ref{functional1}) we can see that we 
have only to perform the Gaussian integral about the nonzero modes 
$\xi^{\mu}$ and $\psi^{\mu}$, which is the same calculation 
as that of the Coleman-Weinberg potential in field theory. The result 
is given by
\beqa
S&=&Z\propto\int d^{p+1}x e^{-|T(x)|^2}
\left[\frac{\prod^{\infty}_{r=\frac{1}{2}}-\det\{(G_{\mu\nu}-B_{\mu\nu})
+\frac{4}{r}\pa_{(\mu}\bar{T}\pa_{\nu)}T\}}
{\prod^{\infty}_{m=1}-\det\{(G_{\mu\nu}-B_{\mu\nu})m
+4\pa_{(\mu}\bar{T}\pa_{\nu)}T\}}\right]\no
&=&\exp\Biggl[-\sum^{\infty}_{m=1}\ln\left[-\det\{(G_{\mu\nu}-B_{\mu\nu})m
+4\pa_{(\mu}\bar{T}\pa_{\nu)}T\}\right]\no
&&~~~~~~~~~~~~
+\sum^{\infty}_{r=\frac{1}{2}}\ln\left[-\det\{(G_{\mu\nu}-B_{\mu\nu})
+\frac{4}{r}\pa_{(\mu}\bar{T}\pa_{\nu)}T\}\right]\Biggr],\no
&=&\exp\left[\ln\{-\det(G_{\mu\nu}-B_{\mu\nu})\}
\left(-\sum^{\infty}_{m=1}e^{-\epsilon m}
+\sum^{\infty}_{r=\frac{1}{2}}e^{-\epsilon r}\right)\right]\no
&&\times~~~
\exp\left[-\sum^{\infty}_{m=1}
{\rm Tr}\{\ln(m\delta^{\mu}_{\nu}+M^{\mu}_{~\nu})\}
+\sum^{\infty}_{r=\frac{1}{2}}{\rm Tr}\left\{\ln\left(\delta^{\mu}_{~\nu}
+\frac{1}{r}M^{\mu}_{~\nu}\right)\right\}\right],
\label{Trln}
\eeqa
where $e^{-\epsilon m}$ and $e^{-\epsilon r}$ are regulators to control the 
divergence, and the matrix $M^{\mu}_{~\nu}$ is given by
\beqa
\label{defM}
M^{\mu}_{~\nu}=4\left(\frac{1}{G-B}\right)^{\mu\rho}
\pa_{(\rho}\bar{T}\pa_{\nu)}T.
\eeqa
Here by using the relation\footnote{The cancellation of the divergence 
with $\epsilon\rightarrow 0$ in eq.(\ref{cancel1}) and eq.(\ref{cancel2}) 
comes from the worldsheet supersymmetry.} 
\beqa
\label{cancel1}
-\sum^{\infty}_{m=1}e^{-2\epsilon m}+\sum^{\infty}_{r=\frac{1}{2}}
e^{-2\epsilon r}=\frac{1}{2}+{\cal O}(\epsilon), 
\eeqa
the first exponential becomes
\beqa
\sqrt{-\det(G_{\mu\nu}-B_{\mu\nu})}.
\eeqa
This part gives the Dirac-Born-Infeld factor if we include the gauge field 
(see section \ref{WV}).

The subtle calculation is in the second exponential of eq.(\ref{Trln}). By 
expanding the logarithm the internal part of this exponential becomes
\beqa
\label{traceformula}
&{\displaystyle \sum^{\infty}_{\ell=1}}& 
\frac{(-1)^{\ell-1}}{\ell}
\left(-\sum^{\infty}_{m=1}\frac{1}{m^{\ell}}
+\sum^{\infty}_{r=\frac{1}{2}}\frac{1}{r^{\ell}}\right){\rm Tr}M^{\ell}\no
&=&\left(-\sum^{\infty}_{m=1}\frac{e^{-\epsilon m}}{m}
+\sum^{\infty}_{r=\frac{1}{2}}\frac{e^{-\epsilon r}}{r}\right){\rm Tr}M
+\sum^{\infty}_{\ell=2}\frac{(-1)^{\ell-1}}{\ell}
\left(-\sum^{\infty}_{m=1}\frac{1}{m^{\ell}}+\sum^{\infty}_{r=\frac{1}{2}}
\frac{1}{r^{\ell}}\right){\rm Tr}M^{\ell}.\no
\eeqa
Here notice that the following relations hold:
\beqa
\label{cancel2}
-\sum^{\infty}_{m=1}\frac{e^{-\epsilon m}}{m}
+\sum^{\infty}_{r=\frac{1}{2}}\frac{e^{-\epsilon r}}{r}
&=&\lim_{\epsilon\rightarrow 0}\left[\ln(1-e^{-\epsilon})
-\ln\frac{1-e^{-\frac{\epsilon}{2}}}{1+e^{-\frac{\epsilon}{2}}}\right]=\ln 4,
\\
\sum^{\infty}_{m=1}\frac{1}{(m-\frac{1}{2})^{\ell}}
&=&2^{\ell}\left(\sum^{\infty}_{m=1}\frac{1}{m^{\ell}}-\sum^{\infty}_{m=1}
\frac{1}{(2m)^{\ell}}\right)=(2^{\ell}-1)\zeta(\ell),\\
\ln \Gamma(z)&=&\sum^{\infty}_{\ell=2}\frac{(-1)^{\ell}}{\ell}\zeta(\ell)
z^{\ell}-\gamma z-\ln z \label{gammaformula},
\eeqa
where $\zeta(\ell)$, $\Gamma(z)$ and $\gamma$ are the zeta function, the gamma 
function and the Euler's constant, respectively. By using these relations we 
can obtain the final expression\footnote{In the paper \cite{JoTy} the authors 
assume that the action for a brane-antibrane includes only the Lorentz 
invariant factors ${\cal G}^{\mu\nu}\pa_{\mu}\bar{T}\pa_{\nu}T$, 
${\cal G}^{\mu\nu}\pa_{\mu}T\pa_{\nu}T$ and 
${\cal G}^{\mu\nu}\pa_{\mu}\bar{T}\pa_{\nu}\bar{T}$, where ${\cal G}$ is the 
open string metric \cite{SeWi} defined by ${\cal G}^{\mu\nu}
=\left(\frac{1}{G-B}\right)^{(\mu\nu)}$ in eq.(\ref{openstringmetric}). 
However, if we calculate ${\rm Tr}M^{\ell}$ in eq.(\ref{traceformula}) by 
using the explicit definition of $M$ in eq.(\ref{defM}), it seems that  
the antisymmetric part of $\left(\frac{1}{G-B}\right)^{\mu\nu}$ appears in the 
action.} of the action of string field 
theory:
\beqa
\label{DDbaraction10}
S=\frac{2\tau_p}{g}\int d^{p+1}x e^{-|T(x)|^2}
\sqrt{-\det(G_{\mu\nu}-B_{\mu\nu})}Z(M),
\eeqa
where the total normalization is determined in the same way as 
eq.(\ref{ponormal}), and $Z(M)$ is given by  
\beqa
\label{specialfunc}
Z(M)&=&4^{{\rm Tr}M}\frac{Z_1(M)^2}{Z_1(2M)},\\
Z_1(M)&=&\det \Gamma(M)e^{\gamma{\rm Tr}M}\det M.
\eeqa
If $\det M$ and $\det\Gamma(M)$ look singular, we have to 
go back to the definition by the trace formula (\ref{traceformula}), where 
they are well-defined.

Incidentally, 
to obtain this result\footnote{The method of the calculation of the disk 
partition function here is different from the one in our original paper 
\cite{TaTeUe} and others \cite{Wi2,KuMaMo1,KuMaMo2}. It is similar to the 
method in the papers \cite{KrLa,FrTs,An,AnTs1,Ts1,Ts2}. However, the result is 
completely the same.} we assumed that the tachyon field $T(X)$ has  
linear dependence of $X^{\mu}$. 
The linear configuration of the tachyon is generally written by
\beqa
T(x)=a+\sum_{\mu=0}^{p}b_\mu x^\mu,\ \ (a,b_\mu \in {\mb{C}}).
\eeqa
This can be put into the following form by a Poincar\'e transformation 
and U(1) gauge transformation in eq.(\ref{eq2.1})  
\ba
\label{tak}
T(x)=\f{1}{2}(iu_1x_1+u_2x_2),\ \ (u_1,u_2 \in {\bf R}).
\ea
Indeed, this configuration of the tachyon field is desirable to obtain 
lower dimensional D-branes. For non-zero $u_i$ this represents a vortex in 
the $x^1$-$x^2$ plane. If $u_1=0$ or $u_2=0$, this 
corresponds to a kink\footnote{The schematic picture of a kink in 
figure \ref{kink} looks different from eq.(\ref{tak}), while both have the 
same feature that the energy localizes around $x^1=0$ or $x^2=0$ 
like eq.(\ref{localize}).
If we consider the appropriate field redefinition $T^{\prime}=f(T)$ on the 
effective field theory of the tachyon field, 
we can transform equation (\ref{tak}) into the 
configuration in figure \ref{kink} \cite{MiZw,MiZw2,HaNa}.} configuration. 
This fact is 
easy to see if we insert this configuration into the tachyon potential $V(T)$ 
\ba
\label{localize}
V(T)\propto e^{-|T(x_1,x_2)|^2}=e^{-\f{1}{4}(u_1x_1)^2-\f{1}{4}(u_2x_2)^2}.
\ea
From this we can see that if both $u_1$ and $u_2$ are nonzero the energy is 
almost zero in the $x^1$-$x^2$ plane except around $x^1=x^2=0$. In other words 
the energy localizes at $x^1=x^2=0$ with width 
$\left(\frac{1}{u_1},~\frac{1}{u_2}\right)$. 
This means that a codimension two 
soliton appears at $x^1=x^2=0$, which can be regarded as a BPS D$(p-2)$ brane. 
In the same way if $u_1$ is nonzero and $u_2$ is zero, a non-BPS D$(p-1)$ 
brane appears as a codimension one soliton at $x^1=0$.   

Let us now condense the tachyon field under the simplest\footnote{This 
restriction is due 
to technical difficulty of the algebra of $M$ in eq.(\ref{defM}).} background 
$(G_{\mu\nu}=\eta_{\mu\nu},~B_{\mu\nu}=0)$. First we insert the tachyon 
profile (\ref{tak}) in the matrix $M$ in eq.(\ref{defM}). It is 
given by $M=\mbox{diag}(u_1^2,~u_2^2)$ and $Z(M)$ in eq.(\ref{specialfunc}) 
becomes $Z(u_1^2)Z(u_2^2)$.
Now according to the g-theorem, which was 
explained in section \ref{BVformalism}, the action $S$ of string field theory 
should 
monotonically decrease along the one-dimensional renormalization group flow. 
Indeed the function $\sqrt{\frac{2}{y}}Z(y)$ is a monotonically decreasing 
function of $y$, and it takes the minimum value at $y=\infty$ with its value
\ba
\sqrt{\frac{2}{y}}Z(y)\to \s{2\pi}\ \ \ (y\to\infty).
\ea
Therefore there exist three decay modes 
$(u_1,u_2)=(\infty,0),\ (0,\infty),\mbox{and}\ (\infty,\infty)$, where the 
conformal invariance is restored. The first two cases represent kink 
configurations and the width $\frac{1}{u_1}$ (or $\frac{1}{u_2}$) of 
the kink goes to zero, which corresponds to the fact that the width of 
D-branes is zero\footnote{In cubic string field theory the width of solitons 
is not zero \cite{lump}, though it might become zero by some field 
redefinition.}. Therefore, we expect a non-BPS D$(p-1)$-brane will be 
generated at $x^1=0$ or $x^2=0$, respectively. This speculation is verified 
if one computes the tension and see that the correct value is 
reproduced as follows
\ba
S&=&\frac{2\tau_p}{g}\int d^{p}x 
\left(\int dx_1 e^{-\frac{u_1^2}{4}x_1^2}\right)Z(u_1^2)\no
&{\displaystyle \mathop{\longrightarrow}^{u_1\rightarrow\infty}}& 
\frac{4\pi \tau_p}{g}\int d^{p}x=\frac{\sqrt{2}\tau_{p-1}}{g}\int d^p x
~~~~~~~~(\mbox{for}~(u_1,u_2)=(\infty,0)).
\ea
Here we have used the relation 
$(2\sqrt{2}\pi)\tau_p=\tau_{p-1}$. The tension of a 
non-BPS D$(p-1)$ brane appears in the coefficient of the action. 

More intuitive way to see the generation of a non-BPS D$(p-1)$-brane is
to discuss the boundary interaction (\ref{eq6}). Let us shift the
original tachyon field by a real constant $T_0$ along $x^1$ as follows:
\ba
T(X)=\f{1}{2}T_0+\f{i}{2}u_1X_1. \label{tachyon} 
\ea
The boundary interaction (\ref{eq6}) after the 
condensation of the tachyon field (\ref{tachyon}) becomes
\ba
I_{B}&=&\int_{\de\Sigma} d\tau \left[\bar{\eta}\dot{\eta}+\f{1}{\s{2\pi}}u_1\psi^1 
(\eta-\bar{\eta})+\frac{1}{8\pi}T_0^2+\frac{1}{8\pi}u_1^2(X^1)^2+\ddd\right], 
\label{ib}
\ea
where the new tachyon field $T_0$ depends only 
on $X^a\ \ (a=0,2,\ddd,p)$. 
{}From this expression it is easy to see that 
in the limit $u_1\to\infty$ we can set $\eta=\bar{\eta}$ after we perform 
the path integral of the fermion $\psi^1$. The term $\ddd$ which 
depends on $A_{\mu}^{(-)}$ vanishes because it is proportional to 
$\bar{\eta}\eta\sim 0$. On the other hand, the gauge field
$A_{\mu}^{(+)}$ is not sensitive to the tachyon condensation except
that the element $A_{1}^{(+)}$ is no longer a gauge field and becomes a 
transverse scalar field since the boundary condition along $x^1$ becomes the 
Dirichlet one. Thus the final boundary action after integrating out the 
fields $X^1$ and $\psi^1$ is identified with that of a non-BPS D$(p-1)$-brane 
(\ref{bin}).

Next we turn to the case $(u_1,u_2)=(\infty,\infty)$. This
corresponds to the vortex-type configuration and the width of the vortex 
$\left(\frac{1}{u_1},~\frac{1}{u_2}\right)$ goes to zero. Therefore, a BPS 
D$(p-2)$-brane is 
expected to be generated at $(x^1,x^2)=(0,0)$. This fact can also be checked
by computing the tension as follows:
\ba
S&=&\frac{2\tau_p}{g}\int d^{p-1}x \left(\int dx_1 dx^2 
e^{-\frac{u_1^2}{4}x_1^2-\frac{u_2^2}{4}x_2^2}\right)Z(u_1^2)Z(u_2^2)\no
&{\displaystyle \mathop{\longrightarrow}^{u_1,u_2\rightarrow\infty}}& 
\frac{8\pi^2 \tau_p}{g}\int d^{p-1}x=\frac{\tau_{p-2}}{g}\int d^{p-1}x,
\ea
matching with the known result. 

It is also interesting to consider multiple brane-antibrane systems. 
Let us consider $2^{k-1}$ pairs of brane-antibrane and condense the tachyon 
field (see the statement (\ref{multiple})). 
As an extension of the tachyon profile (\ref{tak}) the following 
configuration describes the tachyon condensation on multiple 
$Dp+\overline{Dp}$ systems:
\ba
T(x)=i\f{u}{2}\sum_{\mu=1}^{2k}\Gamma^\mu x_\mu,\label{abs-1}
\ea
where $\Gamma^\mu$ denotes $2^{k-1}\times 2^{k-1}$ $\Gamma$-matrix which 
belongs to $SO(2k)$ Clifford algebra. This configuration matches with the 
K-theory argument \cite{Wi1}, which is a 
mathematical description of generation of D-brane charges. Equation 
(\ref{abs-1}) is known as the 
Atiyah-Bott-Shapiro construction \cite{ABS} and a 
BPS D$(p-2k)$-brane is expected to be generated. 
This fact will become more explicit when we investigate the Wess-Zumino term 
in section \ref{RRcouple}. 
Moreover, in the same way as previous cases we can verify that the correct 
tension is produced if one starts with worldsheet action (\ref{eq9}) 
for multiple brane-antibrane systems. Similarly one can 
also see that the condensation of the tachyon field 
\ba
\label{abs-11}
T(x)=i\f{u}{2}\sum_{\mu=1}^{2k-1}\Gamma^\mu x_\mu, 
\ea
on $2^{k-1}$ $Dp+\overline{Dp}$ pairs produces a non-BPS D$(p-2k+1)$-brane.

\subsection{On non-BPS D-branes}
\hspace*{4.5mm}
In this subsection we will 
consider the tachyon condensation on non-BPS D-branes \cite{KuMaMo2}. 
It turns out to be very easy to describe it if we use the descent relation 
(\ref{descentfield}). 

First the tachyon potential on a non-BPS D$p$-brane is obtained by using the 
result (\ref{ponormal}) for a $Dp+\overline{Dp}$ system and the descent 
relation (\ref{descentfield}). It is given by 
\beqa
V(T)=\frac{\sqrt{2}\tau_p}{g}e^{-\frac{T^2}{4}}.
\eeqa
The shape of potential is also the same as the one (fig. \ref{potential3}) for 
a $Dp+\overline{Dp}$ system except the normalization, and it does not depend 
on which kind of closed string background we choose \cite{sen23}.

Next we consider the condensation of a topologically nontrivial tachyon on a 
non-BPS D$p$-brane. To do this we need to obtain the action of string field 
theory for a non-BPS D$p$-brane. It is obtained from the result 
(\ref{DDbaraction10}) for a $Dp+\overline{Dp}$ system by using the descent 
relation (\ref{descentfield}). For a non-BPS D-brane the result 
becomes simpler\footnote{The reason why the expression becomes simple can 
be seen in eq.(\ref{traceformula}). The trace of $M^{\ell}$ simply becomes 
$({\cal G}^{\mu\nu}\pa_{\mu}T\pa_{\nu}T)^{\ell}$.} than that 
for a brane-antibrane system and it is given by
\beqa
\label{nonBPSaction10}
S=\frac{\sqrt{2}\tau_p}{g}\int d^{p+1}x e^{-\frac{T(x)^2}{4}}
\sqrt{-\det(G_{\mu\nu}-B_{\mu\nu})}Z(y),
\eeqa
where
\beqa
\label{openstringmetric}
y&=&{\cal G}^{\mu\nu}\pa_{\mu}T\pa_{\nu}T~~~\mbox{with}~~~
{\cal G}^{\mu\nu}=\left(\frac{1}{G-B}\right)^{(\mu\nu)},\\
\label{specialfunc1}
Z(y)&=&4^y\frac{Z_1(y)^2}{Z_1(2y)},\\
Z_1(y)&=&\Gamma(y)e^{\gamma y}y.
\eeqa
Here ${\cal G}^{\mu\nu}$ is called open string metric \cite{SeWi}, which is 
familiar to noncommutative theorists. 

In the same way as a $Dp+\overline{Dp}$ system we consider the linear profile 
of the tachyon field $T(X)$ under the simplest background 
$(G_{\mu\nu}=\eta_{\mu\nu},~B_{\mu\nu}=0)$. Indeed the 
action (\ref{nonBPSaction10}) can also be obtained by restricting the tachyon 
field $T(X)$ to the 
linear configuration in the worldsheet action (\ref{eq1}) and performing the 
Gaussian path integral to calculate the disk amplitude. By using the 
Poincar\'e transformation the tachyon field becomes
\beqa
\label{tachconfig}
T(x)=u_1x_1~~~(u_1\in{\bf R}).
\eeqa
Now we insert this into the action (\ref{nonBPSaction10}) and minimize it by 
taking the limit $u_1\rightarrow\infty$, which corresponds to the RG-flow 
going into the IR fixed point. The result becomes
\ba
S&=&\frac{\sqrt{2}\tau_p}{g}\int d^{p}x \left(\int dx_1 
e^{-\frac{u_1^2}{4}x_1^2}\right)Z(u_1^2)\no
&{\displaystyle \mathop{\longrightarrow}^{u_1\rightarrow\infty}}& 
2\sqrt{2}\pi\frac{\tau_p}{g}\int d^{p}x=\frac{\tau_{p-1}}{g}\int d^p x.
\ea  
From the coefficient we can see that the correct tension of a BPS D$(p-1)$ 
brane is generated. 

Moreover because the action (\ref{nonBPSaction10}) is simpler than that for a 
brane-antibrane we are able to 
calculate the tachyon condensation under the nontrivial constant background 
$(G_{\mu\nu},~B_{\mu\nu})$. The parameter $y$ in eq.(\ref{openstringmetric}) 
becomes $y={\cal G}^{11}u_1^2$. Inserting this into the action 
(\ref{nonBPSaction10}) it becomes
\ba
S&=&\frac{\sqrt{2}\tau_p}{g}\int d^{p}x \left(\int dx_1 
e^{-\frac{u_1^2}{4}x_1^2}\right)
\sqrt{-\det(G_{\mu\nu}-B_{\mu\nu})}Z({\cal G}^{11} u_1^2)\no
&{\displaystyle \mathop{\longrightarrow}^{u_1\rightarrow\infty}}& 
\frac{\tau_{p-1}}{g}\int d^{p}x \sqrt{-\det(G_{\mu\nu}-B_{\mu\nu})}
\sqrt{{\cal G}^{11}}\no
&=&\frac{\tau_{p-1}}{g}\int d^{p}x 
\sqrt{-\det(G_{\alpha\beta}-B_{\alpha\beta})},
\ea  
where indices $\alpha$ and $\beta$ do not include $1$. This means that 
the tension including the DBI factor is correctly reproduced after the tachyon 
condensation.

One can also see that the condensation of the tachyon field \cite{Ho1} 
\ba
T(x)=i\f{u}{2}\sum_{\mu=1}^{2k-1}\Gamma^\mu x_\mu, 
\ea
on $2^{k-1}$ non-BPS D$p$-branes produces a BPS D$(p-2k+1)$-brane.

In this way we have described all decay modes which can be handled in BSFT by 
free field calculations and they are all consistent with K-theoretic 
arguments \cite{Wi1,Ho1}. 

\section{The Dirac-Born-Infeld Actions for Non-BPS Systems \label{WV}}

\setcounter{equation}{0}
\hspace*{4.5mm}

As we already know, the effective action for a BPS D$p$-brane consists of two 
parts \cite{BeTo}. 
One is the Dirac-Born-Infeld term \cite{AgPoSc,Tsey,FrTs} and the other is the 
Wess-Zumino term \cite{Do1,GrHaMo,Li}. 
The explicit form of these terms can be written as
\beqa
\label{BPSeff}
S=\frac{\tau_p}{g}\int d^{p+1} x
\sqrt{-\det(G_{\mu\nu}-B_{\mu\nu}+4\pi F_{\mu\nu})}+\tau_p
\int C\wedge e^{4\pi F},
\eeqa
where $g$ is the string coupling constant. On the other hand the effective 
actions for non-BPS systems also have the same structure as eq.(\ref{BPSeff}), 
and boundary string field theory enables us to calculate their 
explicit forms. The meaning of ``effective" will be explained in 
section \ref{meaning}. 
In this section we will compute the Dirac-Born-Infeld type action 
with tachyons and gauge fields on non-BPS systems 
\cite{TaTeUe,KrLa,An,KuMaMo2,Ts2}, and in section \ref{RRcouple} we will 
calculate the Wess-Zumino terms. 

In section \ref{exacttach} we obtained the 
target space actions (\ref{DDbaraction10}) and (\ref{nonBPSaction10}) for a 
$Dp+\overline{Dp}$ system and for a non-BPS D$p$-brane without the gauge 
fields. Here we will obtain the 
world volume actions including the gauge fields on these systems. 

First we will discuss the worldvolume action on a non-BPS D$9$-brane 
\cite{KuMaMo1,An,HaHi}. 
If we want to consider a non-BPS D$p$-brane we have only to perform the 
T-dual transformation to the worldvolume action by replacing one component 
of the gauge field with a transverse scalar.  
In the worldsheet action (\ref{eq1}) for a non-BPS D9-brane the gauge field 
appears in the following way
\beqa
I_B&=&\int_{\pa\Sigma}d\tau d\theta\left[\cdots 
-iD_{\theta}{\bf X}^{\mu}A_{\mu}({\bf X})+\cdots\cdots\right]\no
&=&\int_{\pa\Sigma}d\tau\left[\cdots 
-i\dot{X}^{\mu}A_{\mu}(X)+2i F_{\mu\nu}(X)\psi^{\mu}\psi^{\nu}+
\cdots\cdots\right].
\eeqa
Since the fields $A_{\mu}(X)$ and 
$F_{\mu\nu}(X)$ nontrivially depend on the worldsheet field $X$, 
we can not exactly 
perform the path integral. However, if we restrict the field configuration of 
the gauge field to the constant field strength $F_{\mu\nu}$, this worldsheet 
action includes only the second order of the fields $X^{\mu}$ and $\psi^{\mu}$ 
as follows
\beqa
I_B=-i\int_{\pa\Sigma} d\tau \left[\cdots -\frac{1}{2}X^{\mu}\dot{X}^{\nu}
F_{\mu\nu}-2\psi^{\mu}\psi^{\nu}F_{\mu\nu}+\cdots\cdots\right].
\eeqa
From this expression we can see that by adding these terms to 
eq.(\ref{functional0}) or eq.(\ref{functional1}) the path integral 
(\ref{pathint}) is equivalent to the description of replacing 
$B_{\mu\nu}$ with $\ti{B}_{\mu\nu}\equiv B_{\mu\nu}-4\pi F_{\mu\nu}$ in 
eq.(\ref{inverseprop0}) or eq.(\ref{inverseprop1}) \cite{FrTs}. 
Indeed this combination satisfies the desired 
gauge invariance : 
$B_{\mu\nu}\rightarrow B_{\mu\nu}+\pa_{\mu}\zeta_{\nu}-\pa_{\nu}\zeta_{\mu},~
A_{\mu}\rightarrow A_{\mu}+\frac{1}{4\pi}\zeta_{\mu}$. 
 
Therefore the target space action for a non-BPS 
D9-brane\footnote{This is different from another proposal of the action 
for a non-BPS D-brane \cite{Ga,BeRoWiEyPa}.
There might be some relation between these 
two actions by some field redefinition.} is given by 
\beqa
S&=&\frac{\sqrt{2}\tau_9}{g}\int d^{10}x e^{-\frac{T(x)^2}{4}}
\sqrt{-\det(G_{\mu\nu}-B_{\mu\nu}+4\pi F_{\mu\nu})}
Z({\cal G}^{\mu\nu}\pa_{\mu}T\pa_{\nu}T)\no
&=&\frac{\sqrt{2}\tau_9}{g}\int d^{10}x e^{-\frac{T(x)^2}{4}}
\sqrt{-\det(G_{\mu\nu}-B_{\mu\nu}+4\pi F_{\mu\nu})}
\left(1+2(\ln 2){\cal G}^{\mu\nu}\pa_{\mu}T\pa_{\nu}T+\cdots\cdots\right),\no
\label{nonBPSaction11}
\eeqa
where the definition of $Z(y)$ is given in eq.(\ref{specialfunc1}), and in the
second line we have used its Taylor expansion given by
\beqa
Z(y)=1+2(\ln 2)y+{\cal O}(y^2).
\eeqa
Here we can see the Dirac-Born-Infeld factor in the action \cite{sen22}. 
We can apply this calculation to that for a BPS D-brane only if we set the 
tachyon field to zero (see appendix \ref{Expression} or the paper \cite{FrTs}).

Note that this action is available only when 
$\pa_{\mu}\pa_{\nu}T=0$ and $\pa_{\rho}F_{\mu\nu}=0$. The tachyon 
profile (\ref{tachconfig}) satisfies this condition, and the 
following configuration is indeed a solution to the equation of motion of the 
above action:
\beqa
\label{sol1}
T(x)=u_1 x_1~(u_1\rightarrow\infty),~~~A_{\mu}(x)=0.
\eeqa 

Next we move on to the worldvolume action for a $D9+\overline{D9}$ system. 
In this case the situation is more complicated because two kinds 
of gauge fields appear in the action. In fact as we can see in the worldsheet 
action (\ref{eq6}) for a brane-antibrane system, even if we consider the 
constant field strength $F^{+}_{\mu\nu}$ and $F^{-}_{\mu\nu}$ the sigma model 
is not an action with only free fields. The terms with $A^{+}_{\mu}$ 
includes only the second order of fields, while the terms with 
$A^{-}_{\mu}$ can not avoid including higher powers of the fields 
$X^{\mu}$ and $\psi^{\mu}$. 
If we include only the constant field strength $F^{+}_{\mu\nu}$ and set 
the other gauge field $A^{-}$ to zero, we have only to replace 
NSNS B-field $B_{\mu\nu}$ with $B_{\mu\nu}-4\pi F^{+}_{\mu\nu}$ in action 
(\ref{DDbaraction10}). 

On the other hand, 
we can show the general structure of the target space action for a 
brane-antibrane system. Here we set all the massive fields to zero in the 
worldsheet action (\ref{eq6}) and consider the most general 
configuration, not the linear one, of the tachyon field and the gauge fields, 
which is different from the analysis in section \ref{exacttach}. 
It is given by
\beqa
S &=& \frac{\tau_9}{g} \int d^{10}x e^{- T \bar{T}(x)} 
\left[\sqrt{-\det\{G_{\mu\nu}-B_{\mu\nu}
+4\pi F^{(1)}_{\mu\nu} (x) \}}\right.\no
&&~~~~~~~\left.+\sqrt{-\det\{G_{\mu\nu}-B_{\mu\nu}
+4\pi F^{(2)}_{\mu\nu} (x) \}}+ \sum_{n=1}^\infty 
{\cal{P}}_{2n} ( F^{(i)}_{\mu \nu}, T,\bar{T} ,D_{\rho})
\right],
\label{Z1}
\eeqa
where ${\cal{P}}_{2n}$ is the $2n$-derivative term constructed from
$F^{(i)}_{\mu \nu}, T, \bar{T}$ and $2n$ covariant derivatives $D_{\mu}$.
Note that, for example, 
$F_{\mu \nu}^{(-)} T$ is regarded as a 2-derivative term
and should be included in ${\cal{P}}_{2}$
because of the identity $ [D_\mu, D_\nu] T = -i F_{\mu \nu}^{(-)} T$.
This ambiguity is reminiscent of
the case of the non-Abelian Dirac-Born-Infeld action \cite{Ts4}
in which 
$[F_{\mu \nu} ,F_{\gamma \sigma}]=
i [D_\mu, D_\nu] F_{\gamma \sigma}$ should be regarded as a derivative term.

Here note that from the analysis in section \ref{exacttach} this action 
should have the following configurations of $T(x)$ and $A^{(1),(2)}_{\mu}(x)$ 
as exact solutions to the equation of motion of eq.(\ref{Z1}):
\ba
\label{sol2.1}
&&T(x)=\frac{i}{2}u_1x_1~~~~~(u_1\rightarrow\infty),
~~~~~~~~~~~~~~~~A^{(1)}_{\mu}=A^{(2)}_{\mu}=0,\\
\label{sol2.2}
&&T(x)=\frac{1}{2}(iu_1x_1+u_2x_2)~(u_1,u_2\rightarrow\infty),
~~~A^{(1)}_{\mu}=A^{(2)}_{\mu}=0.
\ea
These solitonic solutions and eq.(\ref{sol1}) can be interpreted as the field 
theoretical realization of the famous statement that 
``the D-branes are solitons in string theory".

Now let us prove eq.(\ref{Z1}). We use
the matrix form of the world-sheet action (\ref{eq9}):
\beq
S=Z=\int DX D \psi~{\rm exp} [ -I_0(X,\psi) ]
~{\rm Tr \, P \, exp} \int^{\pi}_{-\pi}d\tau M(\tau),
\eeq
where
\beqa
M(\tau)=\left[\begin{array}{c c }
i\dot{X}^{\mu}A^{(1)}_{\mu}-2i\psi^{\mu}\psi^{\nu}F^{(1)}_{\mu\nu}
-\frac{ 1 }{2 \pi} T \bar{T}
& ~ i\sqrt{\frac{2}{\pi}}\psi^{\mu}D_{\mu}T
\nonumber\\
-i\sqrt{\frac{2}{\pi}}\psi^{\mu}\overline{D_{\mu}T}~ 
& i\dot{X}^{\mu}
A^{(2)}_{\mu}-2i\psi^{\mu}\psi^{\nu}F^{(2)}_{\mu\nu}
-\frac{ 1 }{2 \pi} \bar{T} T
\nonumber
\end{array}\right]. \\
\eeqa
First we expand $X^{\mu}$ around the zero mode $x^{\mu}$ 
as $X^{\mu}(\tau)=x^{\mu}+\xi^{\mu}(\tau)$
and path integrate $\xi^{\mu}(\tau)$ and $\psi^{\mu}(\tau)$. 
From the expansion
$T \bar{T}(X) =
T \bar{T}(x) + \xi^{\mu} (\pa_\mu (T \bar{T}) )(x)+\cdots=
T \bar{T}(x) 
+ \xi^{\mu} (D_\mu (T \bar{T}) )(x)+\cdots$,
we can replace $T\bar{T} (X) $ in $Z$ by $T \bar{T}(x)$
if we remove
the derivative terms which can be included in ${\cal{P}}_{2n}$.
The $F_{\mu \nu}^{(i)} (X)$ terms can also be
replaced by $F_{\mu \nu}^{(i)} (x)$. 
Furthermore we see that 
the contributions from the off diagonal part of the matrix $M(\tau)$ 
have the form of ${\cal{P}}_{2n}$
since they can be expanded as 
$D_{\mu} T(X) =
D_{\mu} T(x) + \xi^{\nu} (\pa_\nu (D_{\mu} T) )(x)+\cdots$.
The terms $i \dot{X}^{\mu} A_{\mu}^{(i)}(X) $ 
in the diagonal part 
are combined with 
the other non-gauge covariant terms
to give gauge covariant ones.
Therefore we conclude that the action for
a brane-antibrane system becomes eq.(\ref{Z1}).

Similarly, we can show that the action for $n$ $D9$-branes and $m$ 
$\overline{D9}$-branes becomes
\beqa
S &=& \frac{\tau_9}{g} \int d^{10}x  \left(  {\rm SymTr} \left[
e^{- T \bar{T}  }
\sqrt{-\det\{G_{\mu\nu}-B_{\mu\nu}
+4\pi F^{(1)}_{\mu\nu}  \}}\right.\right.\no 
&&\left.\left.+ e^{- \bar{T} T  }
\sqrt{-\det\{G_{\mu\nu}-B_{\mu\nu}
+4\pi F^{(2)}_{\mu\nu}  \} }
+ \sum_{n=1}^\infty {\cal{P}}_{2n} ( F^{(i)}, T,\bar{T} ,D_{\mu})\right)
\right],
\label{Z2}
\eeqa
where ${\rm SymTr}$ denotes the symmetrized trace 
for $T \bar{T}, \bar{T} T$ and $F^{(i)}_{\mu \nu}$ \cite{Ts4}. Note that 
if we consider multiple $Dp+\overline{Dp}$ branes transverse scalars appear. 
Moreover it is known that complicated 
terms appear according to T-duality argument 
\cite{My} in multi $Dp+\overline{Dp}$ systems. 

At first glance these results may look false because it seems that massive 
modes which is exchanged 
between a D-brane and an anti D-brane modify the sum of the 
Dirac-Born-Infeld actions. However this effect comes from an open string 
one-loop diagram (cylinder amplitude) 
and in the disk amplitude this effect does not cause any modification.

Now we make another comment about the target space actions 
(\ref{nonBPSaction11}) and (\ref{Z1}) (or (\ref{Z2})). From these actions we 
can see that the classical action completely vanishes after the constant 
tachyon condensation $T(x)=T\rightarrow \infty$. 
This fact indicates that open string degrees of freedom completely disappear 
after the tachyon condensation, and it coincides 
with the physical picture that 
D-branes disappear. On the other hand the vanishing of the action can not be 
explained by the simple field theory picture like the scalar QED or the 
Yang-Mills-Higgs theory, which we saw in section \ref{tachcon}. 
In the usual field theory action \cite{pe} like 
$S=\int d^{10}x \left[ 
F^{\mu\nu}F_{\mu\nu}+\pa^{\mu}T\pa_{\mu}T+V(T)\right]$ the kinetic terms 
remain\footnote{This was called U(1) problem before the string 
field theory was applied to tachyon condensation.} after the tachyon 
condensation. In string field theory the kinetic term of the gauge 
field is given by 
$\frac{1}{g_{\ss YM}^2}e^{-\frac{T^2}{4}}F^{\mu\nu}F_{\mu\nu}$ and it really 
vanishes. 
Here note that the effective coupling constant is given by
\beqa
\frac{1}{g_{\ss eff}^2}=\frac{1}{g_{\ss YM}^2}e^{-\frac{T^2}{4}},
\eeqa
thus at $T\rightarrow\infty$ this becomes strong. 
Because of this property we could consider that the flux of the gauge field 
is confined to become closed string degrees of freedom 
\cite{sen100,KlLaSh,GiHoYi,BeHoYi,Yi,sen22}.

Moreover we can find another intriguing fact which the action like\\
$S=\int d^{10}x~
{\rm Tr}\left[F^{\mu\nu}F_{\mu\nu}+D^{\mu}TD_{\mu}T+V(T)\right]$ can not 
explain. First we already know that the action (\ref{Z2}) has the 
soliton solutions given by the Atiyah-Bott-Shapiro construction (\ref{abs-1}) 
and (\ref{abs-11}) with zero gauge fields:
\beqa
&&T(x)=i\f{u}{2}\sum_{\mu=1}^{2k-1}\Gamma^\mu x_\mu~~~~(u\rightarrow\infty),
~~~~~~~~~~~~~~~~A^{(1)}_{\mu}=A^{(2)}_{\mu}=0,\\
&&T(x)=i\f{u}{2}\sum_{\mu=1}^{2k}\Gamma^\mu x_\mu~~~~~(u\rightarrow\infty),
~~~~~~~~~~~~~~~~A^{(1)}_{\mu}=A^{(2)}_{\mu}=0.
\eeqa
These solutions describe the decay 
of $D9+\overline{D9}$ systems to various lower dimensional D-branes. 
On the other hand, it is known that 
in usual field theories these kinds of solutions do not exist. 
In the field theory picture a non-BPS D$8$-brane can be regarded as a kink 
solution in the field theory on $D9+\overline{D9}$ systems. A BPS D$7$-brane 
is a vortex in it. In fact 
these solutions 
can be found as approximate solutions 
in the truncated action which ignores terms with more than three derivatives 
\cite{MiZw,MiZw2}. 
However, the Atiyah-Bott-Shapiro type solution is generally prohibited in 
standard local field theories. Indeed, according to the Derrick's theorem 
\cite{Derrick} in local field theories, there do not exist more 
than codimension $5$ solitons because the total energy of these solutions 
often diverge. Codimension $4$ solitons are instantons, 
which are BPS-solutions in the four-dimensional Yang-Mills theory, and they 
can be realized as BPS D0-branes on BPS D4-branes \cite{Witten99,Do1}. 
However, the Atiyah-Bott-Shapiro type solutions represent not only less than 
codimension $4$ but also more than codimension 5 solitons like the decay of 
multi $D9+\overline{D9}$ systems to a BPS D3-brane (codimension 6) or to a BPS 
D1-brane (codimension 8). These solutions can not be found in standard 
local field theories. The existence of higher 
codimensional solitons is one of features in string field theories, 
which are nonlocal.

Finally we will compute the action for a brane-antibrane system as 
a sigma model partition function perturbatively up to $\alpha'^2$ 
under the simplest background 
$(G_{\mu\nu}=\eta_{\mu\nu},~B_{\mu\nu}=0)$. Hereafter we will restore the 
dimension-full parameter $\A'$ by including a factor $\A'/2$. 
{}From eq.(\ref{functional0}) and eq.(\ref{inverseprop0}) we can obtain the 
Green functions of $\xi^{\mu}(\tau)$ and $\psi^{\mu}(\tau)$ as follows:
\ba
\langle \xi^\mu(\tau)\xi^\mu(\tau ') \rangle &=& \A' \!\!\!
\sum_{m\in {\mb{Z} \ne 0}}
\!\! \f{1}{|m|}e^{im(\tau-\tau ')-\ep|m|}
=2 \A' \!\!\!
\sum_{m\in {\mb{Z} > 0}}\!\! 
\frac{1}{m} {\rm cos} (m(\tau-\tau')) e^{-\ep|m|}
, \label{gb1}\\
\langle \psi^\mu(\tau) \psi^\mu(\tau ') \rangle &=&
-\f{i}{2}\sum_{r\in {\mb{Z}}+\f12}\f{r}{|r|}e^{ir(\tau-\tau')-\ep|r|}=
\sum_{m\in {\mb{Z}+\f12 > 0}} {\rm sin} (r(\tau-\tau')) e^{-\ep|r|}. 
\label{gf1}
\ea
The regularization used here keeps the world-sheet supersymmetry and 
the spacetime gauge invariance (\ref{eq2.1}) or (\ref{eq2.11}), 
which corresponds to world-sheet global symmetry \cite{Ts1, AnTs1}.
The expansion of fields in eq.(\ref{eq8}) are given by
\beqa
T \bar{T}(X) &=&
T \bar{T}(x) + \xi^{\mu} (D_\mu (T \bar{T}) )(x)+\cdots
+\frac{1}{4!} \xi^{\sigma} \xi^{\rho} \xi^{\nu} \xi^{\mu} 
(D_{\sigma} D_{\rho} D_{\nu} D_{\mu} (T \bar{T}) )(x)+\cdots,
\nonumber \\
A^{(i)}_{\mu}(X) &=&
A^{(i)}_{\mu} (x) + \xi^{\nu} \pa_\nu A^{(i)}_\mu (x)+\cdots, \hspace{.5cm}
F^{(i)}_{\mu \nu}(X) = 
F^{(i)}_{\mu \nu}(x) + \xi^{\rho} (D_\rho F^{(i)}_{\mu \nu} )(x)+\cdots,
\nonumber\\
D_{\mu} T(X) &=&
D_{\mu} T(x) + \xi^{\nu} (\pa_\nu (D_{\mu} T) )(x)+ 
\frac{1}{2} \xi^{\rho} \xi^{\nu} (\pa_{\rho} \pa_\nu (D_{\mu} T) )(x)+\cdots.
\eeqa

Now we can compute the partition function by the perturbation about $\alpha'$.
Since the concrete computation is somewhat complicate,
we will show only the result: 
\beqa
S=Z&=&  \frac{\tau_9}{g}\int d^{10}x e^{-\bar{T_{R}}T_{R} }
\left[ 2+8 \alpha' (\log 2) D^{\mu}T_{R} \overline{D_{\mu} T_{R}} +
\alpha'^2 \pi^2 \left( F^{(1)\mu\nu}_R F^{(1)}_{R\mu\nu}
+F^{(2)\mu\nu}_R F^{(2)}_{R\mu\nu}\right) 
\right. \nonumber \\
&& \hspace{2.0cm} +
4 \alpha'^2 \gamma_0 (D^{\mu} D^{\nu} T_{R}) (D_{\mu} D_{\nu} \overline{T}_{R})
+32 \alpha'^2 i (\log 2)^2 F^{(-)\mu\nu}_R D_{\mu} T_{R} D_{\nu}
\overline{T}_{R}
\nonumber \\
&& \hspace{2.0cm} +
2 \alpha'^2 \left( 8 (\log2)^2 -\frac{1}{3} \pi^2 \right)
(D^{\mu} T_{R} D_{\mu} \overline{T}_{R})^2\no
&& \hspace{2.0cm} - \alpha'^2 \frac{2}{3} \pi^2 (D^{\mu}T_RD_{\mu}T_R)
(D^{\nu}\overline{T}_{R}D_{\nu}\overline{T}_{R})
\nonumber \\
&& \hspace{2.0cm}  +
\frac{\pi^2}{6}\alpha'^2 \left( (D^{\mu} D^{\nu}T_{R}) \overline {T}_{R}
+ T_{R} (D^{\mu} D^{\nu} \overline {T}_{R})   \right)
\nonumber \\
&& \hspace{2.0cm} \left. 
 \;\;\;~~ \times \left( D_{\mu} D_{\nu} (\overline{T}_{R}T_{R}) 
+ D_{\nu} \overline{T}_{R} D_{\mu} T_{R} 
+ D_{\mu} \overline{T}_{R} D_{\nu} T_{R} 
\right)+\cdots
\right], 
\label{a2}
\eeqa
where $\gamma_0$ is a constant which is defined by
\beq
\gamma_0={\rm \lim_{\epsilon \rightarrow 0}} \left(
\sum_{r,m>0} \frac{1}{m} 
\left( \frac{1}{r+m}+\frac{1}{r-m} \right) e^{-(r+m)\epsilon}
-(\log \epsilon)^2 \right),
\eeq
and we have renormalized the tachyon field and the gauge field as 
\ba
\label{fa}
T&=&T_{R}+\alpha' (\log \epsilon) D^{\mu} D_{\mu} T_{R} 
+ \frac{1}{2} \alpha'^2  (\log \epsilon)^2 
D^{\mu} D_{\mu} D^{\nu} D_{\nu} T_{R}\nonumber\\
&&~~~~~~~~~~~~~~~~ + i \alpha^{\prime 2}
(\log \epsilon)^2 D^{\mu}F_{R\mu\nu }^{(-)}D^{\nu}T_{R},\nonumber\\
A_{\mu}^{(-)}&=&A_{R\mu }^{(-)}+\alpha^{\prime} (\log \epsilon) 
D^{\nu}F_{R\nu\mu}^{(-)}.
\ea
Note that the above equation can also be regarded as a field redefinition of 
the target space fields from $T(x),~A_{\mu}(x)$ to $T_R(x),~A_{R\mu}(x)$. 
We can see that the renormalization in the worldsheet corresponds to the field 
redefinition in the target space \cite{SeWi}.

\section{The Wess-Zumino Terms for Non-BPS Systems \label{RRcouple}}
\setcounter{equation}{0}
\hspace*{4.5mm}
In this section we will compute the Wess-Zumino terms (RR-couplings or 
Chern-Simon terms) on non-BPS D-branes and brane-antibrane systems 
within the framework of BSFT \cite{TaTeUe,KrLa}. 
These terms for non-BPS systems correspond to the second term of the 
effective action for a BPS D-brane in eq.(\ref{BPSeff}). The boundary string 
field theory enable us to calculate these terms in the same
way as the calculation of the Dirac-Born-Infeld type actions. 

The difference in the computations from the previous sections is that 
the result is completely exact without any assumptions that the tachyon is the 
linear order or something. 
Furthermore the resulting expression has a intriguing 
mathematical structure known as superconnection \cite{Qu}. This fact was 
already conjectured in \cite{Wi1,KeWi} for brane-antibrane systems. Here we 
can see the explicit proof of this in BSFT and we point out that this 
structure can also be found in the Wess-Zumino terms on non-BPS D-branes. 
These results give another evidence of classification 
of D-brane charges by K-theory\cite{MiMo,HaMi,Wi1,Ho1}.

In this section we consider $D9+\overline{D9}$ systems and non-BPS D$9$-branes.
If we would like to obtain 
the result for $Dp+\overline{Dp}$ systems or non-BPS D$p$-branes, 
we have to replace several components of 
the gauge field with transverse scalars due to the T-duality. However, we have 
to note that they also include other nontrivial terms called Myers terms 
\cite{My}. In our original paper \cite{TaTeUe} we also obtained the complete 
forms of Myers terms, while in this thesis we omit them. 
 
\subsection{Calculations and results}
\hspace*{4.5mm}
Here we start with the simplest flat background 
$(G_{\mu\nu}=\eta_{\mu\nu},~B_{\mu\nu}=0)$, and instead we put a little 
RR-flux in the background. In this case we can deal with the shift of the 
background by perturbation from the flat background. In BSFT this shift 
can be realized by inserting a RR vertex operator inside the disk. 

The important thing here is that the picture \cite{FrMaSh} of the 
RR-vertex operator should be $(-\f{1}{2},-\f{3}{2})$. This is 
because we consider the insertion of only one RR-vertex operator and  
the total picture number on the disk should be $-2$. The RR vertex operator 
with $(-\frac{1}{2},-\frac{3}{2})$ picture is given by \cite{BiDiFrLePeRuSc}
\ba
\label{rr0}
V^{(-\f{1}{2},-\f{3}{2})}&=&
e^{-\f12\phi_L-\f32\phi_R}(P_{-}\hat{C})^{AB}S_{LA}S_{RB},  \no
\hat{C}^{AB}&=&\sum_{p:{\ss odd~(even)}} 
\f{1}{p!}(\Gamma^{\mu_1\ddd\mu_p})^{AB}C_{\mu_1\ddd\mu_p}, 
\ea
where $\phi_L$ and $\phi_R$ denote the bosonized superconformal ghosts of
left-moving and right-moving sectors, respectively. We define $S_{LA}$ and 
$S_{RB}$ as spin-fields constructed from world-sheet fermions and $P_{-}$
 denotes the projection of the chirality $\frac{1-\Gamma_{11}}{2}$. 
The index $p$ in $\sum_{p}$ runs odd integers for Type IIA and even integers 
for Type IIB. 
For more details of the notation for spinors see appendix \ref{Notation}. 

According to the construction of BSFT in section \ref{BVformalism} we can 
separate the matter part from the ghost part and we have only to consider the 
contribution from the matter to calculate the disk partition function. 
Therefore from now on we will ignore the ghost part. Anyway the most important 
thing is that the coefficient $C_{\mu_1\ddd\mu_p}$ of the RR-vertex operator 
represents a RR p-form gauge field \cite{BiDiFrLePeRuSc}, not the field 
strength $F_{\mu_1\ddd\mu_p}$, since the coefficient of a RR-vertex operator 
with standard $(-\frac{1}{2},-\frac{1}{2})$ picture is the field strength 
\cite{Po1} and this does not appear in the Wess-Zumino terms.

\begin{figure} [tbhp]
\begin{center}
\epsfxsize=70mm
\epsfbox{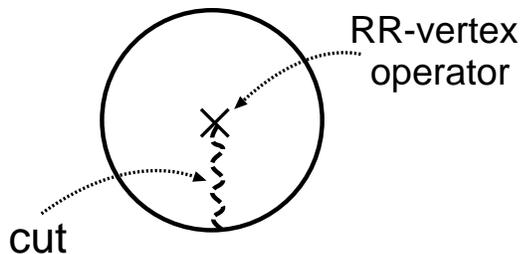}
\caption{A RR-vertex operator 
inserted inside the disk generates the cut. Because of 
this cut worldsheet fermions on the boundary of disk take 
the periodic boundary condition 
$\psi^{\mu}(\tau+2\pi)=\psi^{\mu}(\tau)$. \label{RRvertex}}
\end{center}
\end{figure}

First we turn to a $D9+\overline{D9}$ system and determine the Wess-Zumino 
term up to its overall normalization. 
The important thing we have to notice is that the 
contribution from nonzero modes of $X^{\mu}$ completely cancels out that 
from nonzero modes of $\psi^{\mu}$, and only the contribution from 
zero modes remains. This fact comes from the worldsheet supersymmetry. Indeed 
all fermions at the boundary of disk obey the periodic boundary condition 
$\psi^{\mu}(\tau+2\pi)=\psi^{\mu}(\tau)$ due to 
the cut generated by a RR vertex operator (see figure \ref{RRvertex}), and the 
worldsheet supersymmetry is completely preserved in the one-dimensional
boundary theory. Therefore one can believe that the contribution from 
fermions and bosons is canceled with each other for nonzero-modes 
\cite{TaTeUe,KrLa}. 

As an example, we can see the cancellation of the contribution from nonzero 
modes in eq.(\ref{Trln}). Since the worldsheet fermions take the 
periodic boundary condition in this case, its mode-index runs positive integer 
numbers. If we replace the positive half-integer index $r$ with the positive 
integer index $n$ in eq.(\ref{Trln}), we can see that the contribution from 
bosons completely cancels out that from fermions in eq.(\ref{cancel1}) and 
eq.(\ref{traceformula}). 

However, in eq.(\ref{Trln}) we assumed that the tachyon field $T(X)$ is linear 
and we also set the gauge field strength $F_{\mu\nu}(X)$ 
to a constant or zero. In this section we do not even assume 
that approximation. What we would like to insist is that even under the 
general dependence of $X$ in $T(X)$ and $F_{\mu\nu}(X)$, the contribution from 
non-zero modes disappears. 
Unfortunately its explicit proof is very difficult.

If we admit this fact, the calculation of the Wess-Zumino terms reduces to 
that of only the zero modes of the boson ($x^{\mu}$) and 
the fermions ($\psi_0^{\mu}$ and $\eta,\bar{\eta}$). 
In this calculation it is 
better to change the path integral formalism into the Hamiltonian formalism. 
The procedure is the same as that for obtaining equation (\ref{eq8}) in 
subsection \ref{matrixform}. The path integral of the boundary 
fermions $\eta$ and $\bar{\eta}$ can be represented 
by the RR-sector analog of the important formula (\ref{eq9}):
\ba
& &\int D\eta D\bar{\eta}\ e^{-I_{B}}=\mbox{Tr P}\ (-1)^F\ \exp\ \int_{-\pi}^{\pi}d\tau M(\tau), \label{rr2} \\
& & M(\tau)=
\left(
 	\begin{array}{cc}
 	i\dot{X}^\mu A^{(1)}_{\mu}-2i\psi^{\mu}\psi^{\nu}F^{(1)}_{\mu\nu}
 	-\f{1}{2\pi}T\bar{T} & i\s{\f{2}{\pi}}\psi^{\mu}D_{\mu}T \\
 	-i\s{\f{2}{\pi}}\psi^{\mu}\overline{D_{\mu}T} & i\dot{X}^\mu A^{(2)}_{\mu}
 	-2i\psi^{\mu}\psi^{\nu}F^{(2)}_{\mu\nu}-\f{1}{2\pi}\bar{T}T
 	\end{array}
 \right), \nonumber
\ea
where the insertion of $(-1)^F (=[\bar{\eta},\eta])$ is due to the
periodic boundary condition\footnote{Remember that a path-integral of 
fermions with periodic boundary condition on a circle becomes 
${\rm Tr}(-1)^Fe^{-H}$ in the Hamiltonian formalism.} of $\eta$ and 
$\bar{\eta}$, 
and it can be replaced with the Pauli matrix $\sigma_3$ in the same 
reason as the statement above eq.(\ref{eq9}). 
Since we have only to take the zero
modes into account, we can perform the following 
replacement\footnote{Strictly speaking in eq.(\ref{co}) we should include 
cocycle factors in front of $\Gamma^{\mu}$ in order to guarantee the 
correct worldsheet statistics. Moreover, another kind of cocycle factors is 
also needed in eq.(\ref{rr2}) because we have replaced the boundary fermions 
$\eta,\bar{\eta}$ with the Pauli matrices.} in eq.(\ref{rr2})
\ba
X^{\mu}(\tau)\rightarrow x^{\mu}~~,~~\psi^{\mu}(\tau)\rightarrow
\psi^{\mu}_0=\f{1}{\s{2}}i^{\f12}\Gamma^{\mu},
\label{co}
\ea
where the factor $i^{\f12}$ is due to the conformal map from the open string 
picture to the closed string 
picture \cite{CaLaNaYp}. Because of this 
replacement $\tau$ dependence is dropped out of eq.(\ref{rr2}) and we can 
regard the path-ordered trace Tr P as an ordinary trace.  Thus, we 
have obtained
\ba
\int D\eta D\bar{\eta}DXD\psi~ e^{-I_{0}-I_{B}}=\mbox{Str}:\exp
\left(
	\begin{array}{cc}
	2\pi F_{\mu\nu}^{(1)}\Gamma^{\mu}\Gamma^{\nu}-T\bar{T} & 2i^{\f32}\s{\pi}\Gamma^{\mu}D_{\mu}T \\
	-2i^{\f32}\s{\pi}\Gamma^{\mu}\overline{D_{\mu}T} & 2\pi F_{\mu\nu}^{(2)}\Gamma^{\mu}\Gamma^{\nu}-\bar{T}T
	\end{array}
\right):,
\ea
where the supertrace Str is defined to be a trace with insertion of 
$\sigma_3$, and the symbol $:\ :$ means that $\Gamma$-matrices are 
antisymmetrized because operators should be normal-ordered in the 
Hamiltonian formalism.

After we insert the RR-vertex operator (\ref{rr0}) 
and take the trace using the famous 
relation (\ref{c3}) between the Clifford algebra (\ref{c2}) and the 
differential forms (\ref{c1}) (see appendix \ref{Notation}), we can 
easily obtain the final expression\footnote{Even though one may
 think the factor $i^{\f32}$ strange at first sight, it is an easy task
to show that the action is indeed real by expanding the exponential.} of the 
Wess-Zumino term on a $D9+\overline{D9}$ system \cite{KrLa,TaTeUe}:
\ba
S=\tau_9 \ \mbox{Str}\ \int C\wedge \exp
\left(
	\begin{array}{cc}
	2\pi\al F^{(1)}-T\bar{T} & i^{\f32}\s{2\pi\al}\ DT \\
	-i^{\f32}\s{2\pi\al}\ \overline{DT} & 2\pi\al F^{(2)}-\bar{T}T
	\end{array}
\right), \label{wz}
\ea
where we have recovered $\alpha^{\prime}=2$. Here the total normalization was
determined as follows. If we set the tachyon field $T$ to zero the above 
action becomes
\beqa
\label{RRdiff}
\tau_9\ \int C\wedge e^{2\pi\al F^{(1)}}
-\tau_9\ \int C\wedge e^{2\pi\al F^{(2)}}.
\eeqa 
This should be the sum of the Wess-Zumino terms for a BPS D9-brane and 
an BPS antiD9-brane, and the total normalization should be equal to that for 
a BPS D9-brane, which is given in eq.(\ref{BPSeff}). 

The non-abelian generalization is also straightforward if the above
abelian supertrace is replaced with the non-abelian one:
\ba
\mbox{Str diag}(a_1,a_2,\ddd,a_N,b_1,b_2,\ddd,b_M)=\sum_{i=1}^N a_i-
\sum_{j=1}^M b_j,
\ea
where we have assumed that there are $N$ D$9$-branes and $M$ antiD$9$-branes. 

Next we will 
calculate the Wess-Zumino term for non-BPS D$9$-branes. In the same 
way as the previous sections it is easily obtained by using the 
descent relation (\ref{descentfield}) and (\ref{CPtrans}). According to this 
rule the Chan-Paton degrees of freedom $1,\sigma_1,\sigma_2$ and $\sigma_3$ 
reduce to only $1$ and $\sigma_1$. 
The other difference appears in the fermion number 
operator $(-1)^F$ in eq.(\ref{rr2}). It is represented by 
$(-1)^F=\eta\simeq\frac{1}{\sqrt{2}}\sigma_1$ in the case of a non-BPS 
D9-brane because in the Hamiltonian formalism the quantization of the 
boundary fermion $\eta$ is given by $\eta^2=\frac{1}{2}$. 
Thus up to the overall factor $\mu '$ the result is given by:
\ba
\label{wz10}
S&=&\mu '\ \mbox{Tr}\ \left[\ \int C\wedge \exp{\left[-\f{1}{4}T^2-\s{\pi\al}
i^{\f32}DT\sigma_1+2\pi\al F\right]}\sigma_1\right], 
\no \label{wz-n}
 &=&i\mu '\ \mbox{Tr}\int C\wedge \exp{\left[-\f{1}{4}T^2-\s{\pi\al}
i^{\f12}DT+2\pi\al F\right]}_{odd},
\ea
where the trace $\mbox{Tr}$ in the first equation involves the degree of 
freedom of the Pauli matrix $\sigma_1$. 
The covariant derivative of a Hermitian tachyon
field on non-BPS D-branes is denoted by $DT=dT-i[A,T]$. Also note that in 
the second expression only the terms which include odd powers of $DT$ 
should be picked up because of the trace with respect to $\sigma_1$, and 
therefore we have represented this prescription by $]_{odd}$ . 

The normalization $\mu '$ can not be determined in the same way as 
eq.(\ref{wz}) because if we set the tachyon field to zero this action 
completely vanishes. Instead it can be determined by the calculation of 
the S-matrix between 
a RR-field and a tachyon field since eq.(\ref{wz10}) includes the 
coupling $\int C \wedge dT$ \cite{BiCrRo}, and the result is given by
\beqa
\mu '=i^{\frac{1}{2}}\tau_9.
\eeqa

In this way we have derived the explicit Wess-Zumino terms on 
$\mbox{D9}+\overline{\mbox{D9}}$ systems and non-BPS D$9$-branes in BSFT. 
The important point is that one can 
read off the Wess-Zumino terms if one extracts the fermionic zero modes from
the boundary action $I_{B}$. This may be said as a boundary topological
model which naturally leads to the notion of superconnection as we
will see in the next subsection. Note also that the above results can be
applied to general D$p$-branes if the transverse scalars are set to zero. 
If we would like to include transverse scalars, the new terms called Myers 
terms appear in the action. For details see our original paper \cite{TaTeUe}.

\subsection{Superconnection and K-theory charge}
\hspace*{4.5mm}
Here we will discuss the interpretation of the Wess-Zumino terms on non-BPS
D-branes and brane-antibrane systems as superconnections 
\cite{Qu}. For brane-antibrane systems this fact was first suggested in
\cite{Wi1}. A definite relation between the Wess-Zumino terms and the Chern
character of the superconnection was proposed in the paper
\cite{KeWi}. Our calculations in the previous subsection show that this
interpretation indeed holds within the framework of BSFT as we will see
below. Moreover we will argue that such an interpretation can be applied to
non-BPS D-branes and our previous calculations give evidence for
this. If one is interested in the index theorem and the anomaly related to the 
superconnections, see the papers \cite{KrLa,Sc5,Sz}.

Let us first review the definition and properties of superconnections 
following \cite{Qu}. 
There are two kinds of superconnections: one is for even-cohomology
 and the other is for odd-cohomology. In the K-theoretic language 
 the former is related to $K^{0}(M)$ and the latter to $K^{1}(M)$, where $M$ is  a manifold regarded as the world-volume of D-branes. Both are defined 
 as follows\footnote{We have included 
the explicit factor $\f{i}{2\pi}$ in front of 
 the field strength which was omitted for simplicity in the original paper \cite{Qu}. This is the reason why the factor $i^{\f12}$ does appear in the 
 expressions below.}:

\begin{flushleft}
\noindent{\bf Superconnection for $K^{0}(M)$}
\end{flushleft}
\hspace*{4.5mm}
In this case we consider a $\zz$-graded vector bundle $E=E^{(0)}\oplus
E^{(1)}$, which can be directly applied to a brane-antibrane system if
one identifies $E^{(0)}$ and $E^{(1)}$ as vector bundles on the branes
and antibranes, respectively. The endomorphism $X\in \mbox{End}\ E$ 
of this superbundle has the following $\zz$-grading:
\ba
\mbox{deg}(X)=\left\{\begin{array}{c}
\ 0 \ \ \mbox{if}\ X:E^{(0)}\to E^{(0)}\ \mbox{or}\ E^{(1)}\to E^{(1)},  \\
\ 1 \ \ \mbox{if}\ X:E^{(0)}\to E^{(1)}\ \mbox{or}\ E^{(1)}\to E^{(0)}.
\end{array} \right .
\ea

In addition, there is also a natural $\z$-grading if one considers
the algebra of differential forms $\Omega(M)=\oplus\Omega^{p}(M)$,
where $\Omega^{p}(M)$ denotes the algebra of $p$-forms on $M$. The
crucial observation is to mix these two gradings and to define the
$\zz$-grading for 
$\alpha\in \Omega^{p}(M,\mbox{End}\ E)=
\Omega^{p}(M)\otimes\Omega^{0}(M,\mbox{End}\ E)$ as follows: 
\ba
\alpha=\omega\otimes X\in\Omega^{p}(M)
\otimes\Omega^{0}(M,\mbox{End}\ E),\ \ 
\mbox{deg}(\alpha)\equiv p+\mbox{deg}(X), \label{super1}
\ea
where $\Omega^{0}(M,\mbox{End}\ E)$ denotes the space of sections of 
$\mbox{End}\ E$. Under this definition its superalgebra is given by the 
following rule:
\ba
(\omega\otimes X)(\eta\otimes Y)=(-1)^{\mbox{deg}(X)\cdot\mbox{deg}(\eta)}\ 
(\omega\eta\otimes XY), \label{super2}
\ea
and the supercommutator can be defined as:
\ba
[\ap,\beta]=\ap\beta-(-1)^{\mbox{deg}(\alpha)\cdot\mbox{deg}(\beta)}\beta\alpha.\label{super3}
\ea

An element of $\Omega^{p}(M,\mbox{End}\ E)$ can be written as a $2\times
2$ matrix, where the diagonal elements and off-diagonal elements have
even and odd degree of $\Omega^{0}(M,\mbox{End}\ E)$, respectively. We also define the supertrace as
\ba
\ap\in\Omega(M,\mbox{End}\ E)=
\left(
	\begin{array}{cc}
	\ap_1 & \ap_2 \\
	\ap_3 & \ap_4
	\end{array}
\right), \ \ \ \mbox{Str}(\ap)=\mbox{Tr}(\ap_1)-\mbox{Tr}(\ap_4)\in\Omega(M) ,
\ea
where $\mbox{Tr}$ denotes the ordinary trace on vector bundles. Note that the 
supertrace of supercommutators vanishes.

 Let us now define a superconnection on $E$ as an operator 
${\ca{D}}=d+{\ca{A}}$ on $\Omega(M,\mbox{End}\ E)$ with odd degree satisfying 
the derivation property:
\ba
{\ca{D}}(\omega\phi)=(d\omega)\phi+(-1)^{deg(\omega)}\omega 
({\ca{D}}\phi),\ \ \ \ \omega\in\Omega(M),\ \ \phi\in \Omega(M,\mbox{End}\ E).
\ea

For local calculations familiar to physicists one can regard ${\ca{A}}$ as a degree odd element of $\Omega(M,\mbox{End}\ E)$:
\ba
{\ca{D}}=d+{\ca{A}}=
\left(
	\begin{array}{cc}
	d+A^{(1)} & \s{2\pi i}T \\
	\s{2\pi i}\bar{T} & d+A^{(2)}
	\end{array}
\right), 
\ea
where the factor $\s{2\pi i}$ has been included for later convenience. The
diagonal parts $d+A^{(1)}$ and $d+A^{(2)}$ denote the ordinary gauge
connections of vector bundles $E^{(1)}$ and $E^{(2)}$, respectively. 
$T$ denotes an odd degree
 endomorphism of $E$. Notice that in this definition the exterior
derivative $d$ does {\it anti-commute} with any odd elements in
$\mbox{End}\ E$.

The curvature ${\ca{F}}$ of the superconnection ${\ca{D}}$ is defined as 
an even degree element of $\Omega(M,\mbox{End}\ E)$:
\ba
{\ca{F}}&=&{\ca{D}}^2=d{\ca{A}}+{\ca{A}}^2 \no
&=&
\left(
	\begin{array}{cc}
	F^{(1)}+2\pi iT\bar{T} & \s{2\pi i}DT \\
	\s{2\pi i}D\bar{T} & F^{(2)}+2\pi i\bar{T}T
	\end{array}
\right), \label{su-k0}
\ea
where we have defined\footnote{Note that since $T$ is an odd element, it anti-commutes with any one-forms. Therefore we can say that $T$ does couple to the relative gauge field $A^{(1)}-A^{(2)}$.} $DT=dT+A^{(1)}T+TA^{(2)}$.

The ``even Chern character'' of this superconnection is given by
\ba
\mbox{Str}\exp\left(\f{i}{2\pi}{\ca{D}}^2\right)
=\mbox{Str}\exp\left(\f{i}{2\pi}{\ca{F}}\right).\label{ch-0}
\ea
It is easy to see that this is closed because
\ba
d\ (\mbox{Str}{\ca{D}}^{2n})=\mbox{Str}[{\ca{D}},{\ca{D}}^{2n}]=0.
\ea
Furthermore as shown in the main theorem in the paper \cite{Qu}, its 
cohomology class does not depend on the choice of $T$. Therefore that Chern 
character becomes 
\ba
\mbox{Str}\exp\left(\f{i}{2\pi}{\ca{D}}^2\right)\simeq 
\mbox{ch}(E_1)-\mbox{ch}(E_2)\in H^{even}(M,{\mb{Q}})\cong K^0(M),\label{kk-0}
\ea
where $\mbox{ch}(E)$ denotes the ordinary Chern character and we have applied 
the Chern isomorphism, which states that the even cohomology and the 
K-group $K^0(M)$ are equivalent if ${\mb{Q}}$ is tensored \cite{Wi1,Ol}. 
Note that the difference of the Chern characters is seen in eq.(\ref{RRdiff}).

\begin{flushleft}
\noindent{\bf Superconnection for $K^{1}(M)$}
\end{flushleft}
\hspace*{4.5mm}
The first step to define the second superconnection is to regard a bundle $E$ as
 a module over the Clifford algebra $C_1={\mb{C}}\oplus {\mb{C}}\sigma_1$.
 In other words, we define the endomorphism of this superbundle as
 $\mbox{End}_{\sigma}\ E=\mbox{End}\ E\otimes C_1$. Let us call all
 elements which include $\sigma_1$ degree odd and the others degree
 even. In the physical context these correspond to the fields on non-BPS
 D-branes which belong to the GSO odd and even sector, respectively. The 
supertrace on $\mbox{End}_{\sigma} E$ is defined as follows:
\ba
\mbox{Tr}_{\sigma}(X+Y\sigma_1)=2 \mbox{Tr}(Y),\label{trs1}
\ea
where $X,Y\in \mbox{End}_{\sigma}\ E$ are degree even elements. 
Further we mix the degree of differential forms in the same way as in 
the previous case (\ref{super1}),(\ref{super2}) and (\ref{super3}).

The superconnection on $E$ is locally defined as an odd element as follows
\ba
{\ca{D}}=d+{\ca{A}}=d+A-\s{\f{i\pi}{2}}T\sigma_1,
\ea
where $A$ is an ordinary connection and $T$ is a self-adjoint 
endomorphism.
The curvature of this is also defined as
\ba
{\ca{F}}&=&{\ca{D}}^2=d{\ca{A}}+{\ca{A}}^2 \no
        &=&F-\s{\f{i\pi}{2}}DT\sigma_1-\f{i\pi}{2}T^2, \label{su-k1}
\ea
where we have defined $DT=dT-i[A,T]$. The ``odd Chern character'' is given by
\ba
\mbox{Tr}_{\sigma}\exp\left(\f{i}{2\pi}{\ca{D}}^2\right)
=\mbox{Tr}_{\sigma}\exp\left(\f{i}{2\pi}{\ca{F}}\right).\label{ch-1}
\ea
The main theorem in \cite{Qu} again tells us that this character is closed and its cohomology class does not depend on the choice of $T$. Further we can regard this as an element of K-theory group $K^1(M)$:
\ba
\mbox{Tr}_{\sigma}\exp\left(\f{i}{2\pi}{\ca{D}}^2\right)\ 
\in\ H^{odd}(M,{\mb{Q}})\cong K^1(M).
\ea

\begin{flushleft}
\noindent{\bf Physical interpretation}
\end{flushleft}

The role of superconnections in the D-brane physics is obvious if one
notes that the Wess-Zumino terms 
on $D9+\overline{\mbox{D9}}$ systems and non-BPS
D9-branes can be expressed as wedge products of RR-fields and the Chern
characters of superconnections:
\ba
S&=&\tau_9\ \mbox{Str}\ \int_{M}\ C\wedge 
\exp\left({\f{i}{2\pi}\ca{F}}\right)\ \ \ \ \ \ 
\mbox{(for a D9-$\overline{\mbox{D9}}$ )},
\label{surr-1} 
\\
S&=&i^{\frac{1}{2}}\tau_9\ \mbox{Tr}_{\sigma}\ \int_{M}\ C\wedge
\exp\left({\f{i}{2\pi}\ca{F}}\right)\ \ \ \mbox{(for a non-BPS
D9-brane)},\label{surr-2}
\ea
where the curvature ${\ca{F}}$ in the first equation represents the
superconnection for $K^{0}(M)$ and in the second for $K^{1}(M)$. 
One can transform these mathematical expressions 
(\ref{su-k0}) and (\ref{su-k1}) into the physical ones 
(\ref{wz}) and (\ref{wz-n}) by following the prescription\footnote{Also
note that for a brane-antibrane we need an additional minus sign in
front of $\overline{DT}$. This occurs due to the following reason. The
mathematical definition of the superconnection for $K^{0}(M)$ assumes
that an odd form anti-commutes with an odd degree endomorphism like 
eq.(\ref{super2}). 
On the other hand, in the physical expression (\ref{wz}) it does commute.}:
\ba
D=d+A &\to& 2\pi\s{\al}D=2\pi\s{\al}(d-iA).
\ea

Now, by using the Wess-Zumino terms we can verify the conjecture of 
tachyon condensation. 
If we consider a constant tachyon field the Wess-Zumino terms
completely vanish due to the overall factor $e^{-|T|^2}$ or 
$e^{-\frac{1}{4}T^2}$, and this is consistent with the conjecture. 
The next channel of tachyon condensation is the 
case when the tachyon fields are topologically nontrivial. 
As we have said in section \ref{WV}, the linear tachyon profiles in 
eq.(\ref{sol2.1}) and eq.(\ref{sol2.2}) are exact solutions to the equation 
of motion of the Dirac-Born-Infeld type action 
for a $D9+\overline{D9}$ system, and eq.(\ref{sol1}) 
is an exact solution for a non-BPS D9-brane. By inserting these solutions 
we can see that the correct RR-charges are generated. 
For example by putting the linear tachyon solution (\ref{sol2.2}) into 
the Wess-Zumino term (\ref{wz}) for a $D9+\overline{D9}$ system it becomes
\beqa
\label{chargegenerate}
&&\tau_9\int_{D7} C\wedge \left(\int  
dx_1dx_2~ 2\pi u_1u_2~e^{-\frac{1}{4}(u_1^2x_1^2+u_2^2x_2^2)}\right)\no
&=&\tau_9\int_{D7} C\wedge \left(2\pi u_1u_2~\times 
\frac{(2\sqrt{\pi})^2}{u_1u_2}\right)=\tau_7\int_{D7} C,
\eeqa
where we have used the relation $(2\sqrt{2}\pi)^2 \tau_9=\tau_7$. 
The coefficient 
in front of the last line just represents the charge of a BPS D$7$-brane. 
In the other cases of tachyon condensation we can see the correct 
RR-charges are generated. Especially if we consider the tachyon condensation 
with a higher codimensional soliton configuration like decay of 
$D9+\overline{D9}$ systems to a BPS D$3$ brane, we have only to use the 
Wess-Zumino term for multiple non-BPS systems and its solution given by 
the Atiyah-Bott-Shapiro construction \cite{ABS} like eq.(\ref{abs-1}).
 
By the way, in the above 
calculation the most interesting thing is that we have not taken the 
limit $u_1,~u_2\rightarrow\infty$ to obtain the RR-charge of a D7-brane. 
The finite value of $u_1$ or $u_2$ corresponds to the off-shell region. 
This observation 
means that D-brane charges are quantized even during tachyon 
condensation and are not altered by smooth deformation of the tachyon 
configuration. This is consistent with the fact that D-brane charges are 
determined by the even or odd Chern characters in eq.(\ref{ch-0}) and 
eq.(\ref{ch-1}) and that these values belong to the even or odd cohomology 
element. Note that the result of eq.(\ref{chargegenerate}) holds 
only when both $u_1$ and $u_2$ are nonzero. If either is zero, the 
configuration of the tachyon is trivial and one can not regard $T$ as an 
element in the desirable endomorphism. 

In this way we can relate D-brane charges in the Wess-Zumino terms to other 
D-brane charges due to tachyon condensation. 
Especially we have 
found that all kinds of BPS D$(2p+1)$-brane charges in Type IIB theory are 
generated by starting from only $D9+\overline{D9}$ systems and considering 
its tachyon condensation \cite{Wi1}. 
On the other hand all kinds of BPS D$(2p)$-brane 
charges in Type IIA theory are produced by only non-BPS $D9$-branes 
\cite{Ho1}. 
Another important point is that the Chern characters which appear in the 
Wess-Zumino terms belong to the K-group $K^0(M)$ 
for $D9+\overline{D9}$ branes and $K^1(M)$ for non-BPS D9-branes. 
Therefore, these results are consistent with the already known fact in 
the paper 
\cite{MiMo,HaMi} 
that $K^0(M)$ classifies all kinds of BPS D$(2p+1)$-brane charges 
in Type IIB theory and $K^1(M)$ does all kinds of BPS D$(2p)$-brane charges 
in Type IIA theory. 
In this sense we can even consider that $D9+\overline{D9}$ systems and non-BPS 
D$9$-branes include all kinds of (BPS and non-BPS) D-brane configurations and 
might be regarded as the fundamental objects in string theory. 

On the other hand there exists the other idea that all kinds of D-branes are 
made up of the lowest dimensional D-branes, like BFSS matrix theory. If
we combine this idea with the K-theoretical argument, we could regard 
$D(-1)+\overline{D(-1)}$ systems and non-BPS $D(-1)$-branes as the most 
fundamental objects in Type IIA and IIB theories, and we could even
consider their worldvolume theories as the nonperturbative definition of 
total string theory \cite{K1,K2,K3}. 

Anyway, this is the conclusion obtained from the K-theory argument.
 
\section{The Meaning of Calculations in Boundary String Field Theory 
\label{meaning}}
\hspace*{4.5mm}

In this section we will explain the meaning of the calculations of BSFT 
we have done. As we said in section \ref{calcu}, in general we can not 
perform calculations of BSFT if we include massive terms in 
the sigma model action. Thus, we have restricted the terms only to 
renormalizable ones. Moreover, when we construct BSFT in section 
\ref{BVformalism}, we set all terms with worldsheet ghosts to zero in the 
boundary action $I_B$. From these a natural question occurs : what exactly 
does the action $S$ which we have computed mean? 

Firstly we will answer the question by considering the result (\ref{Z1}) as an 
example. What we can assert is that the action (\ref{Z1}) is regarded as an 
on-shell effective action for $T$ and $A_{\mu}$ in which the massive modes are 
integrated out, or an off-shell action obtained by setting the fields other 
than $T$ and $A_{\mu}$ to zero. However, if the one-point function of massive
fields vanishes in all the off-shell region, we can regard this
action as the off-shell effective action in which the massive fields are
integrated out. 

Here we will explain the above statement. First we will introduce the 
notion of integrating out massive fields. Let us imagine that we 
included all massive fields in the sigma model action and could 
obtain the classical action $S[T,A_{\mu},\upsilon^{\alpha}]$, where 
$\upsilon^{\alpha}~(\alpha=1,2,\cdots)$ denote all massive fields. 
In the quantum string field theory the effective action 
$S_{\ss eff}[T,A_{\mu}]$ in which 
all massive fields are integrated out is defined by
\beqa
e^{-S_{\ss eff}[T,A_{\mu}]}=\int \left(\prod_{\alpha} D\upsilon^{\alpha}\right)
e^{-S[T,A_{\mu},\upsilon^{\alpha}]}.
\eeqa
Of course 
this path integral includes loop calculations, thus to obtain the complete 
effective action we need to know the quantum framework of BSFT. However, if 
we consider the classical theory this path integral reduces to the lowest 
order of WKB approximation, which is the same as inserting a solution to 
equation of motion into the classical action $S[T,A_{\mu},\upsilon^{\alpha}]$ 
and deleting $\upsilon^{\alpha}$. 
Since we are considering the classical BSFT, this is just 
the meaning of integrating out massive fields and obtaining the effective 
action. 

Now let us expand the classical action $S[T,A_{\mu},\upsilon^{\alpha}]$ by 
the massive fields $\upsilon^{\alpha}$: 
\ba
S[T,A_{\mu},\upsilon^{\alpha}]
=S^{(0)}[T,A_{\mu}]+\upsilon^{\alpha} S^{(1)}_{\alpha}[T,A_{\mu}]
+\frac{1}{2}\upsilon^{\alpha}\upsilon^{\beta}S^{(2)}_{\alpha\beta}[T,A_{\mu}]
+\cdots\cdots.
\ea
In general we can not obtain an exact solution to the equation of motion of 
$\upsilon^{\alpha}$. 
However, if the coefficient $S^{(1)}_{\alpha}[T,A_{\mu}]$ in the above 
equation vanishes, we can easily see that $\upsilon^{\alpha}=0$ 
is a solution and we can regard $S^{(0)}[T,A_{\mu}]$ as the effective 
action in which the massive fields are integrated out. On the other hand 
$S^{(0)}[T,A_{\mu}]$ 
is just the action which we are able to calculate from the renormalizable 
sigma model action. Therefore the 
crucial point is whether $S^{(1)}_{\alpha}[T,A_{\mu}]$ does vanish or not. 
Note that $S^{(1)}[T,A_{\mu}]=\frac{\pa S}{\pa \upsilon^i}|_{\upsilon^i=0}$ is 
equal to the one-point functions of the massive fields on the disk.

Indeed if we consider an open string background with the conformal invariance
we can easily see that these one-point functions of massive fields 
vanish\footnote{Note the following identity
\beqa
0=\la 0|[L_0,{\cal V}_i(0)]|0\lb=\Delta_i\la {\cal V}_i(0)\lb,
\eeqa
where $L_0$ is the Virasoro generator and $\Delta_i$ is the conformal 
dimension of ${\cal V}_i$.}. 
Therefore the starting 
prescription of setting all massive fields to zero in the sigma model is 
the same as classically integrating them out in the target space action. This 
is the meaning of ``on-shell effective action" in which massive fields are 
integrated out, although on-shell actions do not include any other 
information than S-matrices and are not useful. 

The important point is whether these one-point functions vanish even in the 
off-shell region or not, which is nontrivial. In the cubic string 
field theory (CSFT) \cite{WiCSFT,Wisu} or Berkovits' formalism \cite{Ber1} 
one point functions do not vanish and we have to include massive modes to 
obtain the correct answers\footnote{This might be related to the fact that 
in CSFT the gauge transformation $\delta \Psi=Q_B\Lambda
+g\left(\Psi*\Lambda-\Lambda*\Psi\right)$ mix the tachyon 
and the gauge field with the massive fields, while in BSFT it is closed 
like eq.(\ref{eq2.1}) and eq.(\ref{eq2.11}) 
even if we consider only the tachyon and the gauge field.}. 
However, in BSFT from the success and 
correctness of the tachyon condensation in section \ref{exacttach} we guess 
that above facts hold in the case of the sigma model action with only 
free fields. Therefore if the above facts hold not only for the free sigma 
model but also for the generally renormalizable one (eq.(\ref{eq3}) or 
eq.(\ref{eq1})), 
$S^{(1)}[T,A_{\mu}]$ vanishes in the off-shell region and we can regard 
eq.(\ref{Z1}) as the off-shell effective action in which all massive modes 
are integrated out. On the other hand the Wess-Zumino terms might be regarded 
as off-shell effective actions because the results are completely exact and 
topological. However, we can not assert that these predictions are actually 
true. Now we have finished explaining the statement in the second paragraph 
of this section.
 
Finally we make a comment about the validity of the Dirac-Born-Infeld 
action (\ref{BPSeff}) 
for a BPS D-brane and the action (\ref{nonBPSaction11}) for a non-BPS 
D-brane. To obtain these actions we assumed $\pa_{\mu}\pa_{\nu}T
=\pa_{\rho}F_{\mu\nu}=0$ and set the closed string background to the flat one 
$(\pa_{\rho}G_{\mu\nu}=\pa_{\rho}B_{\mu\nu}=0)$. These results are based on 
the framework of BSFT. A natural question is what 
occurs if we ignore these constraints and examine these actions as pure field 
theories. Indeed 
there have been a lot of reports that these actions are valid beyond this 
approximation (for example see the papers 
\cite{BeTo,AgPoSc,CoMyTa,My,HaHi}). 
This is one of mysterious facts in the relation between string theory and 
field theory.
  
\chapter{On-Shell Description of Tachyon Condensation}
\hspace*{4.5mm}
In the last chapter we have discussed tachyon condensation in the 
off-shell formalism of string theory - string field theory. In this chapter
we will deal with tachyon condensation by using the on-shell string theory,
which is just the conventional first-quantized string theory. We only need the 
knowledge of the conformal field theory, and especially we use the 
boundary state formalism, which was explained in chapter 2.
  
Historically the method in this chapter had been used 
\cite{NaTaUe,frau,MaSe2,sen14,sen16,sen19,sen20,bergman3,bergman1} 
before the string field theory approach was found. 
The physical situation in this chapter is special and a little complicated, 
thus we think that this chapter is easily understood after the explanation of 
the general off-shell formalism.  

\section{Set Up \label{setup}}
\hspace*{4.5mm}
In this chapter we pick up a case of tachyon condensation which we can 
deal with by the conformally invariant technique called marginal deformation. 
We consider a $D2+\overline{D2}$ system in the flat space background and put 
the $\overline{D2}$-brane on top of the D2-brane. In this situation, as we 
saw in the previous chapter, there appears a tachyon in the spectrum of an 
open string between the D2-brane and the $\overline{D2}$-brane and off-shell 
tachyon condensation starts. However, here we consider a little different 
situation : we compactify the world volumes of the D2-branes on a 
two-dimensional torus $T^2$ and switch on $Z^2$ Wilson line on the 
$\overline{D2}$-brane along each direction of the torus $T^2$. 

Let us express this situation by using the boundary state technique 
which was explained in chapter 2. First note that the world volumes 
of D2-branes are compactified on torus $T^2$ and the structure of the zero 
mode part of boundary states becomes a little complicated as we saw in 
eq.(\ref{Wilsonline}). If we put a D2-brane which is spread in $x^1$ and $x^2$ 
directions at $x^i=0~(i=3\sim9)$, the zero mode part of its boundary state can 
be written as
\beqa
\label{eq2} 
|D2,\gamma\lb^0_{\ss NSNS}&=&\delta^7(x)
\sum_{(w_{\ss X}^1,w_{\ss X}^2)\in{\bf Z}^2}|{\bf w_{\ss X}},{\bf 0}\lb,\no
|D2,\gamma\lb^0_{\ss RR}&=&\delta^7(x)
\sum_{(w_{\ss X}^1,w_{\ss X}^2)\in{\bf Z}^2}|{\bf w_{\ss X}},{\bf 0}\lb
\otimes|\Omega 2_{\ss X},\gamma\lb^0_{\ss RR}.
\eeqa
The second important element in this chapter is to 
add $Z^2$ Wilson line in $x^1$ and $x^2$ directions only on the anti D2-brane. 
The general Wilson line which appears in the boundary state has already been 
explained in eq.(\ref{Wilsonline}), while $Z^2$ Wilson line means that the 
boundary state of the anti D2-brane has the phase factor with $Z^2$ 
periodicity $(-1)^{w_{\ss X}^1+w_{\ss X}^2}$.  

Now we have to check whether there are any tachyonic modes in open strings 
between the D2-brane and the anti D2-brane. Let us see the open string 
spectrum by calculating the vacuum amplitude 
$\la\overline{D2}|\Delta|D2\lb$ and transforming it to the open partition 
function. The result is the same as that in eq.(\ref{DDbaramp2}) except the 
explicit form of open string Hamiltonian, which is written as
\beqa
L_0&=&2\left(p^0p_0+\sum_{i=3}^9 p^ip_i+M^2\right),\no
\mbox{where}&&\no
M^2&=&\sum_{\A=1,2}
\frac{1}{R^2_{\A}}\left(n^{\A}+\frac{1}{2}\right)^2
+(\mbox{nonzero modes})+\left\{\begin{array}{c}
-\frac{1}{4}~~(\mbox{for NS})\\
0~~(\mbox{for R})
\end{array}\right.,
\eeqa
and $R_1,R_2$ are radii of the torus $T^2$. From this expression we can 
see that the momenta in $x^1$ and $x^2$ directions are discretized in such a 
way that they are labeled by half integer numbers 
$n^{\A}+\frac{1}{2}~~(n^{\A}\in{\bf Z})$. This is because the boundary 
condition of the open string at one end obeys the unusual antiperiodic 
boundary condition, which is just caused by the effect of $Z^2$ Wilson line. 
From this mass spectrum $M^2$ of open strings with the opposite 
GSO-projection (see eq.(\ref{DDbaramp2})) we can see that the ``tachyon" 
which we called before corresponds to a particle with its mass square 
\beqa
M^2&=&\left(\frac{n^1+\frac{1}{2}}{R_1}\right)^2
+\left(\frac{n^2+\frac{1}{2}}{R_2}\right)^2-\frac{1}{4}.
\eeqa  
From this equation the tachyon has the following Fourier modes
\beqa
\label{TFourier}
T(x^1,x^2)=\sum_{n^1,n^2\in{\bf Z}^2}T_{n^1+\frac{1}{2},n^2+\frac{1}{2}}
\exp\left\{{i\left(\frac{n^1+\frac{1}{2}}{R_1}\right)x^1
+i\left(\frac{n^2+\frac{1}{2}}{R_2}\right)x^2}\right\}.
\eeqa
Especially $T_{\pm\frac{1}{2},\pm\frac{1}{2}}$ have the lowest mass square
\beqa
\label{lowestmass}
M^2=\frac{1}{4}\left(\frac{1}{R_1^2}+\frac{1}{R_2^2}-1\right).
\eeqa

From this equation we can predict an interesting fact about stability 
of the brane-antibrane system, which can not be seen unless we switch on 
$Z^2$ Wilson line. It is the fact that whether the modes 
$T_{\pm\frac{1}{2},\pm\frac{1}{2}}$ have the positive or negative mass square 
depends on the size of torus $T^2$ which the D2-branes wind on. If the torus 
$T^2$ is large enough to satisfy $\frac{1}{R_1^2}+\frac{1}{R_2^2}<1$,  
$T_{\pm\frac{1}{2},\pm\frac{1}{2}}$ are tachyonic in the usual sense. 
This system becomes unstable and decays to something. However, if 
$\frac{1}{R_1^2}+\frac{1}{R_2^2}\geq 1$, $T_{\pm\frac{1}{2},\pm\frac{1}{2}}$ 
become massive and tachyonic modes completely disappear. This means 
that the brane-antibrane system becomes perturbatively {\sl stable}.

This situation is quite different from that of the last chapter. Especially 
we can find that in some region of the parameter space $(R_1,R_2)$ the 
conformal invariance is not broken 
even if $|T|$ gets any expectation values. This is the 
case when $\frac{1}{R^2_1}+\frac{1}{R_2^2}$ is equal to $1$ and the modes 
$T_{\pm\frac{1}{2},\pm\frac{1}{2}}$ become massless. In that case the tachyon 
potential becomes completely flat, which means that the energy does not 
decrease or not increase if the tachyon has an expectation value. Therefore, 
we can consider that 
$T_{\pm\frac{1}{2},\pm\frac{1}{2}}$ plays the same role as one 
component of the gauge field $A^{\mu}(x)$ and we can regard the expectation 
value of $T_{\pm\frac{1}{2},\pm\frac{1}{2}}$ as the Wilson line like 
$a^{\alpha}$ in eq.(\ref{Wilsonline}). 
The method of shifting the background in the 
conformally invariant manner by switching on parameters like Wilson lines is 
called marginal deformation. Especially the process of a massless 
``tachyon" getting an expectation value is called marginal tachyon 
condensation.

However, there is a technically subtle point to perform the marginal 
deformation. This is because the vertex operator corresponding to massless 
elements $T_{\pm\frac{1}{2},\pm\frac{1}{2}}$ has a complicated form if we 
express it by using worldsheet fields 
$(X^{\mu}(z,\bar{z}),~\psi^{\mu}(z))$. Therefore, at first glance we can not 
understand how to incorporate the tachyon field as a Wilson line correctly. 

\section{The Vertex Operators of Marginal Tachyons}
\hspace*{4.5mm}
To see the technical subtlety mentioned above and 
to solve its problem let us choose the case $R_1=R_2=\sqrt{2}$ to 
satisfy $\frac{1}{R_1^2}+\frac{1}{R_2^2}=1$ and consider the condensation of 
the following linear combination\footnote{There are also other two marginal 
deformations which represent the tachyon condensation. However, these 
only shift the center of the vortex and this fact is not physically important. 
Thus, we only consider the tachyon fields (\ref{eqn:tachyon1}).} 
of massless elements in eq.(\ref{TFourier})
\ba
\label{-1tach}
T_{1}(x^1,x^2)&\propto&e^{i\frac{1}{2\sqrt{2}}(x^1+x^2)}
-e^{-i\frac{1}{2\sqrt{2}}(x^1+x^2)}, \no
T_{2}(x^1,x^2)&\propto&-ie^{i\frac{1}{2\sqrt{2}}(x^1-x^2)}
+ie^{-i\frac{1}{2\sqrt{2}}(x^1-x^2)}.\label{eqn:tachyon1}
\ea
If we replace $x^1,x^2$ with $X^1(z,\bar{z}),X^2(z,\bar{z})$ these 
become the vertex operators with $(-1)$-picture in terms of worldsheet 
superconformal ghost number \cite{FrMaSh}. 

According to this the corresponding open string vertex operators with 
0-picture are written as
\ba
\label{vertexop}
V_{T1}&\propto&(\psi^1+\psi^2)(e^{i\frac{1}{2\sqrt{2}}(X^1+X^2)}
+e^{-i\frac{1}{2\sqrt{2}}(X^1+X^2)})\otimes 
\sigma_1 \label{eqn:tv1}, \no
V_{T2}&\propto&(\psi^1-\psi^2)(e^{i\frac{1}{2\sqrt{2}}(X^1-X^2)}
+e^{-i\frac{1}{2\sqrt{2}}(X^1-X^2)})\otimes 
\sigma_2 \label{eqn:tv2},
\ea
where we have assigned Chan-Paton factors $\sigma_1$ and $\sigma_2$ to $V_{T1}$
and $V_{T2}$.  

The marginal deformation is to consider the coherent shift from the original 
background in the conformally invariant manner. To explain this let us 
consider the general boundary state (\ref{geneb2}) with the operator  
$e^{\A V_{T1}+\B V_{T2}}$ acting on it. Here $\alpha$ and 
$\beta$ are the parameters which represent the vacuum expectation values of 
the tachyon fields $T_1$ and $T_2$ in eq.(\ref{-1tach}). However, this is a 
complicated form to deal with because as we can see in eq.(\ref{vertexop}) 
both $V_{T1}$ and $V_{T2}$ include exponentialized forms of 
$X^{1,2}(z,\bar{z})$ and double exponentials appear in 
$e^{\A V_{T1}+\B V_{T2}}$. Thus, to make it an easy form let us change the 
worldsheet fields 
$(X^{\mu}(z,\bar{z}),~\psi^{\mu}_L(z),~\psi^{\mu}_R(\bar{z}))$ into more 
convenient ones \cite{sen14,MaSe2}. First let us rotate the coordinates by 
$\f{\pi}{4}$
\beq
\begin{array}{rclrcl}
Y^1&=&\f{1}{\s{2}}(X^1+X^2),&\ \ \ Y^2&=&\f{1}{\s{2}}(X^1-X^2), \no
&&&&&\no
\chi^1_{\ss L(R)}&=&\f{1}{\s{2}}(\psi^1_{\ss L(R)}+\psi^2_{\ss L(R)})
,&\ \ \ \chi^2_{\ss L(R)}&=&\f{1}{\s{2}}(\psi^1_{\ss L(R)}-\psi^2_{\ss L(R)}). 
\label{eqn:45do}
\end{array}
\eeq
In this coordinate the radii of the torus $T^2$ become 1 in 
both $Y^1$ and $Y^2$ directions. This special radius enables us to use the 
method of bosonization and fermionization as follows
\begin{eqnarray}
\label{eqn:bos}
\tau_i\otimes C^i_{\ss Y}(\pm 1,0)~e^{\pm iY^i_{\ss L}}
=\frac{1}{\sqrt{2}}(\xi^i_{\ss L}\pm i\eta^i_{\ss L}),& 
\tau_i\otimes C^i_{\ss Y}(0,\pm 1)~e^{\pm iY^i_{\ss R}}
=\frac{1}{\sqrt{2}}(\xi^i_{\ss R}\pm i\eta^i_{\ss R}),\no
\ti{\tau}_i\otimes C^i_{\phi}(\pm 1,0)~e^{\pm i\phi^i_{\ss L}}
=\frac{1}{\sqrt{2}}(\xi^i_{\ss L}\pm i\chi^i_{\ss L}),&
\ti{\tau}_i\otimes C^i_{\phi}(0,\pm 1)~e^{\pm i\phi^i_{\ss R}}
=\frac{1}{\sqrt{2}}(\xi^i_{\ss R}\pm i\chi^i_{\ss R}),
\end{eqnarray}
where there appear two types of cocycle factors. These cocycle factors must 
be put in to guarantee their correct statistics and correct (anti)commutation 
relations to various fields. One type is $\tau_i$ and 
$\ti{\tau}_i~(i=1,2)$ \cite{MaSe2,sen16} and we also assume that 
$\chi^i_{L,R}$ and $\eta^i_{L,R}$ have the cocycle factors $\tau_{3}$ and 
$\ti{\tau}_3$. The other type of cocycle factor is 
$C^i_{\ss Y}(k^i_{\ss YL},k^i_{\ss YR})$ and 
$C^i_{\phi}(k^i_{\phi{\ss L}},k^i_{\phi{\ss R}})$, 
which are defined by \cite{Po1}
\ba
C^i_{\ss Y}(k^i_{\ss YL},k^i_{\ss YR})\equiv\exp\left[\frac{1}{2}\pi i
(k^i_{\ss YL}-k^i_{\ss YR})(\hat{p}^i_{\ss YL}+\hat{p}^i_{\ss YR})\right].
\ea
Especially in the later calculations and proofs this second type of cocycle 
factor becomes crucial. 
 
\begin{figure} [tbhp]
\begin{center}
\epsfxsize=100mm
\epsfbox{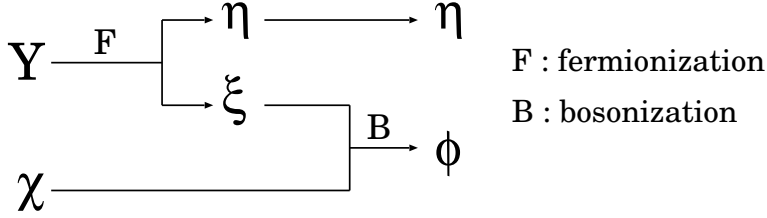}
\caption{Two steps of bosonization \label{variable}}
\end{center}
\end{figure}

Now we have changed the coordinate system from $(Y^i,\chi^i_{L(R)})$ to 
$(\phi^i,\eta^i_{L(R)})$ by the bosonization technique 
(see figure \ref{variable}). The operator product expansions (OPE) among 
these new fields can be obtained from those of old ones:\footnote{The factor 
$i$ which appears in the OPEs and identities (\ref{eqn:ope}) comes from the 
second-type cocycle factors.}
\ba
Y^i_{L}(z)Y^j_{L}(0)\sim -\delta_{ij}{\mbox{ln}}z&,&
Y^i_{R}(\bar{z})Y^j_{R}(0)\sim -\delta_{ij}{\mbox{ln}}\bar{z}, \no
\xi_L^i(z)\xi_L^j(0)\sim \delta_{ij}\frac{i}{z}&,
&\xi_R^i(\bar{z})\xi_R^j(0)\sim -\delta_{ij}\frac{i}{\bar{z}}, \no 
\eta_L^i(z)\eta_L^j(0)\sim \delta_{ij}\frac{i}{z}&,
&\eta_R^i(\bar{z})\eta_R^j(0)\sim -\delta_{ij}\frac{i}{\bar{z}}.\ea
Also the following identities are useful:
\ba 
\eta_L^i\xi_L^i=i\de Y^i_{L}&,
&\eta_R^i\xi_R^i=-i\bar{\de} Y^i_{R}, \no
\chi_L^i\xi_L^i=i\de \phi^i_{L}&,
&\chi_R^i\xi_R^i=-i\bar{\de}\phi^i_{R}. \label{eqn:ope}
\ea

By using these relations we can express the tachyon vertex operators 
(\ref{eqn:tv1}) in the following 
way\footnote{To derive these we have to use the 
T-dualized equations of bosonization in eq.(\ref{eqn:bos}).}
\ba
\label{Tvertex}
V_{T1}&\propto&2i\chi^1\xi^1\otimes\tau_2\otimes\sigma_1
=-2\de\phi^1\otimes\tau_2\otimes\sigma_1,  \no            
 V_{T2}&\propto&-2i\chi^2\xi^2\otimes\tau_1\otimes\sigma_2
=2\de\phi^2\otimes\tau_1\otimes\sigma_2,
\ea
where $\de$ denotes a tangential derivative along the boundary. The form of 
eq.(\ref{eqn:tv2}) in the old coordinate $(X^{1,2}(z,\bar{z}),~\psi^{1,2}(z))$ 
becomes very simple in the new coordinate 
$(\phi^{1,2}(z,\bar{z}),~\eta^{1,2}(z))$.

The tachyon condensation is thus represented as insertion of the 
following operator like the Wilson line in terms of the field $\phi^i$
\ba 
M(\alpha,\beta)
=\exp\left(\f{i\ap}{4}\oint \de\phi^1\otimes\tau_2\otimes\sigma_1
+\f{i\beta}{4}\oint\de\phi^2\otimes\tau_1\otimes\sigma_2\right), 
\label{eqn:wilson}
\ea
where $\oint$ denotes integration along the boundary and $\ap,\beta$ mean by 
parameters of tachyon condensation, which represent expectation
values of coefficients of tachyons $T_1$ and $T_2$ in 
eq.(\ref{eqn:tachyon1}). Notice that $\tau_2\otimes\sigma_1$ 
commutes with $\tau_1\otimes\sigma_2$ 
and the above operator is well-defined without path ordering. 

\section{The Boundary State Analysis of the Tachyon Condensation} 
\hspace*{4.5mm}
Now let us go back \cite{NaTaUe,frau} to the argument of boundary states 
and construct the 
total boundary state of a $D2+\overline{D2}$ system. This is given by the 
superposition of boundary states for a $D2$-brane and a $\overline{D2}$-brane: 
\begin{eqnarray}
\label{eq6.1}
\left\{\begin{array}{rcl}
|D2+\overline{D2},\gamma\lb_{\ss NSNS}&=&
|D2,\gamma\lb_{\ss NSNS}+|D2^{\prime},\gamma\lb_{\ss NSNS}, \\
|D2+\overline{D2},\gamma\lb_{\ss RR}&=&|D2,\gamma\lb_{\ss RR}
-|D2^{\prime},\gamma\lb_{\ss RR},
\end{array}\right.
\end{eqnarray}
where the prime ($\prime$) indicates that there is $Z^2$ Wilson line on the 
D-brane. Note that the sign in front of the RR-sector of $\overline{D2}$-brane 
is minus, which was seen in the definition of anti D-branes 
(see eq.(\ref{DDbar})). 

From this we can see that the explicit form of this boundary state is almost 
the same as eq.(\ref{nonzerosol}), although the important difference appears 
only in the zero mode part of it as   
\beqa
\label{bstzero0}
\begin{array}{lcl}
{\displaystyle |D2+\overline{D2},\gamma\lb^0_{\ss NSNS}}
&=&{\displaystyle \delta^7({\bf x})
\sum_{(w^1_{\ss X},w^2_{\ss X})\in Z^2}
\left(|{\bf w_{\ss X}},{\bf 0}\lb+(-1)^{w^1_{\ss X}+w^2_{\ss X}} 
|{\bf w_{\ss X}},{\bf 0}\lb\right),} \no
|D2+\overline{D2},\gamma\lb^0_{\ss RR}&&
\end{array}\\
=\delta^7({\bf x})\sum_{(w^1_{\ss X},w^2_{\ss X})\in Z^2}
\left(|{\bf w_{\ss X}},{\bf 0}\lb-(-1)^{w^1_{\ss X}+w^2_{\ss X}}
|{\bf w_{\ss X}},{\bf 0}\lb\right)
\otimes|\Omega 2_{\ss X},\gamma\lb^0_{\ss RR}.
\eeqa

Here let us rewrite this boundary state by using the new coordinate 
$(\phi^{1,2},~\eta^{1,2}_L,~\eta^{1,2}_R)$ in 
eq.(\ref{eqn:bos}). As a first step the boundary state can easily be rewritten 
by using the coordinate $(Y^{1,2},~\chi^{1,2}_L,~\chi^{1,2}_R)$ in 
eq.(\ref{eqn:45do}) as 
\begin{eqnarray}
\label{eq12}
|D2+\overline{D2},\gamma\lb_{\ss NSNS}&=&\frac{T_2}{2}
\exp\left[\sum^{\infty}_{m=1}
\sum^2_{i=1}\left\{-\frac{1}{m}(\alpha_{\ss Y})^i_{-m}
(\tilde{\alpha}_{\ss Y})^i_{-m}+i\gamma\chi^i_{-m+\frac{1}{2}}
\tilde{\chi}^i_{-m+\frac{1}{2}}\right\}\right]\nonumber\\
&&\times 2\sum_{(w^1_{\ss Y},w^2_{\ss Y})\in {\bf Z}^2}
|2{\bf w_{\ss Y}},{\bf 0}\lb,   \\
\label{eq13}
|D2+\overline{D2},\gamma\lb_{\ss RR}&=&\frac{T_2}{2}
\exp\left[\sum^{\infty}_{m=1}
\sum^2_{i=1}\left\{-\frac{1}{m}(\alpha_{\ss Y})^i_{-m}
(\tilde{\alpha}_{\ss Y})^i_{-m}+i\gamma\chi^i_{-m}
\tilde{\chi}^i_{-m}\right\}\right]\nonumber\\
&&\times 2\sum_{(w^1_{\ss Y},w^2_{\ss Y})\in {\bf Z}^2}
|2{\bf w_{\ss Y}}+{\bf 1},{\bf 0}\lb\otimes
|\Omega 2_{\ss Y},\gamma\lb^0_{\ss RR},
\end{eqnarray}
where we have defined ${\bf 1}=(1,1)$ and we have omitted various 
contributions\footnote{These include $\delta^7({\bf x})$ in 
eq.(\ref{eq2}) and eq.(\ref{bstzero0}).} which come from $(X^0,X^3\sim X^9)$ 
and $(\psi^0,\psi^3\sim\psi^9)$. This is because these contributions are not 
related to the later calculations of tachyon condensation at all. 
Therefore we will continue to abbreviate these parts from now on.

Next we would like to rewrite the above expression by using the coordinate 
$(\phi^{1,2},~\eta^{1,2}_L,~\eta^{1,2}_R)$, while that 
coordinate is defined in the complicated way by bosonization in 
eq.(\ref{eqn:bos}). So, the question is how boundary states (\ref{eq12}) and 
(\ref{eq13}) are expressed. 
Here we suppose that the following boundary states are equivalent 
to eq.(\ref{eq12}) and eq.(\ref{eq13}) respectively for $\gamma=+1$ 
\cite{NaTaUe,frau}:
\begin{eqnarray}
\label{eq20}
|D2+\overline{D2},+\lb_{\ss NSNS}&=&
T_2~{\rm exp}\left[\sum^{\infty}_{m=1}\sum^2_{i=1}
\left\{-\frac{1}{m}\phi^i_{-m}
\tilde{\phi}^i_{-m}+i\eta^i_{-m+\frac{1}{2}}\tilde{\eta}^i_{-m+\frac{1}{2}}
\right\}\right]\nonumber\\
&&~~~~~~~~~~~~~~~~~~~~~~~~~~~
\times\sum_{(w^1_{\phi}w^2_{\phi})\in{\bf Z}^2}|2{\bf w_{\phi}},{\bf 0}\lb,\\
\label{eq21}
|D2+\overline{D2},+\lb_{\ss RR}&=&T_2~{\rm exp}\left[\sum^{\infty}_{m=1}
\sum^2_{i=1}\left\{-\frac{1}{m}\phi^i_{-m}\tilde{\phi}^i_{-m}+i\eta^i_{-m}
\tilde{\eta}^i_{-m}\right\}\right]\nonumber\\
&&~~~~~~~~~~~~
\times\sum_{(w^1_{\phi}w^2_{\phi})\in{\bf Z}^2}
|2{\bf w_{\phi}}+{\bf 1},{\bf 0}\lb\otimes|\Omega2_{\phi},+\lb^0_{\ss RR}.
\end{eqnarray}
These are simply obtained by replacing $(\alpha_{\ss Y})^i_m,~\chi^i_m,~
w^i_{\ss Y}$ in eq.(\ref{eq12}),(\ref{eq13}) with 
$\phi^i_m,~\eta^i_m,~w^i_{\phi}$. Indeed they satisfy the following desirable 
boundary conditions\footnote{Of course equations (\ref{eq22}) are not 
enough for the proof of equivalence. This is because these constraints do not 
determine the detailed structures of zero modes like Wilson lines. On the 
other hand we can see that for several closed string states eq.(\ref{eq20}) 
and eq.(\ref{eq21}) have the same overlap as eq.(\ref{eq12}) and 
eq.(\ref{eq13}) do, which is almost the same calculation as that in 
appendix B of the paper\cite{frau}. 
The other evidence is the equality of their 
partition functions, which we will see in the next section. We propose 
that these three arguments are enough for the proof of the equivalence.} 
\cite{NaTaUe}:
\begin{eqnarray}
\label{eq22}
\partial_2 Y^i(w)|_{\sigma_2=0}|D2+\overline{D2},+\lb_{\ss NSNS(RR)}=0, \no
(\chi^i_{\ss L}(w)-i\chi^i_{\ss R}(\bar{w}))|_{\sigma_2=0}
|D2+\overline{D2},+\lb_{\ss NSNS(RR)}=0,
\end{eqnarray}
with $i=1,2$. 

The proof of eq.(\ref{eq22}) is as follows. 
First from the explicit form of boundary state we can 
easily see that eq.(\ref{eq20}) and eq.(\ref{eq21}) satisfy the following 
equations:
\begin{eqnarray}
\label{1000}
(\eta^i_{\ss L}(w)-i\eta^i_{\ss R}(\bar{w}))|_{\sigma_2=0}
|D2+\overline{D2},+\lb_{\ss NSNS(RR)}=0,\\
\label{1001}
(\phi^i_{m}+\ti{\phi}^i_{-m})|D2+\overline{D2},+\lb_{\ss NSNS(RR)}=0,
\end{eqnarray}
If we replace $\pa_2 Y^i(w)|_{\sigma_2=0}$ with $(\phi^i,\eta^i)$ 
variables and use equation (\ref{1000}), then the first equation of 
eq.(\ref{eq22}) can be rewritten as
\begin{eqnarray}
\label{eq24.1}
& &\partial_2 Y^i(w)|_{\sigma_2=0}|D2+\overline{D2},+\lb_{\ss NSNS(RR)}
\nonumber\\
& &=\ti{\tau}_i\otimes\frac{\sqrt{z}}{\sqrt{2}i}\eta^i_{\ss L}(z)
\Biggl[\sqrt{z}\left\{C^i_{\phi}(1,0):e^{i\phi^i_{\ss L}}(z):
+C^i_{\phi}(-1,0):e^{-i\phi^i_{\ss L}}(z):\right\}\nonumber\\
& &+\sqrt{\bar{z}}\left\{C^i_{\phi}(0,1):e^{i\phi^i_{\ss R}}
(\bar{z}):+C^i_{\phi}(0,-1):e^{-i\phi^i_{\ss R}}(\bar{z}):\right\}
\Biggl]~|D2+\overline{D2},+\lb_{\ss NSNS(RR)}, \no
\end{eqnarray}
where $:\sim:$ denotes the conformal normal ordering. If we use equation 
(\ref{1001}) it is easy to see that the first and the third, the second and 
the fourth term in eq.(\ref{eq24.1}) cancel respectively. The second equation 
of eq.(\ref{eq22}) can be proven in the same way. 
 
On the other hand, the explicit form of $|D2+\overline{D2},-\lb_{\ss NSNS(RR)}$
is obtained by using eq.(\ref{GSONSNS}) and eq.(\ref{GSORR}).
\begin{eqnarray}
|D2+\overline{D2},-\lb_{\ss NSNS(RR)}=(-1)^{F_{\ss Y}}
|D2+\overline{D2},+\lb_{\ss NSNS(RR)},
\end{eqnarray}
where we have removed the ghost contribution, which is the minus sign 
in eq.(\ref{GSONSNS}).
Note that the action of $(-1)^{F_{\ss Y}}$ on $(\phi^i,\eta^i)$ can be seen 
from eq.(\ref{eqn:bos}) as
\begin{eqnarray}
(-1)^{F_{\ss Y}}~:~&(\alpha_{\ss Y})^i_m
\rightarrow (\alpha_{\ss Y})^i_m~~,~~\chi^i_m\rightarrow 
-\chi^i_m,\nonumber\\
&\phi^i_m\rightarrow -\phi^i_m~~,~~\eta^i_m
\rightarrow \eta^i_m,
\end{eqnarray}
and to zero modes
\begin{eqnarray}
(-1)^{F_{\ss Y}}~:~|{\bf w_{\phi}},{\bf n_{\phi}}\lb
\rightarrow |2{\bf n_{\phi}},\frac{\bf w_{\phi}}{2}\lb.
\end{eqnarray}
Therefore, 
$|D2+\overline{D2},-\lb_{\ss NSNS}$ is given by\footnote{The phase factor 
$(-1)^{w^1_{\phi}+w^2_{\phi}}$ comes from the cocycle factor.}
\begin{eqnarray}
\label{eq24}
|D2+\overline{D2},-\lb_{\ss NSNS}&=&
T_2~{\rm exp}\left[\sum^{\infty}_{m=1}\sum^2_{i=1}\left\{\frac{1}{m}\phi^i_{-m}
\tilde{\phi}^i_{-m}+i\eta^i_{-m+\frac{1}{2}}\tilde{\eta}^i_{-m+\frac{1}{2}}
\right\}\right]\nonumber\\
&&~~~~~~~~~~~~~
\times\sum_{(w^1_{\phi},w^2_{\phi})\in {\bf Z}^2}(-1)^{w^1_{\phi}+w^2_{\phi}}
|{\bf 0},{\bf w_{\phi}}\lb,
\end{eqnarray}
and we can obtain $|D2+\overline{D2},-\lb_{\ss RR}$ in the same way. Here we 
have finished expressing the boundary state of a $D2+\overline{D2}$ system 
in the new coordinate $(\phi^{1,2},\eta^{1,2}_L,\eta^{1,2}_R)$.

Now we are ready to condense the tachyon by using the Wilson line-like 
operator (\ref{eqn:wilson}). To perform this let us remember the definition 
of the general boundary state in eq.(\ref{geneb2}). 
In the last section we have 
obtained the operator (\ref{eqn:wilson}) which acts on the boundary 
state for a $D2+\overline{D2}$ system. However, the operator 
(\ref{eqn:wilson}) includes Chan-Paton factors, while the operator in 
eq.(\ref{geneb2}) belongs to the closed string Hilbert space. Since closed 
strings do not usually have Chan-Paton factors, the tachyon condensation in 
the closed string viewpoint corresponds to the insertion of the trace of 
$M(\alpha,\beta)$ in front of the boundary state. It is given by
\begin{eqnarray}
\label{eq26}
W_1(\alpha,\beta)\propto{\rm Tr}~M(\alpha,\beta)
=\cos\left(\frac{\pi\alpha \hat{w}_{\phi}^1}{2}\right) 
\cos\left(\frac{\pi\beta \hat{w}_{\phi}^2}{2}\right),
\end{eqnarray}
where $\hat{w}^1_{\phi}$ and $\hat{w}^2_{\phi}$ are the operators which count 
winding numbers in the Hilbert space of closed strings and act on the boundary 
state for a $D2+\overline{D2}$ system. 

However, this is not sufficient \cite{NaTaUe}. As we saw in eq.(\ref{wz}) the 
Wess-Zumino term appears in the effective action of a $D2+\overline{D2}$ 
system, and it includes the following coupling
\begin{eqnarray}
\label{Wess-Zumino}
\int_{D2+\overline{D2}} C_1\wedge dT \wedge d\bar{T},
\end{eqnarray}
where $T$ and $\bar{T}$ are complex tachyon fields and $C_1$ is the R-R
1-form, which couples to D0-branes. At first glance, this coupling 
looks strange because the complex tachyon fields 
$T$ and $\bar{T}$ have Chan-Paton 
factors $\sigma_{+},\sigma_{-}$ and its trace around the boundary of the 
disk is zero. This paradox can be solved if we remember the calculation of 
the Wess-Zumino term in the last chapter. As we saw in eq.(\ref{rr2}) an 
extra Chan-Paton factor $\sigma_3$ appears as $(-1)^F$ from the branch cut 
generated by a RR-vertex operator, and as a result the above coupling turns 
out to be nonzero. 

{}From the above consideration we claim that the tachyon condensation 
switches not only eq.(\ref{eq26}) but also 
\begin{eqnarray}
\label{eq27}
W_{\sigma_3}(\alpha,\beta)\propto{\rm Tr}[\sigma_3M(\alpha,\beta)]
=\sin\left(\frac{\pi\alpha 
\hat{w}_{\phi}^1}{2}\right) \sin\left(\frac{\pi\beta \hat{w}_{\phi}^2}{2}
\right).
\end{eqnarray}
This kind of operator is needed in order to obtain the physically 
reasonable result after the tachyon condensation. Moreover, 
this operator should be included to reproduce the correct spectrum of open 
strings and to satisfy Cardy condition \cite{cardy} as we will see in the next 
section. 

After all the general formula in eq.(\ref{geneb2}) gives the boundary state 
$|B(\alpha,\beta)\lb_{\ss NSNS(RR)}$ after switching on the expectation values 
of tachyons ($\alpha$, $\beta$) as follows:\footnote{As we saw in the last 
chapter, the extra Chan-Paton factor in eq.(\ref{eq27}) comes from a 
{\sl RR}-vertex operator inserted inside the disk. Therefore, in the context 
of the boundary string field theory we have not yet understood why we should 
act an extra operator like eq.(\ref{eq27}) even on {\sl NSNS} sector of the 
boundary state too. 
However, Cardy condition which we will verify in the next section becomes 
inconsistent if we do not consider this.}\footnote{Strictly speaking, from the 
above explanation we can not decide the relative normalization between 
$W_1(\alpha,\beta)$ and $W_{\sigma_3}(\alpha,\beta)$. This is determined by 
the vacuum energy calculation in the next section.}
\beqa
|B(\alpha,\beta),\gamma\lb_{\ss NSNS(RR)}
=\left[W_1(\alpha,\beta)+W_{\sigma_3}(\alpha,\beta)\right]
|D2+\overline{D2},\gamma\lb_{\ss NSNS(RR)}~~~~~(\gamma=\pm 1).\no
\eeqa
Because of $W_1(0,0)=1$ and $W_{\sigma_3}(0,0)=0$, we can see that the initial 
condition is satisfied as $|B(0,0),\gamma\lb_{\ss NSNS(RR)}
=|D2+\overline{D2},\gamma\lb_{\ss NSNS(RR)}$. 
After inserting the explicit forms of $W_1(\alpha,\beta)$ and 
$W_{\sigma_3}(\alpha,\beta)$, we can obtain the following result:
\\\\\\
\begin{eqnarray}
\label{eq28}
|B(\alpha,\beta),+\lb_{\ss NSNS}&=&T_2~{\rm exp}\left[\sum^{\infty}_{m=1}
\sum^2_{i=1}\left\{-\frac{1}{m} \phi^i_{-m}\tilde{\phi}^i_{-m}
+i\eta^i_{-m+\frac{1}{2}}\tilde{\eta}^i_{-m+\frac{1}{2}}\right\}\right]
\nonumber\\
&&\times\sum_{(w^1_{\phi},w^2_{\phi})\in{\bf Z}^2}
\Bigl[\cos(\pi\alpha w^1_{\phi})\cos(\pi\beta w^2_{\phi})\no
&&~~~~~~~~~~~~+\sin(\pi\alpha w^1_{\phi})
\sin(\pi\beta w^2_{\phi})\Bigr]|2{\bf w}_{\phi},{\bf 0}\lb, \\ \label{eq29}
|B(\alpha,\beta),+\lb_{\ss RR}
&=&T_2~{\rm exp}\left[\sum^{\infty}_{m=1}\sum^2_{i=1}
\left\{-\frac{1}{m} \phi^i_{-m}\tilde{\phi}^i_{-m}+i\eta^i_{-m}
\tilde{\eta}^i_{-m}\right\}\right]\nonumber\\
&&\times\sum_{(w^1_{\phi},w^2_{\phi})\in {\bf Z}^2}
\Biggl[\cos\left\{\pi\alpha\left(w^1_{\phi}+\frac{1}{2}\right)\right\}
\cos\left\{\pi\beta\left(w^2_{\phi}+\frac{1}{2}\right)\right\}\nonumber\\[-4mm]
&& +\sin\left\{\pi\alpha\left(w^1_{\phi}+\frac{1}{2}\right)\right\}
\sin\left\{\pi\beta\left(w^2_{\phi}+\frac{1}{2}\right)\right\}\Biggr]
|2{\bf w}_{\phi}+{\bf 1},{\bf 0}\lb\otimes|\Omega 2_{\phi},+\lb^{0}_{\ss RR}. 
\no
\end{eqnarray}

Here, as we can see from the above equations, the boundary state 
$|B(\alpha,\beta),+\lb_{\ss NSNS(RR)}$ is invariant under the shift 
$(\alpha,\beta)\rightarrow(\alpha+2,\beta)~\mbox{or}~(\alpha,\beta+2)$ 
up to an irrelevant overall phase factor. Moreover, it is also invariant under 
the reflection $(\alpha,\beta)\rightarrow(2-\alpha,\beta)~\mbox{or}~
(\alpha,2-\beta)$. From these facts the physically distinguishable region in 
the parameter space of $(\alpha,\beta)$ is restricted to $0\leq\alpha\leq 1$ 
and $0\leq\beta\leq 1$. Let us examine what happens at edges of the 
parameter space $(\alpha,\beta)$, which are $(\alpha,\beta)=(1,0),~(0,1)$ and 
$(1,1)$. 

\begin{figure} [tb]
\begin{center}
\epsfxsize=111mm
\epsfbox{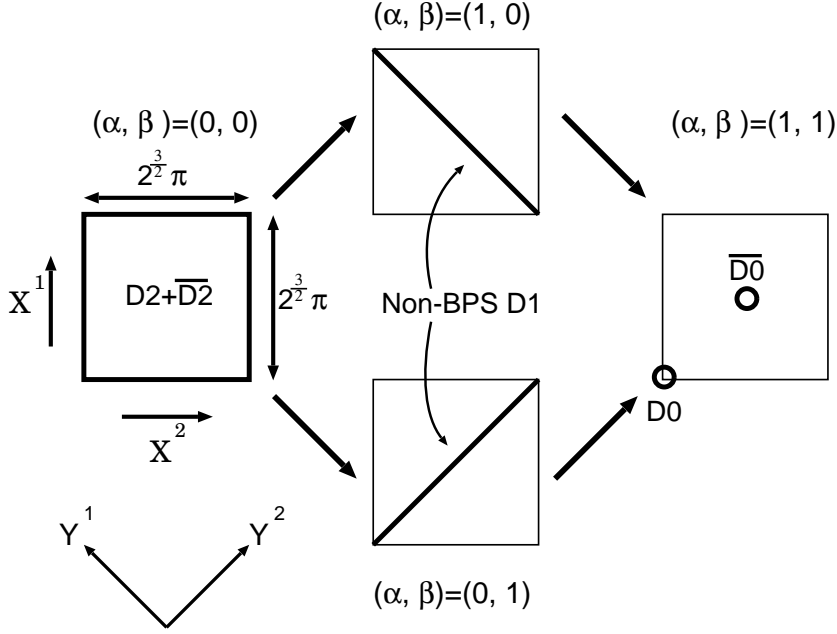}
\caption{The tachyon condensation in $D2+\overline{D2}$ system \label{de}}
\end{center}
\end{figure}

First we consider the tachyon condensation with $(\alpha,\beta)=(1,0)$ 
\cite{frau,sen14}. 
This tachyon condensation corresponds to switching on the expectation 
value of only $T_{1}(x^1,x^2)$ in eq.(\ref{eqn:tachyon1}). 
From eq.(\ref{eqn:tachyon1}) we can see that the tachyon field 
$T_1(x^1,x^2)\propto\sin{\frac{1}{2\sqrt{2}}(x^1+x^2)}$ has a kink type 
profile. Here the kink means by one kind of topologically nontrivial 
configurations of a scalar field whose sign changes at 
the node $x^1+x^2=0$. Moreover, it is almost the same as figure 
\ref{kink} except the existence of periodicity of $x^1$ and $x^2$. 
Therefore, we can predict 
that some kind of 1-brane remains after this tachyon condensation because the 
tachyon makes the energy (mass) 
density of a $D2+\overline{D2}$ system 
vanish almost everywhere except the line 
$x^1+x^2=0$. Moreover, because the RR-sector of the boundary state in 
eq.(\ref{eq29}) completely vanishes, we can 
expect that a non-BPS D1-brane is generated (see figure \ref{de}). 

Indeed, we can prove that the boundary state $|B(1,0),+\lb_{\ss NSNS(RR)}$ 
satisfies the desirable boundary condition \cite{NaTaUe}:
\begin{eqnarray}
\label{eq30}
\partial_2 Y^1(w)|_{\sigma_2=0}|B(1,0),+~\rangle_{\ss NSNS(RR)}&=&0,\no
\partial_1 Y^2(w)|_{\sigma_2=0}|B(1,0),+~\rangle_{\ss NSNS(RR)}&=&0,\no
\left(\chi^1_{\ss L}(w)-i\chi^1_{\ss R}(\bar{w})\right)|_{\sigma_2=0}
|B(1,0),+\lb_{\ss NSNS(RR)}&=&0,\no
\left(\chi^2_{\ss L}(w)+i\chi^2_{\ss R}(\bar{w})\right)|_{\sigma_2=0}
|B(1,0),+\lb_{\ss NSNS(RR)}&=&0.
\end{eqnarray}
The proof of these equations is the same as that of eq.(\ref{eq22}). 
We can also verify that the boundary state $|B(1,0),-\lb_{\ss NSNS(RR)}$ 
with $\gamma=-1$ satisfies the same equations as the above ones.

Next we can go on to the case $(\alpha,\beta)=(0,1)$. The situation 
is almost the same as the previous one. In this case only the vacuum 
expectation value of $T_2(x^1,x^2)$ in eq.(\ref{eqn:tachyon1}) is switched on, 
and we can see that a non-BPS D1-brane appears which stretches on the line 
$x^1-x^2=0$. Therefore, from the results of $(\alpha,\beta)=(1,0)$ 
and $(0,1)$ we can conclude that a tachyon kink on a 
brane-antibrane system produces a codimension $1$ non-BPS D-brane.

The last channel of the tachyon condensation 
is the case with $(\alpha,\beta)=(1,1)$ \cite{NaTaUe,MaSe2}. 
This tachyon condensation
corresponds to switching on the expectation values of both $T_1(x^1,x^2)$ 
and $T_2(x^1,x^2)$ in eq.(\ref{eqn:tachyon1}). Note that this tachyon 
configuration looks like a combination of a vortex and an anti-vortex. This is 
because if we consider the total configuration 
$\left(T_{(1)}+T_{(2)}\right)(x^1,x^2)$ it has two nodes at $x^1=x^2=0$ and 
$x^1=x^2=\sqrt{2}\pi$, around which it becomes
\beqa
\label{vortex}
\left(T_{1}+T_{2}\right)(x^1,x^2)&\sim&\frac{1}{\sqrt{2}}(1+i)(x^1+ix^2)~~
(\mbox{around}~x^1=x^2=0),\no
&\sim&\frac{1}{\sqrt{2}}(1-i)\left((x^1-\sqrt{2}\pi)-i(x^2-\sqrt{2}\pi)
\right)\no
&&~~~~~~~~~~~~~~~~~~~~~~~~~~~~(\mbox{around}~x^1=x^2=\sqrt{2}\pi).
\eeqa
From this we can expect that a D0-brane and a $\overline{D0}$-brane are 
generated after the tachyon condensation because the energy density of 
$D2+\overline{D2}$ vanishes except the nodes 
($x^1=x^2=0$ and $x^1=x^2=\sqrt{2}\pi$) of the tachyon profile. In other word 
the vortex and the antivortex can be regarded as a D0-brane and a 
$\overline{D0}$-brane, respectively. 

Indeed, in the same way as the proof of eq.(\ref{eq22}) 
we can verify that the desirable boundary condition for D0-branes 
holds as follows \cite{NaTaUe}
\begin{eqnarray}
\label{eq31}
\partial_1 Y^i(w)|_{\sigma_2=0}|B(1,1),+\lb_{\ss NSNS(RR)}&=&0,\no
\left(\chi^i_{\ss L}(w)+i\chi^i_{\ss R}(\bar{w})\right)|_{\sigma_2=0}
|B(1,1),+\lb_{\ss NSNS(RR)}&=&0,~~~(i=1,2).
\end{eqnarray}
In other words the tachyon condensation from $\ap=\beta=0$ to $\ap=\beta=1$ 
changes the boundary condition (\ref{eq22}) into (\ref{eq31}). 
At $\ap=\beta=0$ only the first term 
($\cos\times\cos$ part) of RR-sector in eq.(\ref{eq29}) is nonzero and this 
corresponds to 
the RR charge of the D2-brane. As the tachyon is condensed, the second term 
($\sin\times\sin$ part) of RR-sector in eq.(\ref{eq29}) also ceases to be zero 
and this 
means that the RR charge of the D0-brane is generated. Finally at 
$\ap=\beta=1$ only the second term ($\sin\times\sin$ part) of RR-sector is 
nonzero and 
this is the pure RR charge of the D0-brane. Note that if we ignored the 
contribution of eq.(\ref{eq27}), which comes from $\sigma_3$ sector in 
eq.(\ref{Wess-Zumino}), then the RR-sector of the boundary state would 
vanish\footnote{Here note that the two D0-branes are separated from each other 
with some distance, thus like eq.(\ref{bstzero0}) the total RR-sector of their 
boundary state can not be zero.} at $\alpha=\beta=1$ and would be 
inconsistent\footnote{The second term ($\sin\times\sin$ part) of NSNS sector 
in eq.(\ref{eq28}) vanishes at every edge points of the parameter space of 
$(\alpha,\beta)$. As a result that term in NSNS-sector does not cause any 
physical effects, and it is required only for keeping the 
modular invariance as we will see in the next section.}. 
In this way we can explicitly see in the closed string formalism that a 
tachyon vortex on a brane-antibrane system produces a codimension two 
D-brane (see Figure \ref{de}). 
Moreover, if we consider that the tachyon condensation 
consists of two steps : $(\alpha,\beta)=(0,0)\rightarrow(1,0)$ and 
$(1,0)\rightarrow(1,1)$ in figure \ref{de}, we can also conclude that a 
tachyon kink 
on a non-BPS D-brane produces a codimension one BPS D-brane. 
These two facts can be regarded 
as a continuous version of the descent relation \cite{sen16} which was 
previewed in chapter 2. 

Now we have shown the evidence of tachyon condensation by showing that the 
correct boundary conditions (\ref{eq30}) and (\ref{eq31}) hold. 
However, this is 
not sufficient as the complete proof of the tachyon condensation, thus in the 
next section we will calculate the vacuum amplitude and will verify that 
this amplitude reproduces the correct open string spectrum after the tachyon 
condensation.  

\section{Calculation of the Vacuum Amplitude}
\hspace*{4.5mm}
In this section we will calculate the vacuum amplitude of a 
$D2+\overline{D2}$ system for every value of parameters $\ap$ and $\beta$ and 
read off the open string spectrum \cite{NaTaUe,frau}. As a 
result it will be shown that the additional operator in eq.(\ref{eq27}) is 
indeed required in order to satisfy Cardy's condition \cite{cardy}.

First note that in the new coordinate system $(\phi^i,\eta^i_L,\eta^i_R)$ the 
Virasoro generator $L_0$ of closed strings is given by
\beqa
\label{L00}
L_0=\frac{1}{2}\phi^{i}_0\phi_{i0}+\sum^{\infty}_{m=1}\phi^{i}_{-m}\phi_{im}
+\sum^{\infty}_{r\in {\bf Z}+\nu} r\eta^{i}_{-r}\eta_{ir}+a+\cdots\cdots,
\eeqa
where $a$ is the zero point energy which is equal to $-1/2~(0)$ for 
NS (R)-sector, and $\cdots$ part denotes the contribution from directions 
other than $x^1$ and $x^2$. $\ti{L}_0$ can be obtained by adding the 
tilde over all oscillators. We can easily obtain this expression if we notice 
that the explicit form of energy momentum tensor is preserved under 
bosonization (\ref{eqn:bos}). By using these expressions in the closed 
string propagator (\ref{closedprop}) we can calculate the vacuum amplitude 
$Z(\alpha,\beta)=\la B(\alpha,\beta)|\Delta|B(\alpha,\beta)\lb$, which is 
given by 
\ba
Z_{NSNS}(\alpha,\beta)
&\equiv&\la B(\alpha,\beta)|\Delta|B(\alpha,\beta)\lb_{\ss NSNS}\no
&=&\f{(T_2)^2V_3}{2}\int^{\infty}_0 ds (4\pi s)^{-\f72}
\Biggl[\sum_{(w_{\phi}^ 1,w_{\phi}^2)\in{\bf Z}^2}\Bigl\{\cos^2(\pi\ap w_{\phi}^1)\cos^2(\pi\beta w_{\phi}^2)\no 
& & \ +\sin^2(\pi\ap w_{\phi}^1)\sin^2(\pi\beta w_{\phi}^2)\Bigr\}q^{(w_{\phi}^1)^2+
(w_{\phi}^2)^2}\f{f_3(q)^8}{f_1(q)^8}-2\f{f_4(q)^6f_3(q)^2}{f_1(q)^6f_2(q)^2}
\Biggl], \no
&&\no
&&\no
Z_{RR}(\alpha,\beta)
&\equiv&\la B(\alpha,\beta)|\Delta|B(\alpha,\beta)\lb_{\ss RR}\no
&=&-\f{(T_2)^2V_3}{2}\int^{\infty}_0 ds (4\pi s)^{-\f72}
\Biggl[\sum_{(w_{\phi}^1,w_{\phi}^2)\in{\bf Z}^2}\Biggl\{
\cos^2\left(\pi\ap \left(w_{\phi}^1+\f12\right)\right)
\cos^2\left(\pi\beta \left(w_{\phi}^2+\f12\right)\right)\no 
& &\!\! +\sin^2\left(\pi\ap \left(w_{\phi}^1+\f12\right)\right)
\sin^2\left(\pi\beta \left(w_{\phi}^2+\f12\right)\right)
\Biggr\}q^{(w_{\phi}^1+\f12)^2+(w_{\phi}^2+\f12)^2}\f{f_2(q)^8}{f_1(q)^8}
\Biggl],
\ea
where $q=e^{-s}$, and $V_3$ is the volume of a D2-brane. 
$f_i(q)~(i=1,2,3,4)$ are the modular functions defined in eq.(\ref{func}).

Let us transform this expression into the form of one-loop amplitude 
of open strings like eq.(\ref{oneloop}). By defining $t$ by $t=\frac{\pi}{s}$
and using the modular properties (\ref{modular}) of the modular functions 
$f_i(q)$, we can obtain the following open string amplitude:
\\\\\\
\ba
Z^{open}_{NS}(\alpha,\beta)
&=&2^{-2}\pi^{-1}V\int^{\infty}_0 dt\ t^{-\f32}\sum_{(n_1,n_2)
\in{\bf Z}^2}
\Biggl[\f{1}{2}(\ti{q}^{n_1^2+n_2^2}+\ti{q}^{(n_1-\ap)^2+(n_2-\beta)^2)})\f{f_3(\ti{q})^8}{f_1(\ti{q})^8} \no
& & ~~~~~~~~~~~~~~~~~~~~-\f{1}{2}(-1)^{n_1+n_2}(\ti{q}^{n_1^2+n_2^2}+\ti{q}^{(n_1-\ap)^2+(n_2-\beta)^2})\f{f_4(\ti{q})^8}{f_1(\ti{q})^8}\Biggl], \no
Z^{open}_{R}(\alpha,\beta)
&=&-2^{-2}\pi^{-1}V\int^{\infty}_0 dt\ t^{-\f32}\sum_{(n_1,n_2)
\in{\bf Z}^2}
\ti{q}^{n_1^2+n_2^2}\f{f_2(\ti{q})^8}{f_1(\ti{q})^8} \label{eqn:openamp},
\ea
where $\tilde{q}=e^{-\pi t}$ and 
$V\left(\equiv\frac{V_3}{(2\sqrt{2}\pi)^2}\right)$ is the ``volume" of the 
time direction. Here we have used the following identities of $f_i(q)$:
\beqa
\label{eqn:id}
\sum_{n}q^{n^2}=f_1(q)f_3(q)^2 &,& \sum_{n}(-1)^nq^{n^2}=f_1(q)f_4(q)^2, \no
\sum_{n}q^{(n-\f12)^2}=f_1(q)f_2(q)^2 &,& f_2(q)f_3(q)f_4(q)=\s{2}\ . 
\eeqa

Now it is obvious that for each value of $\ap$ and $\beta$ the open string 
spectrum 
is well-defined only if we consider the additional operator (\ref{eq27}) 
defined in the previous section. If we did not include the operator 
(\ref{eq27}) in the boundary state $|B(\alpha,\beta)\lb_{\ss NSNS(RR)}$, the 
number of open 
string states for given $n_1,n_2$ would be fractional. This physically 
important constraint is also called Cardy's condition \cite{cardy}. 

Finally let us see what kind of open string amplitudes appears at 
particular values of $\ap$ and $\beta$. In the 
case of $(\ap,\beta)=(1,0)$ or $(0,1)$ we obtain the following amplitude 
\cite{NaTaUe,frau}
\ba
Z(1,0)=Z(0,1)&=&2^{-2}\pi^{-1}V\int^{\infty}_0 dt\ 
t^{-\f32}\sum_{(n_1,n_2)\in{\bf Z}^2}
\ti{q}^{n_1^2+n_2^2}\f{f_3(\ti{q})^8-f_2(\ti{q})^8}{f_1(\ti{q})^8},\no
&=&4\pi V\int^{\infty}_0 \f{dt}{2t}{\rm Tr}_{\ss NS-R}\ti{q}^{2L_0}.
\ea
Therefore we can identify the system as a non-BPS D1-brane whose length is 
$4\pi$ as we expected in figure \ref{de} (compare the above amplitude 
with eq.(\ref{NDppa})). The other case left is $(\ap,\beta)=(1,1)$ and its 
amplitude can be written as \cite{NaTaUe} 
\ba
Z(1,1)&=&2^{-2}\pi^{-1}V\int^{\infty}_0 dt\
t^{-\f32}\sum_{(n_1,n_2)\in{\bf Z}^2}
\Biggl[\ti{q}^{n_1^2+n_2^2}\f{f_3(\ti{q})^8-f_2(\ti{q})^8}{f_1(\ti{q})^8}
-(-1)^{n_1+n_2}\ti{q}^{n_1^2+n_2^2}\f{f_4(\ti{q})^8}{f_1(\ti{q})^8}\Biggl], \no
&=&2V\int^{\infty}_0 \f{dt}{t}\f{1}{\s{16\pi^2 t}}\sum_{(m_1,m_2)\in{\bf Z}^2}
\Biggl[\ti{q}^{2m_1^2+2m_2^2}
\f{f_3(\ti{q})^8-f_4(\ti{q})^8-f_2(\ti{q})^8}{2f_1(\ti{q})^8}\no 
&&~~~~~~~~~~~~~~~~~~~~~~~~~~~~~~~~~~~~~
+\ti{q}^{2(m_1+\f12)^2+2(m_2+\f12)^2}\f{f_3(\ti{q})^8+
f_4(\ti{q})^8-f_2(\ti{q})^8}{2f_1(\ti{q})^8}\Biggl],\no
&=&2V\int^{\infty}_0 \f{dt}{2t}
{\rm Tr}_{NS-R}
\left[\f{1+(-1)^F}{2}\tilde{q}^{2L_0}\right]
+2V\int^{\infty}_0 \f{dt}{2t}
{\rm Tr}_{NS-R}
\left[\f{1-(-1)^F}{2}\tilde{q}^{2L_0}\right],\no
\ea
where we have defined $m_1=\f{n_1+n_2}{2},m_2=\f{n_1-n_2}{2}$. Note that 
this partition function counts not only the degrees of freedom of open strings 
stretching between a brane and an antibrane but also those of open strings 
both of whose end points are on the same D-branes. This explicitly shows that 
the system is equivalent to a 
D0-brane and an anti D0-brane. Moreover, from the half-integer power of 
$\tilde{q}$ in the second term of the second line we can see that these two 
D-branes are separated from each other by $\Delta x_1=\Delta x_2=\sqrt{2}\pi$. 
This is consistent with figure \ref{de}.

\section{Tachyon Condensation at General Radii of Torus}
\hspace*{4.5mm}

In the previous sections we have proved the tachyon condensation by the 
conformally invariant method called marginal deformation. In this proof the 
crucial ingredient was to fix the radius of the torus $T^2$ at some 
convenient value in order to keep the conformal invariance. 
However, its physical 
situation is too special and it is desirable to generalize the argument of 
tachyon condensation into arbitrary radii of the torus 
\cite{sen14,MaSe2,NaTaUe}.  

Before its analysis let us consider what happens in the internal region of the 
parameter space $(\alpha,\beta)$. In the previous section we have 
considered only edge points $((\alpha,\beta)=(0,0),~(1,0),~(0,1),~(1,1))$ 
of this parameter space. This is because the 
boundary state $|B(\alpha,\beta)\lb$ except those edge points does not 
correspond to any physical objects we know. This fact becomes obvious if we 
consider general radii of the torus $T^2$. At general radii the 
process of tachyon condensation is an off-shell phenomenon and generates 
tachyon potential as we explained in section \ref{setup}. In tachyon 
potential physical objects we already know appear at its extremum points 
because these points satisfy the equation of motion of open strings 
($\alpha^{\prime}$ corrected Yang-Mills-Higgs equation) and preserve the 
conformal invariance. 

Here what we would like 
to show is that if we consider general radii of the torus 
$T^2$ the conformal invariance is only kept at edge points of the parameter 
space $(\alpha,\beta)$. 
Its proof is as follows. First the tachyon potential can be 
regarded as the total energy (mass) of the D-brane system we are considering, 
and the form of potential at the special radii\footnote{Note that these radii 
are measured by the original coordinate $X^1,X^2$.} $(R_1=R_2=\sqrt{2})$ can 
be calculated by the internal product 
$V_{R_i}(\alpha,\beta)=\la 0|B(\alpha,\beta)\lb_{\ss NSNS}$, 
where $\alpha$ and $\beta$ denote the expectation values of the 
tachyons $T_1$ and 
$T_2$ in eq.(\ref{eqn:tachyon1}). Indeed as we can see in eq.(\ref{555}) and 
appendix \ref{Expression} this quantity is proportional to the disk 
partition function, and it is consistent with boundary string field theory. 
Moreover by using equation (\ref{eq28}) we can 
see $V_{R_i}(\alpha,\beta)=T_2=$ constant. This fact just indicates that 
the tachyon potential is flat and we can use the marginal deformation as we 
explained in section \ref{setup}. Next, we consider the deformed potential 
$V_{R_i+\Delta R_i}(\alpha,\beta)$ by $R_i\rightarrow R_i+\Delta R_i~~(i=1,2)$.
As we have explained in the last paragraph the potential ceases to be flat at 
general radii and this deformed potential gets nontrivial dependence of 
$\alpha$ and $\beta$. Here instead of using string field theory we will 
use a simple trick. 
We consider the infinitesimal deformation of this potential, which is given by
\beqa
V_{R_i+\Delta R_i}(\alpha,\beta)\cong V_{R_i}(\alpha,\beta)
+\sum_i\Delta R_i \frac{d}{dR_i}V_{R_i}(\alpha,\beta)\no
=T_2+\sum_i \Delta R_i\times
\la 0|\psi_L^i(0)\psi_R^i(0)|B(\alpha,\beta)\lb_{\ss NSNS}.
\eeqa  
Here $\psi_L^i(0)\psi_R^i(0)$ is a vertex operator with $(-1,-1)$ 
picture, which represents the radius deformation\footnote{This vertex operator 
corresponds to $\pa X^i(0)\bar{\pa} X^i(0)$ with $(0,0)$ picture, which can be 
obtained by
\beqa
\frac{d}{dR^2_i}e^{-S(R)}~~\mbox{with}~~S(R)=\int dz^2\sum^2_{i=1}R_i^2\pa 
X^i\bar{\pa}X^i.
\eeqa
On the other hand it is known that the boundary state belongs to the closed 
string Hilbert space with $(-1,-1)$ picture \cite{Vecchia}. 
Therefore in order to cancel the 
ghost anomaly on the disk we have to use the vertex operator with $(-1,-1)$ 
picture.}. Therefore, to find extremum points of the radius deformed 
tachyon potential we have to solve the following equations\footnote{In the 
open string terminology these equations can be rewritten as
\beqa
\frac{d}{dR_i}\la V_{T1}\lb_{\alpha,\beta,R_1,R_2}
=\frac{d}{dR_i}\la V_{T2}\lb_{\alpha,\beta,R_1,R_2}=0~~~(i=1,2),
\eeqa
where $V_{T1}$ and $V_{T2}$ are the vertex operators in eq.(\ref{Tvertex}) and 
this one point function is evaluated by including the boundary operator 
(\ref{eqn:wilson}) in the sigma model action. These equations show that 
the expectation values of the tachyons T1 and T2 are stable under 
the radius deformation.}
\beqa
\frac{\pa}{\pa\alpha}
\la 0|\psi^i_L(0)\psi^i_R(0)
|B(\alpha,\beta)\lb_{\ss NSNS}=\frac{\pa}{\pa\beta}
\la 0|\psi^i_L(0)\psi^i_R(0)
|B(\alpha,\beta)\lb_{\ss NSNS}=0.
\eeqa
Calculating this by bosonizing $\psi^i_L$ and $\psi^i_R$ we can obtain 
equations $\sin(\pi\alpha)=\sin(\pi\beta)=0$ and the solution is 
$(\alpha,\beta)=(0,0),~(0,1),~(1,0)$ and $(1,1)$, which we have just 
investigated. 

As the next step let us compare the total mass between $D2+\overline{D2}$, 
non-BPS $D1$ and $D0+\overline{D0}$ at general radii 
$(R_1,~R_2)$ of the torus. These values become as 
follows
\beq
\begin{array}{cl}
2T_2(2\pi R_1)(2\pi R_2)~~&~~(\mbox{for}~~ D2+\overline{D2}),\\
\sqrt{2}T_1 (2\pi\sqrt{R_1^2+R_2^2})~~&~~(\mbox{for  non-BPS}~D1),\\
2T_0~~&~~(\mbox{for}~~D0+\overline{D0}),
\end{array}
\eeq
where we have assumed that a non-BPS D1-brane winds along the diagonal line of 
the torus like figure \ref{de}. Note that the tension of a non-BPS D1-brane is 
$\sqrt{2}$ times as large as that of a BPS D1-brane (see equation (\ref{NDp})).
If we write the masses of these D-brane systems as 
$M_{D2+\overline{D2}},~M_{ND1}$ and $M_{D0+\overline{D0}}$, we can obtain the 
following relations
\beq
\label{masscompare}
\begin{array}{cl}
M_{D0+\overline{D0}}\leq M_{ND1}\leq M_{D2+\overline{D2}}~~
&~~(\mbox{when}~\frac{1}{R_1^2}+\frac{1}{R_2^2}\leq 1),\\
&\\
M_{D0+\overline{D0}}\leq M_{D2+\overline{D2}}\leq M_{ND1}~~
&~~(\mbox{when}~\frac{1}{R_1^2}+\frac{1}{R_2^2}\geq 1~~\mbox{and}~~
R_1R_2\geq 2),
\\ 
&\\
M_{D2+\overline{D2}}\leq M_{D0+\overline{D0}}\leq M_{ND1}~~
&~~(\mbox{when}~R_1R_2\leq 2~~\mbox{and}~~R_1^2+R_2^2\geq 4),\\
&\\
M_{D2+\overline{D2}}\leq M_{ND1}\leq M_{D0+\overline{D0}}~~
&~~(\mbox{when}~R_1^2+R_2^2\leq 4).
\end{array}
\eeq
Furthermore, the condition that there are no tachyons on 
a $D2+\overline{D2}$ system can be seen in eq.(\ref{lowestmass}) as
\beqa
\label{D2notac}
\frac{1}{R_1^2}+\frac{1}{R_2^2}\geq 1.
\eeqa 
In the same way we can examine the condition of the absence of tachyons on 
a $D0+\overline{D0}$ system and it is given by
\beqa
\label{D0notac}
R_1^2+R_2^2\geq 4,
\eeqa
which means that a D$0$-brane is separated enough from an anti D$0$-brane 
(see figure \ref{de}). 
Here one interesting thing is that the equations which determine whether 
a tachyon exists are the same as some equations in eq.(\ref{masscompare}). 
Note that there always exists a tachyonic mode on the 
non-BPS D1-brane\footnote{If we would like 
to delete tachyonic modes on a non-BPS 
D1-brane, we have to consider other curved backgrounds like orbifolds 
\cite{NaTaUe,BeGa1,bergman3,Gab1,sen20} 
or orientifolds \cite{sen14,bergman1}}.

\begin{figure} [htbp]
\begin{center}
\epsfxsize=150mm
\epsfbox{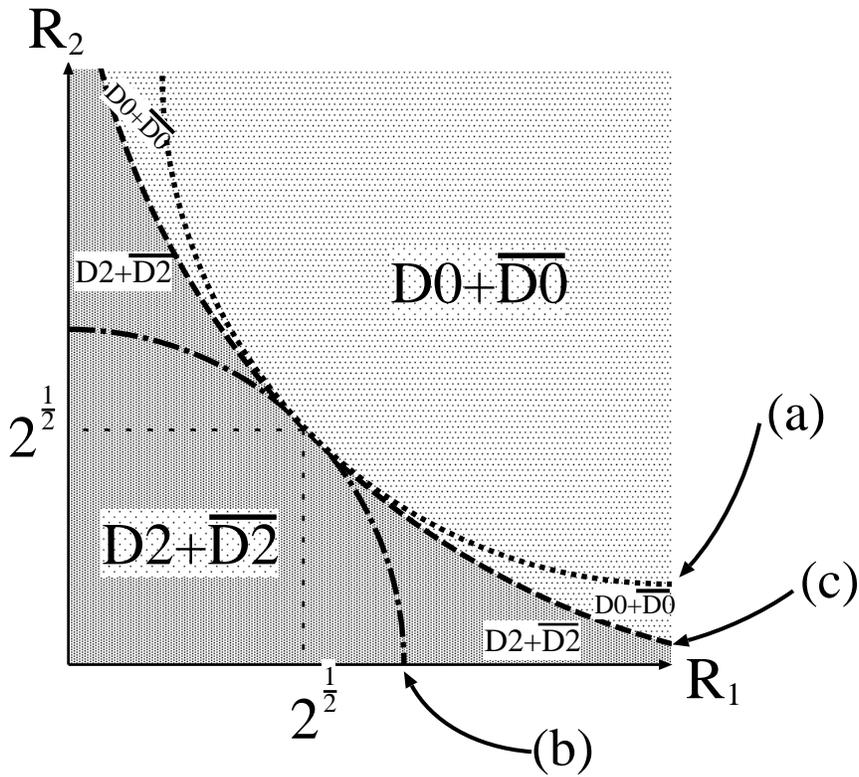}
\caption{The phase diagram about the stability of D-brane systems.\label{phase}
If the torus is large a $D0+\overline{D0}$ system is stabler, and if the torus 
is small a $D2+\overline{D2}$ system is. 
The equations of three curves in the phase diagram are given by
$(a)~\frac{1}{R_1^2}+\frac{1}{R_2^2}=1,~~~(b)~R_1^2+R_2^2=4,~~~
(c)~R_1R_2=2$.}
\end{center}
\end{figure}

From all of the above results ((\ref{masscompare}) $\sim$ (\ref{D0notac})) we 
can understand what the stablest object is among them 
(see figure \ref{phase}). If $R_1$ and 
$R_2$ are very large, the mass of D2-branes is heavier than that of D0-branes 
and a $D2+\overline{D2}$ system decays to a $D0+\overline{D0}$ system. On the 
contrary, if $R_1$ and $R_2$ are very small, the opposite process happens. 
Indeed in 
the region over line (a) in figure \ref{phase} 
a tachyonic mode appears on a $D2+\overline{D2}$ 
system and 
this is consistent with the fact that D2-branes are heavier than D0-branes. 
This kind of fact also holds in the region under line (b). 

The point with $R_1=R_2=\sqrt{2}$ is the only one where all kinds 
of D-brane systems have exactly the same mass and some tachyonic mode on them 
becomes massless. Therefore, only at this point we can use the method of the 
marginal deformation, while in the other regions the conformal invariance is 
broken and the process of tachyon condensation is an off-shell one. For 
example, at the point $R_1=2,~R_2=\frac{2\sqrt{3}}{3}$ the lowest mode on 
a $D2+\overline{D2}$ system 
becomes massless (see equation (\ref{lowestmass})) and it 
looks like we can use the marginal deformation. However, at this point the 
operator (\ref{eqn:wilson}) is not a {\sl truly marginal} operator and 
switching on this operator causes a renormalization group flow. 

The most subtle region in figure \ref{phase} is one which is surrounded 
between line (a) and (b). 
In this region any tachyonic modes do not exist on either $D2+\overline{D2}$ 
or $D0+\overline{D0}$. This fact indicates that both of 
the systems are {\sl perturbatively stable}. 
On the other hand line (c) determines 
which system is heavier, thus we expect that the stabler object is 
determined by its mass. In other words the heavier D-brane system is 
{\sl metastable} and {\sl nonperturbatively} decays to the lighter one by the 
quantum mechanical tunneling effect. 

Note that the argument of comparison of masses among various objects here is 
not an exact result because the masses of unstable systems obtain the quantum 
corrections as we have seen in eq.(\ref{masscorrection}). In this sense 
figure \ref{phase} is correct only in the weak coupling string theory.

\chapter{Conclusion and Future Problems}

\section{Conclusion}
\hspace*{4.5mm}
In this thesis we have considered the tachyon condensation on unstable D-brane 
systems in the flat space. Although various methods are known to analyze the 
tachyon condensation, 
in this thesis we have dealt with the boundary string field 
theory and the marginal deformation. 

The string field theory is a general method to analyze off-shell 
phenomena in string theory. Especially the boundary string field theory 
enables us to describe the tachyon condensation in the 
exact manner. Although we start with the Batalin-Vilkovisky formulation to 
construct it, this theory can simply be regarded as a natural extension 
of the sigma model calculation of the disk partition function. 

This theory also determines 
the action of its effective field theory for non-BPS 
systems. In 
superstring theory the action consists of two parts. One is 
the Dirac-Born-Infeld action 
and the other is the Wess-Zumino term (RR-coupling). 
If we assume some approximation 
scheme, we can write down the explicit form of the 
Dirac-Born-Infeld type action. 
Moreover we have found 
that the Wess-Zumino terms on brane-antibrane systems and non-BPS D-branes can 
exactly be written by the superconnections, which were 
known in the mathematics.
These results are consistent with K-theory arguments and the tachyon 
condensation.  
     
On the other hand in the special situation we can describe the tachyon 
condensation in the conformally invariant manner. 
In this thesis we have used the 
boundary state formalism to describe it. The boundary state formalism is 
convenient to examine the interaction between D-branes and closed strings, 
thus we were able to see how RR-charges of D-branes with lower dimension 
appear during the tachyon condensation. Moreover by changing 
radii of a torus we could understand the phase diagram about stability of 
various non-BPS systems.  

\section{Problems in Boundary String Field Theory}
\hspace*{4.5mm}
We think that the analyses in this thesis have given enough evidence to show 
various aspects of tachyon condensation. 
However, we have several problems which have not 
been solved. The first problem is about 
the relation between the cubic string field 
theory \cite{WiCSFT} (or Berkovits' supersymmetrized version \cite{Ber1}) and 
the boundary string field theory \cite{Wi2}. 
Although the appearance of these two theories is quite different, there 
should be some relationship between two as 
long as they describe the same physics. 
The naive expectation is that they are related to each other 
by a redefinition of target space fields. 
However, we do not know its explicit form. 

The next problem is how we should solve the nonrenormalizability 
of boundary string field theory. 
The calculations we did in chapter 3 
were restricted to those with only renormalizable terms included in the 
boundary part of the sigma model. 
One of principles of boundary string field theory 
is to include all possible terms in the boundary action, while we do not know 
how to calculate it. In this sense the cases where we can use boundary string 
theory are 
rather limited to special situations than cubic string field theory. 

The third problem is related to the second one. We have not yet understood why 
boundary string field theory gives exact answers to problems of tachyon 
condensation even though we do not include massive terms in the boundary 
action \cite{KuMaMo1,TaTeUe}. As we saw in section \ref{meaning} the point is 
whether all the one-point functions 
$\frac{\pa S}{\pa \lambda^i}|_{\lambda^i=0}$ of massive fields $\lambda^i$ 
become zero or not. If this is true we can admit that we have calculated 
the truly exact quantities, however we can not verify it. 
Anyway the reason why we could obtain the desirable results
might be related to the integrability of path integrals. In other words 
this might be because 
the renormalization group is closed in the algebraic sense. For 
example the RG flow does not induce any divergences in coefficients of 
nonrenormalizable terms if we only include renormalizable terms. In this 
case the theory is well-defined and we might be able to regard its answer as 
an exact one. 

The fourth problem is about the quantum theory of boundary string field 
theory. We do not know how to consistently incorporate quantum 
corrections in the sense of target space and how to 
apply it to tachyon condensation. 
The first naive approach is to regard the one-loop correction to 
the action $S$ of string field theory 
as a cylinder amplitude with non-conformal 
worldsheet action $I_B$. However, there are not 
any dualities which connect the 
cylinder with the annulus because the conformal invariance is broken 
\cite{CrKrLa,Ras}. For example, if we consider the general boundary state in 
eq.(\ref{geneb}) with an off-shell boundary action $I_B$, we can not assert 
that the interaction amplitude between two D-branes can be re-expressed as 
some kind of open string amplitude.
Moreover, there is not any support to this idea like the BV-formalism of 
classical boundary string field theory. 

Finally we have a question about the relation between the worldsheet SUSY and 
the target space SUSY. In boundary string field theory we have imposed the 
worldsheet SUSY in the boundary actions for a brane-antibrane 
system and a non-BPS D-brane. Although there do not exist any principles to 
impose the worldsheet SUSY, this description was successful as we saw in 
chapter \ref{OFFSHELL} even without the target space supersymmetry. 
It is mysterious.
 
\section{General Problems in the Tachyon Condensation}
\hspace*{4.5mm}
The first general question is how to include target space fermions in the 
action $S$ of string field theory. 
In boundary string field theory only 
bosonic target space fields appear in the boundary sigma model in RNS 
formalism. There is also this problem in Berkovits' string field theory 
\cite{Ber1}. 
Of course we can include them by constructing spin-fields from RNS variables, 
but this is not applicable to off-shell phenomena like tachyon 
condensation. A naive resolution to this problem is to use Green-Schwarz 
formalism \cite{GrSc} 
or pure-spinor formalism \cite{Ber4} instead of RNS formalism. 
However, we have not yet understood 
how to include the GSO-odd sector like a tachyon in the sigma 
model action, and we do not know what kind of principles determines the
explicit form of boundary action $I_B$. 
Indeed we can find that the insertion of the boundary 
fermions $\eta$ and $\bar{\eta}$ which appeared in eq.(\ref{eq3.5}) can not 
be used in these formulations.  

The next problem is to extend the analysis in the flat space into that in 
the curved space. 
The non-BPS systems and their dynamical phenomena in the curved space are 
also interesting. Although in general these are 
difficult to analyze, the orbifold is one of simple curved backgrounds which 
we can deal with by using the conformal field theory technique. 
Indeed we can argue the tachyon 
condensation by the on-shell description in the orbifold $T^4/{\bf Z}_2$ 
\cite{NaTaUe,BeGa1,bergman3,Gab1,sen20,mot}. 
The concrete calculation by using the boundary state is almost the same
as that in the flat space, except the subtlety coming from appearance of the 
twisted sector in the boundary state \cite{NaTaUe}. The physical 
difference appears in the fact that some non-BPS D1-branes become stable if 
we choose appropriate radii of the torus $T^4$, although they are always 
unstable in the flat space. We can prove this 
fact by examining its open string spectrum. Indeed modes with negative 
mass square are dropped by the orbifold projection which is inserted
in the open string partition function.  Moreover the non-BPS D1-brane can be 
interpreted as 
a BPS D2-brane wound on a nonsupersymmetric cycle of 
$T^4/{\bf Z}_2$ \cite{sen19,sen20,sen16}. Maybe this can be generalized into 
the argument in general K3 surfaces if we regard $T^4/{\bf Z}_2$ as the 
orbifold limit of it. 

How about the string field theory? Of course the analyses in curved space is 
difficult in both of the cubic string field theory 
(or Berkovits' supersymmetrized version) and the 
boundary string field theory except a few backgrounds like 
orbifolds \cite{Takayana2,Takayana1}. However, there 
is a theorem which states that 
the explicit form of tachyon potential does not depend on which closed 
string background we choose. This is called the universality of the tachyon 
potential \cite{sen23}.
What we need to do or would like to do is to find other universal 
properties. As we said in chapter \ref{OFFSHELL}, the original name of 
boundary string field theory was ``the background independent open string field
theory" \cite{Wi2} because bulk action (\ref{action0}) is background 
independent. Thus, the boundary string field theory might be suitable to 
seek for the universality of the tachyon condensation.  
  
The third problem is about the relation between the time of the target 
space and the worldsheet renormalization scale. As we can see in the 
boundary string field theory, when unstable D-brane systems decay there seems 
to be a correspondence between the flow of the time in the target space and 
the one-dimensional RG-flow in the worldsheet. Therefore, it might be 
important to consider time dependent backgrounds in string theory. 

The fourth problem is how we can understand nonperturbative instability of 
non-BPS systems. Until now any nonperturbative corrections of string field
theory have not been known. It might be captured as some nonperturvative 
corrections in the worldvolume field theory on D-branes\cite{Ha100}.

\section{Closed String Dynamics}
\hspace*{4.5mm}
A problem occurs when we pay attention to the vacuum without D-branes 
which appears after tachyon condensation on unstable D-brane systems. 
In boundary string field theory the stable vacuum appears at the point 
$|T|\rightarrow\infty$ and the total action like eq.(\ref{Z1}) vanishes there. 
This seems consistent with the fact that there are no open string degrees of 
freedom. On the other hand, if we expand
the action around\footnote{The expression ``$|T|=\infty$" is 
singular, thus we have to perform some field redefinition $T^{\prime}=f(T)$ 
to move the stable point from $|T|=\infty$ to $|T^{\prime}|=\mbox{finite}$.} 
$|T|=\infty$, some unknown fluctuation modes can appear. 
Indeed if we take the limit $|T|\rightarrow \infty$ after we move to the 
Hamiltonian formalism, the Hamiltonian remains nonzero \cite{GiHoYi}. 
This indicates that some unknown stuff is left 
at the stable vacuum. The recent idea to solve it is to consider them as 
closed strings. In other words, closed strings appear as bound states of 
classical open strings \cite{sen100,KlLaSh,GiHoYi,BeHoYi,Yi}. For the time 
being it is difficult to construct a closed string from open strings in the 
level of the {\sl classical} string field theory or its classical effective 
field theory. Anyway even 
if closed strings do not appear as bound states of classical open strings, we 
already know that they appear as intermediate states in one-loop open string 
diagrams. In this sense we can say that the open string field theory 
includes closed strings in the {\sl quantum} level, and we can expect
that open string field theory already knows the closed string theory 
\cite{Stro,Sred}. 
The open string field theory might be regarded as 
a nonperturbative definition of the {\sl total} string theory in the same way 
as the matrix model \cite{BFSS}. 

Anyway we would like to consider the closed string dynamics as a next step. 
Indeed if we put some non-BPS systems in the flat space, the D-branes 
themselves bend the space because of their masses. 
This is called the back reaction
of these D-branes, while such a closed string effect comes from the quantum 
level of open string field theory. 
If we understand the open string tachyon dynamics in the 
quantum level we might be able to use it as a model of evaporation of a 
blackhole or the cosmological evolution. As we said in chapter 3, the decay of 
non-BPS systems in the context of classical open string field theory has 
correspondence to the one-dimensional 
renormalization group flow in the worldsheet. In the same way we might relate 
the worldsheet renormalization group flow with the complete description of 
decay of these D-branes accompanying the radiation of closed strings. 
Moreover there might exist the dynamical 
version of holography and we might find some correspondence between the open 
string theory and closed string theory. 

This problem will be related to that of the closed string tachyon 
condensation in the theory with a closed string tachyon like type 0 theory. 
We have not yet understood the relation between the two-dimensional 
RG-flow and the dynamics of the target space, which are expected to be 
decay of the spacetime. Although the idea of this relation was inspired by the 
open string tachyon condensation which has been successful, this 
expectation might not always be applied to the closed string side: 

The first reason is that we do not know whether there is a global minimum in 
the tachyon potential of closed strings. Sen's conjecture of open string 
tachyon condensation \cite{sen13} was based on 
it, from which we can say that a theory with a tachyon is not 
inconsistent but only stays in a dynamical phase. Although that was correct 
in the open string theory, 
we do not know whether it is correct in the closed string theory also. 

The second reason comes from the fact that the off-shell 
target space theory of closed strings (closed string field theory) might
not be well-defined. The closed string theory represents the gravitation, 
which is not a usual field theory like gauge theories. Thus we 
might not be able to apply a naive field theoretical idea to gravitational 
theories like the closed string theory. For example we can imagine that the 
expectation value of the dilaton changes during off-shell gravitational 
phenomena, and this means that the string coupling is not a constant 
during those processes. In this case we might not be able to rely on the 
fundamental assumption that systems are described by 
a classical quantity plus its quantum correction. 
In other words we might have to start with a unknown nonperturbative 
quantity, and we might not be able to take the classical limit in off-shell 
gravitational phenomena. If that is the case the closed string field theory 
might become meaningless. 

The third reason is how we can define the quantity which decreases along the 
two-dimensional renormalization group flow. In open string theory this 
quantity is given by the g-function or the action of string field theory. In 
the target space we can interpret this as energy or mass of D-branes. On the 
other hand the c-function is known which decreases along the two-dimensional 
renormalization group flow corresponding to the g-function. This is equal to 
the central charge at conformal fixed points. The problem is that 
theories at IR fixed points become non-critical string theories since the 
values of the c-function at IR-fixed points are always smaller than those at 
UV fixed points. As long as we have not understood whether we can regard 
non-critical string theories as physically meaningful ones, we can not 
conclude that this idea is correct.
 
Moreover, even if we can accept this idea 
the difficult problem is how to define the notion of energy in the target 
space of closed 
string theory corresponding to the c-function in the worldsheet. 
In fact it is known that we can not define the universal notion of 
energy in the general relativity except a few backgrounds like ADM energy 
in the asymptotically flat background. 
Therefore, even if we would like to consider the closed string tachyon 
condensation, we can not define the energy in the universal way and it 
crucially depends on which background we choose \cite{Ada10,Harvey100,Gutp10}. 
In this way this problem comes from the fact that we have not understood the 
background independence of string theory. This is a difficult and important 
problem. 
      
One possible idea to avoid considering the problem of closed 
string tachyon condensation is to restrict consistent closed string 
backgrounds to on-shell ones. This idea comes from the holography 
\cite{tHo} in AdS/CFT correspondence \cite{Mal}. According to the 
holography, a non-local d-dimensional 
gravitational theory, which is difficult to deal with, has some 
correspondence with a 
(d-1)-dimensional field theory, which we know a lot. In this way we can 
interpret the closed string dynamics by using field theories. However, 
the holography is mysterious in the sense that we do not know where it comes 
from.

Anyway the serious problems of the closed string dynamics stem from 
the fact that we have not yet understood the two important 
notions-``the backreation" and ``the background 
independence". To fully solve these problems we might need to go beyond the 
perturbative string theory. Until now the string theory has been developed 
by using similar ideas which come from field theories, like supersymmetry or 
duality or soliton etc. However, we have come to the stage where we have to 
find physically new ideas which can not be seen in field theories. In this 
sense the holography might be one of important ingredients to go beyond old 
ideas and to understand the nonperturbative formulation of string theory or 
quantum gravity.  

\newpage

{\bf Acknowledgements}
\\\\
\hspace*{4.5mm}
I would like to thank Prof. Kazuo Fujikawa for a lot of advice and continuous 
encouragement. I am also grateful to Dr. Seiji Terashima, Dr. Michihiro Naka 
and Dr. Tadashi Takayanagi for fruitful collaborations and many useful 
discussions. Without them I could not have finished writing this manuscript.  
I am thankful to all the staffs and students in the high energy physics 
theory group of University of Tokyo for giving me the opportunity to study 
a lot of things and having discussions. 

The work of the author was supported by JSPS Research Fellowships for Young 
Scientists.

\appendix
\chapter{Conventions}

\section{Conformal Field Theory in the Flat Space Background}
\hspace*{4.5mm}
In this section we fix the convention about the conformal field theory in the 
flat space background. In this thesis we set $\alpha^{\prime}$ to $2$, and the 
convention is almost identical to that of \cite{Po1}. The sigma model 
action in the flat space is given by 
\beqa
\label{action1}
I_0=\frac{1}{4\pi}\int d^2z (\pa X^{\mu}\bar{\pa}X_{\mu}
-i\psi^{\mu}_{L}\bar{\pa}\psi_{L\mu}+i\psi^{\mu}_{R}
\pa\psi_{R\mu}).
\eeqa
Here we have already fixed the worldsheet gauge freedom by taking the 
conformal gauge, while we have omitted the terms with the ghosts 
$(b,c~;~\beta,\gamma)$ because in this thesis we can mostly do without 
considering the ghost contribution.  
\\\\
{\bf Operator product expansions (OPE)}
\\\\
\hspace*{4.5mm}
From this action we can obtain the operator product expansions (OPE) for the 
fields $(X^{\mu}(z,\bar{z}),~\psi^{\mu}_L(z),~\psi^{\mu}_R(\bar{z}))$
\beqa
\label{OPE1}
X^{\mu}(z)X^{\nu}(0)&\sim& -\delta^{\mu\nu}{\mbox{ln}}|z|^2,\no
\psi^{\mu}_L(z)\psi^{\nu}_L(0)&\sim& \delta^{\mu\nu}\frac{i}{z}~,~
\psi^{\mu}_R(\bar{z})\psi^{\nu}_R(0)\sim -\delta^{\mu\nu}\frac{i}{\bar{z}}.
\eeqa
Here we have defined $z=e^{-\sigma_2+i\sigma_1}$ as the radial plane 
coordinate of the world sheet and $w=\sigma_1+i\sigma_2$ as its cylindrical 
coordinate. 
Thus, $\sigma_1$ and $\sigma_2$ represent the worldsheet coordinates of space 
and time, respectively.  
\\\\\\\\
{\bf Mode expansions for closed strings}
\\\\ 
\hspace*{4.5mm}
Next let us list the closed string mode expansions of fields
$(X^{\mu},~\psi^{\mu}_L,~\psi^{\mu}_R)$:
\beqa
\label{modeexpand}
& &X^{\mu}_{L}(z)=x^{\mu}_L-i\A^{\mu}_0\ln z
+i\sum_{m\neq 0}\frac{\alpha^{\mu}_m}{mz^m},\no
& &X^{\mu}_{R}(\bar{z})=x^{\mu}_R-i\ti{\A}^{\mu}_0\ln \bar{z}
+i\sum_{m\neq 0}\frac{\ti{\alpha}^{\mu}_m}{m\bar{z}^m},\no
& &\psi^{\mu}_L(z)=i^{\frac{1}{2}}\sum_{r\in Z+\nu}
\frac{\psi^{\mu}_{r}}{z^{r+\frac{1}{2}}}~,~ \psi^{\mu}_R(\bar{z})
=i^{-\frac{1}{2}}\sum_{r\in Z+\nu}\frac{\ti{\psi}^{\mu}_{r}}
{\bar{z}^{r+\frac{1}{2}}}, 
\eeqa
where $X^{\mu}(z,\bar{z})=X^{\mu}_L(z)+X^{\mu}_R(\bar{z})~,
~p^{\mu}_L=\alpha^{\mu}_0~,~p^{\mu}_R=\ti{\alpha}^{\mu}_0$, and 
$\nu=\frac{1}{2}$ represents NS-sector and $\nu=0$ R-sector. The Virasoro 
generators $L_0$ and $\ti{L}_0$ can be written as
\beqa
\label{L0}
L_0=\frac{1}{2}\A^{\mu}_0\A_{\mu 0}+\sum^{\infty}_{m=1}\A^{\mu}_{-m}\A_{\mu m}
+\sum^{\infty}_{r\in {\bf Z}+\nu} r\psi^{\mu}_{-r}\psi_{\mu r}
+a,
\eeqa
where $a$ is the zero point energy which is equal to $-1/2~(0)$ for 
NS (R)-sector, and $\ti{L}_0$ can be obtained by replacing $\A^{\mu}_{m},
~\psi^{\mu}_r$ with $\ti{\A}^{\mu}_{m},~\ti{\psi}^{\mu}_r$, 
respectively. 

The subtle point is about the zero modes 
$(x^{\mu}_L,p_L^{\mu}~;~x^{\mu}_R,p_R^{\mu})$ of the bosonic fields 
$X^{\mu}(z,\bar{z})$. If we consider a noncompactified direction, 
the valuable $x^{\mu}$ defined by $x^{\mu}\equiv x^{\mu}_L+x^{\mu}_R$ 
represents the physical coordinate and the physical momentum $p^{\mu}$ is 
defined by $p^{\mu}=p^{\mu}_L=p^{\mu}_R$. By the canocial quantization 
these valuables have the usual quantization condition
\beqa
& &[x^{\mu},p^{\nu}]=i\eta^{\mu\nu},
\eeqa
where $\eta^{\mu\nu}=\mbox{diag}(-1,+1,+1,\cdots)$. 

On the other hand if we consider a compactified direction on $S^1$ with 
radius $R_{\mu}$, the crucial difference appears in the the momentum modes 
$p^{\mu}_L$ and $p^{\mu}_R$ which are defined by
\beqa
\label{discretep}
p^{\mu}_L~\equiv~\frac{n^{\mu}}{R_{\mu}}+\frac{w^{\mu}R_{\mu}}{2},\no
p^{\mu}_R~\equiv~\frac{n^{\mu}}{R_{\mu}}-\frac{w^{\mu}R_{\mu}}{2},
\eeqa
where $n^{\mu}$ and $w^{\mu}~$ are integer numbers which represent the 
discrete momentum and the winding number in that direction, respectively. 

The nonzero modes can be treated irrespective of whether we consider a 
compactified direction or a noncompactified direction. The OPE relations 
(\ref{OPE1}) are equivalent to the following (anti)commutation relations for 
modes :
\begin{eqnarray}
\label{com}
& &[\alpha^{\mu}_m,\alpha^{\nu}_n]=[\tilde{\alpha}^{\mu}_m,
\tilde{\alpha}^{\nu}_n]=m\delta_{m,-n}\eta^{\mu\nu},\no
& &\{\psi^{\mu}_r,\psi^{\nu}_s\}=
\{\tilde{\psi}^{\mu}_r,\tilde{\psi}^{\nu}_s\}=\delta_{r,-s}\eta^{\mu\nu}.
\end{eqnarray}
\\\\
{\bf Mode expansions for open strings}
\\\\
\hspace*{4.5mm}
In the open strings there is only left-sector, thus we can set 
$\ti{\A}^{\mu}_m,~\ti{\psi}^{\mu}_r,~\ti{L}_0$ equal to 
$\A^{\mu}_m,~\psi^{\mu}_r,~L_0$, respectively, in eq.(\ref{modeexpand}) and 
eq.(\ref{L0}). 
The only subtle point is that the relation between 
${\A}^{\mu}_0$ and $p^{\mu}$ is different from that of closed strings. It is 
given by 
\beqa
p^{\mu}\equiv \frac{1}{2}\A^{\mu}_0,
\eeqa
and note that if we consider a compactified direction there are no winding 
modes for open strings. 

\section{Notation for SO(1,9) Spinors and Some Formulas \label{Notation}}
\hspace*{4.5mm}
Here we summarize our notation for SO(1,9) spinors and after that we show the 
calculation which is needed for the derivation of the Wess-Zumino term in 
chapter \ref{OFFSHELL}.

The ten-dimensional $\Gamma$-matrices 
$\Gamma^{\mu}_A\ ^B,\ ( A,B=1,2,\ddd,32)$ 
are defined by the following Clifford algebra:
\ba
 \{ \Gamma^\mu,\Gamma^\nu \}=2\eta^{\mu\nu}.
\ea 
Note that we distinguish spinor and adjoint spinor indices as subscripts 
and superscripts, respectively. The charge conjugation matrix $C^{AB},
C^{-1}_{AB}$, which obeys the relation:
\ba
C\Gamma^\mu C^{-1}=-(\Gamma^\mu)^{T},
\ea
can raise or lower these spinor indices. Therefore we can omit the matrix $C$ as\ba
(\Gamma^\mu)_{AB}=C^{-1}_{BC}(\Gamma^{\mu})_A\ ^C, \ \ \
(\Gamma^\mu)^{AB}
=C^{AC}(\Gamma^{\mu})_C\ ^B.
\ea
Note also from the above equations it is easy to see
\ba
C_{AB}=-C_{BA},\ \ \ (\Gamma^\mu)_{AB}=(\Gamma^\mu)_{BA}.
\ea
We also define $(\Gamma_{11})_A\ ^B$ as
\ba
\Gamma_{11}=\Gamma_{0}\Gamma_{1}\ddd \Gamma_{9},
\ea
and the chirality projection matrix $P_{\pm}$ is defined by
\ba
P_{\pm}=\f{1}{2}(1\pm \Gamma_{11}).
\ea
The matrix $\Gamma_{11}$ satisfies the following identities
\ba
(\Gamma_{11})_{AB}=(\Gamma_{11})_{BA}, \ \ (\Gamma_{11})^2=1, \ \ \{\Gamma_{11},\Gamma^{\mu}\}=0.
\ea

Before we discuss the calculation of the Wess-Zumino terms, 
let us show some useful 
formulae. The first one is about the trace of $\Gamma$-matrices:
\ba
\mbox{Tr}[\Gamma^{\mu_0\mu_1\ddd\mu_p}\Gamma^{01\ddd p}]=
32\ (-1)^{\f{p(p+1)}{2}}\ \ep^{\mu_0\mu_1\ddd\mu_p}\ \ (0\le\mu_i\le p),
\label{trace1}
\ea
where $\Gamma^{\mu_0\mu_1\ddd\mu_p}=1/p_!(\Gamma^{\mu_0}\Gamma^{\mu_1}\dd\Gamma^{\mu_p}-\Gamma^{\mu_1}\Gamma^{\mu_0}\dd \Gamma^{\mu_p}+\ddd )$ denotes the antisymmetrized $\Gamma$-matrices.
The second one is the famous relation between the $\Gamma$-matrices and the 
differential forms. More explicitly, a $r$-form in the ten dimension:
\ba
C=\f{1}{r!}C_{\mu_1\mu_2\ddd\mu_r}dx^{\mu_1}dx^{\mu_2}\ddd dx^{\mu_r},\label{c1}
\ea
corresponds to the following $32\times 32$ matrix:
\ba
\hat{C}=\f{1}{r!}C_{\mu_1\mu_2\ddd\mu_r}\Gamma^{\mu_1\mu_2\ddd\mu_r}.\label{c2}
\ea
This correspondence preserve the multiplication as 
\ba
\label{c3}
:\hat{C_1}\hat{C_2}:=\f{1}{(r_1+r_2)!}(C_1\wedge C_2)_{\mu_1\mu_2\ddd\mu_{(r_1+r_2)}}\Gamma^{\mu_1\mu_2\ddd\mu_{(r_1+r_2)}},
\ea
where $:\ :$ denotes the antisymmetrization.

Let us turn to the derivation of the Wess-Zumino terms. It involves 
computations of correlation functions on a disk whose boundary is on a 
D$p$-brane. We assume its world-volume extends in the direction 
$x^0,x^1,\ddd,x^p$. It is easier to calculate the suitable correlation 
functions by performing T-duality transformation. 
This transformation is given with respect to the spin 
operators by
\ba
S_{LA}\ \to\ S_{LA} \ &,&\ S_{RA}\ \to\ M_{A}\ ^{B}S_{RB}, \no
M_{A}\ ^{B}&=&\left\{\begin{array}{c}
\pm i\Gamma^{0}\Gamma^{1}\ddd\Gamma^{p}\ \ \ (p=\mbox{even})    \\
\pm \Gamma^{0}\Gamma^{1}\ddd\Gamma^{p}\Gamma_{11}\ \ \ (p=\mbox{odd})
\end{array} \right .,
\ea
where $S_{LA}$ and $S_{RB}$ denote the left-moving and right-moving spin 
operators, respectively; the sign ambiguity $\pm$ depends on the convention 
and we choose the plus sign.
 The above rule can be derived by requiring that the OPEs of 
left-moving fermions $\psi^{\mu}_{L}$ and spin operators $S_{LA}$ have the same
structure as those of right-moving ones $\psi^\mu_R$ and $S_{RB}$
even after one performs the T-duality transformation. Using these facts, the
Wess-Zumino terms on both a non-BPS D-brane and a brane-antibrane system are summarized as the following form up to the overall normalization:
\ba
S=\sum_{r=0}^{p+1}\f{1}{r!}K_{\mu_1\mu_2\ddd\mu_r}\ \mbox{Tr}[\ P_{-}\hat{C}M\Gamma^{\mu_1\mu_2\ddd\mu_r}\ ],
\ea  
where $K_{\mu_1\mu_2\ddd\mu_r}\ \ (0\le \mu_i \le p)$ is a $r$-form which depends on the gauge field-strength
 and the tachyon field; $\hat{C}$ denotes the RR-sector vertex as
defined in eq.(\ref{rr0}). If one takes transverse scalars into
account, one should also discuss $K_{\mu_1\mu_2\ddd\mu_r}$ for $\mu_i
\ge p$. However such a case can also be treated similarly and we omit this.

Now we can write down the Wess-Zumino term explicitly up to the overall 
normalization which is independent of $r$ as follows:
\ba
S=\sum_{q,r=0}^{p+1}\f{1}{q!r!}\ \delta_{p+1,q+r}\ \ep^{\mu_1\ddd\mu_q\nu_1\ddd\nu_r}C_{\mu_1\ddd\mu_q}K_{\nu_1\nu_2\ddd\nu_r},
\ea
where we have used the formula (\ref{trace1}). Finally these couplings are 
written in the language of the differential forms as
\ba
S=\int_{\Sigma_{(p+1)}} C\wedge K,
\ea
where $\Sigma_{(p+1)}$ denotes the world volume of the D$p$-brane.

\chapter{The General Expression of the Disk Partition Function 
\label{Expression}}
\hspace*{4.5mm}
Here we will give the general expression of the disk partition function which 
we use in the calculations of boundary string field theory (BSFT). First 
the disk partition function $Z$ is defined by
\begin{eqnarray}
\label{eq5}
Z &=&
\int [DX D\psi_L D\psi_R] \exp[-I_0(X,\psi_L,\psi_R)-I_B(X,\psi)],
\end{eqnarray}
where $I_0$ is given by eq.(\ref{action0}) and $I_B$ depends on which kind of 
D-brane we would like to consider. It is given by eq.(\ref{BPSboundary}), 
eq.(\ref{eq3}) and eq.(\ref{eq1}) for a BPS D$9$-brane, a $D9+\overline{D9}$ 
system and a non-BPS D$9$-brane, respectively.

In this thesis we consider only the flat space as a closed string background. 
Here the term ``flat" means that we set the target space metric $G_{\mu\nu}$ 
and the antisymmetric tensor $B_{\mu\nu}$ to some constant values. 
In this case 
the complicated expression of the action $I_0$ in eq.(\ref{action0}) becomes 
a simpler form given by
\beqa
I_0&=&\frac{1}{4\pi}\int d^2z\Bigl[(G_{\mu\nu}+B_{\mu\nu})\pa X^{\mu}
\bar{\pa}X^{\nu}\no
&&~~~~~~~~~~~~~~-i(G_{\mu\nu}+B_{\mu\nu})\psi^{\mu}_L\bar{\pa}\psi^{\nu}_L
+i(G_{\mu\nu}-B_{\mu\nu})\psi^{\mu}_R\pa\psi^{\nu}_R\Bigr].
\eeqa
Here by considering the flat background more than third powers of the fields 
$X^{\mu}$ and $\psi_{L,R}$ disappear from the action and we can see that the 
path integral inside the disk turns out 
to be calculable by an Gaussian integral.

Now if we write the boundary values of the fields $X^{\mu}$ and 
$\psi_{L}+\psi_{R}$ as $X^{\mu}_b$ and $\psi_b$, we can rewrite the path 
integral (\ref{eq5}) as
\beqa
\label{split}
Z &=&\int [DX_b D\psi_b]\Psi[X_b,\psi_b]\exp[-I_B(X_b,\psi_b)],
\eeqa
where 
\beqa
\label{bulkpath}
\Psi[X_b,\psi_b]=\int [DX D\psi_L D\psi_R]\exp[-I_0(X,\psi_L,\psi_R)].
\eeqa
The last path integral 
is evaluated inside the disk with the boundary condition 
$X^{\mu}|_{\ss boundary}=X^{\mu}_b,
~(\psi^{\mu}_L+\psi^{\mu}_R)|_{\ss boundary}=\psi^{\mu}_b$. 

$\Psi[X_b,\psi_b]$ is calculated in the following way. 
Here we will show only the path integral of $X^{\mu}$ explicitly, while it is 
almost\footnote{There is a little subtle point. For example see the paper 
\cite{LaWi}.} the same 
procedure about the path integral of the fermions $\psi^{\mu}_L$ and 
$\psi^{\mu}_R$. First because of 
the periodicity $X^{\mu}_b(\tau+2\pi)=X^{\mu}_b(\tau)$ 
of the boundary field we can obtain a Fourier series
\beqa
\label{clbst}
X^{\mu}_b(\tau)=\sum_{m\in{\bf Z}}X^{\mu}_m e^{im\tau},
\eeqa
with its reality condition $(X^{\mu}_m)^*=X^{\mu}_{-m}$. Next, we split the 
valuable $X^{\mu}$ in the path integral into the classical part 
$(X_{cl}^{\mu})$ and the quantum part $(X^{\prime\mu})$ like
\beqa
X^{\mu}&=&X^{\mu}_{cl}+X^{\prime\mu},\no
X^{\mu}_{cl}&=&X^{\mu}_0
+\sum_{m=1}^{\infty}(z^mX^{\mu}_m+\bar{z}^mX^{\mu}_{-m}).
\eeqa
Here the classical part $X^{\mu}_{cl}$ is the solution to the equation of 
motion $\pa\bar{\pa}X^{\mu}=0$ with the boundary condition (\ref{clbst}). 
Inserting this into the path integral (\ref{bulkpath}) it becomes
\beq
\begin{array}{l}
{\displaystyle \exp\left[-\frac{G_{\mu\nu}-B_{\mu\nu}}{2}
\sum_{m=1}^{\infty}mX^{\mu}_{-m} X^{\nu}_m\right]}\no
\end{array}
\\
\times\int DX^{\prime}|_{X^{\prime}_b=0}\exp\left[-\frac{1}{4\pi}\int dz^2 
(G_{\mu\nu}+B_{\mu\nu})\pa X^{\prime\mu}\bar{\pa}X^{\prime\nu}\right].
\eeq
Here the dependence of the boundary value $X^{\mu}_m$ is included only in the 
first exponential and we can ignore the second exponential as an irrelevant 
overall normalization. Therefore, after rewriting this by 
$X^{\mu}_b(\tau)$ its final result is given by
\beqa
\label{functional}
\Psi[X_b,\psi_b]&\propto&\exp\left[\left(G_{\mu\nu}-B_{\mu\nu}\right)
\left(-\sum^{\infty}_{m=1}\frac{m}{2}X^{\mu}_{-m}X^{\nu}_m-2i
\sum^{\infty}_{r=1/2}\psi^{\mu}_{-r}\psi^{\nu}_r\right)\right]\no
&=&\exp\Biggl[-\frac{1}{8\pi}\int d\tau_1 \int d\tau_2 
X^{\mu}_b(\tau_1)X^{\nu}_b(\tau_2)K^{-1}_{\mu\nu}(\tau_1,\tau_2)\no
&&~~~~~~~~-\frac{1}{2\pi}\int d\tau_1 \int d\tau_2 \psi^{\mu}_b(\tau_1)
\psi^{\nu}_b(\tau_2)L^{-1}_{\mu\nu}(\tau_1,\tau_2)\Biggr],
\eeqa
where
\beqa
\label{inverseprop}
K^{-1}_{\mu\nu}(\tau_1,\tau_2)&=&\frac{1}{2\pi}(G_{\mu\nu}-B_{\mu\nu})
\sum^{\infty}_{m=1}me^{im(\tau_1-\tau_2)}+
\frac{1}{2\pi}(G_{\mu\nu}+B_{\mu\nu})
\sum^{\infty}_{m=1}me^{-im(\tau_1-\tau_2)},\no
L^{-1}_{\mu\nu}(\tau_1,\tau_2)&=&\frac{i}{2\pi}(G_{\mu\nu}-B_{\mu\nu})
\sum^{\infty}_{r=\frac{1}{2}}e^{ir(\tau_1-\tau_2)}-
\frac{i}{2\pi}(G_{\mu\nu}+B_{\mu\nu})
\sum^{\infty}_{r=\frac{1}{2}}e^{-ir(\tau_1-\tau_2)}.\no
\eeqa
Here we have included the result from the path integral of the fermions 
$\psi^{\mu}_L$ and $\psi^{\mu}_R$. This is the most general expression which 
is useful in this thesis. 

By the way, from this expression we can prove the equivalence of the internal 
product $\la 0|B\lb_{\ss NSNS}$ and the disk partition function, which 
appeared in eq.(\ref{555}). If we compute the former quantity by using the 
general expression of $|B\lb$ in eq.(\ref{geneb}), it becomes as
\beqa
\la 0|B\lb_{\ss NSNS}=\int {\cal D}x{\cal D}\varphi~ e^{-I_B(x,\varphi)}
\exp\left[-\sum^{\infty}_{m=1}\frac{m}{2}x^{\mu}_{-m}x_{m\mu}
-\sum^{\infty}_{r=1/2}\frac{1}{2}\varphi^{\mu}_{-r}\varphi_{r\mu}\right],
\eeqa
and we can see that this expression is completely equivalent to 
eq.(\ref{split}) and eq.(\ref{functional}) if we rescale $\psi^{\mu}_{r}$ 
as $\frac{i^{-\frac{1}{2}}}{2}\varphi^{\mu}_r$.

Now, in the rest of this appendix we will derive the Dirac-Born-Infeld 
action for a BPS D$9$-brane \cite{FrTs}. 
The calculation of eq.(\ref{split}) is possible if we consider the 
constant field strength $F_{\mu\nu}$ in the boundary sigma model action 
(\ref{BPSboundary}) and set all massive fields ($\cdots$ part in 
eq.(\ref{BPSboundary})) to zero. In this case the boundary action $I_B$ is 
given by 
\beqa
I_B=-i\int d\tau \left[-\frac{1}{2}X^{\mu}_b\dot{X}^{\nu}_bF_{\mu\nu}
-2\psi^{\mu}_b\psi^{\nu}_bF_{\mu\nu}\right].
\eeqa
{}From this we can see that by inserting this action into eq.(\ref{split}) and 
combining it with eq.(\ref{functional}), the path integral (\ref{split}) is 
equivalent to the description of replacing $B_{\mu\nu}$ with 
$\ti{B}_{\mu\nu}\equiv B_{\mu\nu}-4\pi F_{\mu\nu}$ in 
eq.(\ref{inverseprop}) without any boundary terms. Indeed this combination 
satisfies the desired gauge invariance : 
$B_{\mu\nu}\rightarrow B_{\mu\nu}+\pa_{\mu}\zeta_{\nu}-\pa_{\nu}\zeta_{\mu},~
A_{\mu}\rightarrow A_{\mu}+\frac{1}{4\pi}\zeta_{\mu}$. 

Now the calculation of eq.(\ref{split}) with eq.(\ref{functional}) is easy 
because it is a simple Gaussian path integral given by
\beqa
Z&=&\int [DX_b D\psi_b] \Psi[X_b,\psi_b]\no
&\propto&\int d^{10}x [\det~K^{-1}_{\mu\nu}(\tau_1,\tau_2)]^{-\frac{1}{2}}
[\det~L^{-1}_{\mu\nu}(\tau_1,\tau_2)]^{+\frac{1}{2}}. 
\eeqa 
By diagonizing the matrices $K^{-1}(\tau_1,\tau_2)$ and 
$L^{-1}(\tau_1,\tau_2)$ we 
can rewrite the above expression in the following way:
\beqa
Z&\propto&\int d^{10}x
\left(\frac{\prod^{\infty}_{r=\frac{1}{2}}
\det~(G_{\mu\nu}-\ti{B}_{\mu\nu})~
\prod^{\infty}_{r=\frac{1}{2}}\det~(G_{\mu\nu}+\ti{B}_{\mu\nu})}
{\prod^{\infty}_{m=1}\det~(G_{\mu\nu}-\ti{B}_{\mu\nu})m~
\prod^{\infty}_{m=1}\det~(G_{\mu\nu}+\ti{B}_{\mu\nu})m}\right)^{\frac{1}{2}}
\no
&=&\int d^{10}x 
\left(\frac{\prod^{\infty}_{r=\frac{1}{2}}\det~(G_{\mu\nu}-\ti{B}_{\mu\nu})}
{\prod^{\infty}_{m=1}\det~(G_{\mu\nu}-\ti{B}_{\mu\nu})m}\right)\no
&=&\int d^{10}x 
\exp\left[-\sum^{\infty}_{m=1}\ln\{\det~(G_{\mu\nu}-\ti{B}_{\mu\nu})m\}
+\sum^{\infty}_{r=\frac{1}{2}}\ln\{\det~(G_{\mu\nu}-\ti{B}_{\mu\nu})\}\right]
\no
&\propto&\int d^{10}x
\exp\left[\ln\{\det(G_{\mu\nu}-\ti{B}_{\mu\nu})\}
\left(-\sum^{\infty}_{m=1}e^{-2\epsilon m}+\sum^{\infty}_{r=\frac{1}{2}} 
e^{-2\epsilon r}\right)\right].
\eeqa
Here the summation in the fourth line obviously diverges, thus we have 
put the dumping factors $e^{-2\epsilon m}$ and $e^{-2\epsilon r}$.
By using the relation\footnote{Here the cancellation of the divergence 
with $\epsilon\rightarrow 0$ comes from the worldsheet supersymmetry.} 
\beqa
-\sum^{\infty}_{m=1}e^{-2\epsilon m}+\sum^{\infty}_{r=\frac{1}{2}}
e^{-2\epsilon r}=\frac{1}{2}+{\cal O}(\epsilon), 
\eeqa
we can identify the disk partition function with the Dirac-Born-Infeld action
\beqa
Z\propto \int d^{10}x\sqrt{-\det(G_{\mu\nu}-B_{\mu\nu}+4\pi F_{\mu\nu})}.
\eeqa

\end{document}